\documentclass[12pt]{article}
\pdfoutput=1
\usepackage{comment}
\usepackage{amsmath}
\usepackage{amssymb}
\usepackage{nccmath}
\usepackage{graphicx}
\usepackage{animate}
\usepackage{here}
\usepackage{subcaption}
\usepackage{url}
\usepackage[sort&compress, numbers, merge]{natbib}
\usepackage{braket}
\usepackage{physics}
\usepackage{tikz}
\usepackage[compat=1.1.0]{tikz-feynhand}
\usepackage{slashed}
\usepackage{mathtools}
\usepackage{bbold}
\usepackage{pgfplots}
\usepackage{tcolorbox}
\usepackage{bm}
\usepackage{bbm}
\usepackage{standalone}

\tcbuselibrary{raster,skins}
\tcbuselibrary{theorems}

\usepackage{pgfplots}  
\usetikzlibrary{arrows.meta,arrows,angles,calc,perspective} 
\usetikzlibrary{shapes.geometric}

\usepackage{tikz-3dplot}
 \tdplotsetmaincoords{60}{115}
\pgfplotsset{compat=newest}

\usetikzlibrary{3d}
\usetikzlibrary{calc}
\pgfplotsset{compat=newest}

\usetikzlibrary{decorations.markings}

\usepackage{todonotes}
\newcommand{\SU}{\mathrm{SU}}

\def\SU{\mathrm{SU}}

\def\bartau{\bar{\tau}}

\def\boldtau{\boldsymbol{\tau}}

\def\simgt{\mathrel{\lower2.5pt\vbox{\lineskip=0pt\baselineskip=0pt
           \hbox{$>$}\hbox{$\sim$}}}}
\def\simlt{\mathrel{\lower2.5pt\vbox{\lineskip=0pt\baselineskip=0pt
           \hbox{$<$}\hbox{$\sim$}}}}
\def\simprop{\mathrel{\lower3.0pt\vbox{\lineskip=1.0pt\baselineskip=0pt
             \hbox{$\propto$}\hbox{$\sim$}}}}

\newcommand{\version}{arXiv2}
\DeclareRobustCommand{\modifiedat}[2]{%
\ifthenelse{\equal{\version}{#1}}{\textcolor{red}{#2}}{#2}%
}

\numberwithin{equation}{section}

\usepackage[colorlinks=true, linkcolor=blue, citecolor=blue,
urlcolor=black]{hyperref} 
\usepackage[capitalize]{cleveref}
\crefformat{section}{Sec.\,#2#1#3}
\Crefformat{section}{Section~#2#1#3}
\crefformat{equation}{Eq.\,(#2#1#3)}
\Crefformat{equation}{Equation~(#2#1#3)}
\crefformat{figure}{Fig.\,#2#1#3}
\Crefformat{figure}{Figure~#2#1#3}

\usepackage{diagbox}

\begin{document}
\def\ps{\mathbf{p}}
\def\PS{\mathbf{P}}
\baselineskip 0.6cm
\def\simgt{\mathrel{\lower2.5pt\vbox{\lineskip=0pt\baselineskip=0pt
           \hbox{$>$}\hbox{$\sim$}}}}
\def\simlt{\mathrel{\lower2.5pt\vbox{\lineskip=0pt\baselineskip=0pt
           \hbox{$<$}\hbox{$\sim$}}}}
\def\simprop{\mathrel{\lower3.0pt\vbox{\lineskip=1.0pt\baselineskip=0pt
             \hbox{$\propto$}\hbox{$\sim$}}}}
\def\tr{\mathop{\rm tr}}
\def\SU{\mathop{\rm SU}}

\def\uoneG{\mathrm{U}(1)_{G}}
\def\VG{V}
\def\VH{v}
\def\azimuth{\theta}
\newcommand{\az}{\azimuth}
\def\zenith{\theta_s}
\def\higgsprofile{\xi}
\def\Newton{G_{\mathrm{N}}}

\begin{titlepage}

\begin{flushright}
IPMU26-0012
\end{flushright}

\vskip 1.1cm

{\centering
{\Large \bf 
A Closer Look at Constrained Instantons
}

\vskip 1.2cm
Takafumi Aoki$^{a}$, 
Masahiro Ibe$^{a,b}$ and
Satoshi Shirai$^{b}$
\vskip 0.5cm

{\it

$^a$ {ICRR, The University of Tokyo, Kashiwa, Chiba 277-8582, Japan}

$^b$ {Kavli Institute for the Physics and Mathematics of the Universe
 (WPI), \\The University of Tokyo Institutes for Advanced Study, \\ The
 University of Tokyo, Kashiwa 277-8583, Japan}

}
}

\begin{abstract}
Instantons play a crucial role in understanding non-perturbative dynamics in quantum field theories, including those with spontaneously broken gauge symmetries. In the broken phase, finite-size instanton-like configurations are no longer exact stationary points of the Euclidean action, in contrast to the symmetric phase. Non-perturbative effects in this setting are therefore typically studied within the constrained instanton framework. However, a previous study pointed out a possible difficulty in constructing consistent constrained instanton solutions based on conventional gauge-invariant constraints. 
In this work, we revisit the asymptotic structure of constrained instantons and re-examine the claimed difficulty. By carefully tracking the behavior of the solutions near the spatial origin and at infinity, we show that the required boundary conditions can be satisfied without encountering the inconsistency. We explicitly construct consistent constrained instantons in both massive $\phi^4$
theory and Yang--Mills theory with spontaneous symmetry breaking, and we support our analytic matching procedure with numerical solutions. 
Our results establish that conventional gauge-invariant constraints can be consistently employed in semiclassical computations when asymptotic expansions are treated properly.
\end{abstract}
\end{titlepage}
\newpage

\tableofcontents

\section{Introduction}
\label{sec:intro}

Instantons
have played a central role in the analysis of quantum field theories~\cite{Belavin:1975fg,tHooft:1976snw}, providing a semiclassical understanding of the dynamics beyond perturbative analysis.
For example, instantons help to clarify the
nontrivial vacuum structure of QCD, most notably the $\theta$-dependence and the $\mathrm{U}(1)_A$ anomaly~\cite{tHooft:1976rip,Callan:1976je,Callan:1977gz}.
More generally, instantons are also essential for understanding strong gauge dynamics~\cite{Novikov:1983ee,Affleck:1984xz,Novikov:1985rd,Seiberg:1994bz,Seiberg:1994pq}.

In many of the applications mentioned above, instantons are considered in gauge theories with spontaneously broken gauge symmetries.
When the symmetry is spontaneously broken, however, finite-size instanton solutions are no longer minima of the Euclidean action.
Nevertheless, non-perturbative effects remain important even in such settings.
To analyze them semiclassically, the framework of constrained instantons~\cite{Affleck:1980mp} has been widely employed, for instance, in computations of baryon- and lepton-number violating rates in the electroweak theory~\cite{Ringwald:1989ee,Espinosa:1989qn,Arnold:1990pe,Khoze:1990bm,Silvestrov:1992ct}.
In addition, the relevance of instantons from gauge sectors in the broken phase to axion properties has been investigated in a variety of models~\cite{Elliott:1992ut,Redi:2016esr,Agrawal:2017ksf,Agrawal:2017evu,Gaillard:2018xgk,Csaki:2019vte,Aoki:2024usv}.

The constrained instanton framework is designed to obtain extrema of the Euclidean
action subject to a constraint that fixes the instanton size~\cite{Frishman:1978xs,Affleck:1980mp}. 
The properties of constrained instantons and related instanton-like objects in similar frameworks have been studied extensively in the literature, and the subject remains an active area of ongoing interest~\cite{Klinkhamer:1991pq,Klinkhamer:1993kn,Aoyama:1995zt,Aoyama:1995ca,Nielsen:1999vq,Elder:2025kya}.
Nielsen and Nielsen (N\&N)~\cite{Nielsen:1999vq} investigated the detailed properties of constrained instantons. They examined the asymptotic conditions required for the finiteness of the action and pointed out a possible obstruction to constructing a consistent solution. They further argued that this issue severely restricts the allowed form of the constraint and may exclude the conventional gauge-invariant choices commonly used in the literature. As explicit gauge-dependent examples, N\&N identified viable constraints that satisfy the required asymptotic conditions.

Despite these observations, conventional gauge-invariant constraints have nevertheless remained common in the literature (see, e.g., Ref.~\cite{Vandoren:2008xg} for a review), suggesting that the asymptotic structure of constrained instantons merits closer examination.
In this work, we revisit the asymptotic structure of constrained instanton solutions and re-examine the origin of the issue raised by N\&N. 
By carefully tracking the behavior of the solution both near the spatial origin and at infinity, we show that the required asymptotic conditions can in fact be satisfied once the asymptotic expansions are treated properly. Consequently, conventional gauge-invariant constraints can be employed consistently in the construction of constrained instantons.

The organization of the paper is as follows.
In Sec.\,\ref{sec: phi4 case},
we revisit the constrained instanton in $\phi^4$ theory
and analyze its asymptotic behavior as a preparation for the case of Yang--Mills theory with spontaneous symmetry breaking.
In Sec.\,\ref{sec: revisit YM constrained instanton}, we review the instanton configuration in Yang--Mills theory with spontaneous symmetry breaking and explain how to
incorporate the constraint into the semiclassical computation.
In Sec.\,\ref{sec: explicit construction in YM}, we study the asymptotic behavior of the constrained instanton solution in detail and compare the analytic construction with numerical solutions.  
We summarize our findings in the final section.

\section{Prelude: Constrained Instanton in \texorpdfstring{$\boldsymbol{\phi^4}$}{} Theory}
\label{sec: phi4 case}

Before turning to spontaneously broken gauge theories, we first examine the constrained instanton appearing in the semiclassical approximation of scalar $\phi^4$ theory. This model provides a particularly transparent setting in which the essential features of constrained instantons can be understood. Following the foundational work of Affleck~\cite{Affleck:1980mp}, the constrained instanton was first formulated in the context of this $\phi^4$ theory. Moreover, the criticisms raised by N\&N~\cite{Nielsen:1999vq} concerning the asymptotic inconsistency of constrained instantons were also directed at this $\phi^4$ setup. By carefully addressing these issues in this simpler framework, we can extract the key physical and mathematical lessons. With this preparation, we will then proceed to spontaneously broken gauge theories in the subsequent sections.

\subsection{Instanton and Constrained Instanton}

We consider the instanton, which is the classical solution, in a scalar field theory in four-dimensional Euclidean space with a negative quartic interaction. The Euclidean Lagrangian density is
\begin{align}
    \mathcal{L}
    =
    \frac{1}{g_\phi}
    \left[
        \frac{1}{2}(\partial_\mu \phi)^2
        + \frac{m^2}{2}\phi^2
        - \frac{1}{4!}\phi^4
    \right] \,,
\end{align}
where $g_\phi>0$ is the coupling constant and $m>0$ is the mass parameter. The quartic coupling $g_\phi^2$ has been scaled out by a field redefinition.
Implicit summation of the indices is understood.
See Appendix~\ref{app:notations} 
for our notation in Euclidean space.
We focus on the $\mathrm{O}(4)$-symmetric configuration, $\phi = \phi(r)$ with $r = |x|$. The Euler--Lagrange equation then becomes
\begin{align}
    \phi''(r) + \frac{3}{r}\phi'(r) - V'(\phi) = 0 \,,
    \qquad
    V(\phi) = \frac{m^2}{2}\phi^2 - \frac{1}{4!}\phi^4 \,.
    \label{eq: Euler--Lagrange eq in phi4}
\end{align}
Here, primes denote derivatives with respect to the argument of each function; in particular, $\phi'(r)=\dd\phi/\dd r$ and $V'(\phi)=\dd V/\dd\phi$.

In the massless case $m=0$, 
Eq.\,\eqref{eq: Euler--Lagrange eq in phi4} admits
the finite-action solution under the boundary conditions,
\begin{align}
    \phi(r=0)=\phi_\mathrm{ini} \ ,  \quad \phi'(r=0)  = 0 \ ,
    \label{eq: phi4 BC}
\end{align}
which is the instanton,
\begin{align}
\label{eq:phi instanton}
    \phi_0(r) = \frac{4\sqrt{3}\rho}{r^2+\rho^2} \,,
    \quad
    \rho=\frac{4\sqrt{3}}{\phi_\mathrm{ini}}\,.
\end{align}
Here, the parameter $\rho$ characterizes the size of the instanton. The corresponding action is
\begin{align}
\label{eq: phi4 instanton action}
    S = \int \mathrm{d}^4x\,\mathcal{L} = \frac{16\pi^2}{g_\phi} \,.
\end{align}
Since the action is independent of $\rho$, the associated scale mode is a zero mode.

\begin{figure}[t]
    \centering
    \begin{subfigure}{0.48\textwidth}
        \centering
        \begin{tikzpicture}
  \begin{axis}[
    width=9cm,height=5.2cm,
    axis lines=middle,
    axis line style={semithick},
    xmin=-3.2,xmax=3.2,
    ymin=-0.6,ymax=3.2,
    ticks=none,
    xlabel={$\,\phi$},
    ylabel={$-V$},
    xlabel style={at={(xticklabel* cs:1)},anchor=west},
    ylabel style={at={(yticklabel* cs:1)},anchor=west},
    clip=false
  ]
    \pgfmathsetmacro{\xstart}{2.5}
    \addplot[domain=-2.8:2.8, samples=400, very thick]
      {(1/24)*x^4};
      \addplot[domain=0:\xstart, samples=400, very thick,<-,blue]
      {(1/24)*(x+.1)^4+.15};
    \addplot[mark=*,only marks,mark size=0.8pt] coordinates {(0,0)};
    \addplot+[only marks,mark=*,mark size=2.2pt, mark options={fill=blue,draw=blue}]
      coordinates {(\xstart,{(1/24)*(\xstart+.1)^4+.15})}
      node[left,align=left,yshift=10pt]{\small start \\[-3pt] $r=0$}; 
      \node[anchor=west , align=left] at (axis cs:-0.0,0.7) {\small end \\[-3pt] $r=\infty$}; 
  \end{axis}
\end{tikzpicture}
        \caption{$V = -\dfrac{1}{4!}\phi^4$}
        \label{fig:pot1}
    \end{subfigure}
    \hfill
    \begin{subfigure}{0.48\textwidth}
        \centering
        \begin{tikzpicture}
  \begin{axis}[
    width=9cm,height=5.2cm,
    axis lines=middle,
    axis line style={semithick},
    xmin=-5,xmax=5,
    ymin=-2.2,ymax=3.2,
    ticks=none,
    xlabel={$\,\phi$},
    ylabel={$-V$},
    xlabel style={at={(xticklabel* cs:1)},anchor=west},
    ylabel style={at={(yticklabel* cs:1)},anchor=west},
    clip=false,
    declare function={mu=1.0;}
  ]
    \pgfmathsetmacro{\xstart}{3.7}
    \addplot[domain=-4:4, samples=400, very thick]
      {-(0.5*mu^2)*x^2 + (1/24)*x^4};
      \addplot[domain=1.5:3.7, samples=400, very thick,<-,blue]
      {-1.2*(0.5*mu^2)*(x+.1)^2 + 1.2*(1/24)*(x+.1)^4+.6};
      \addplot[domain=1.9:2.9, samples=400, very thick,<->]
      {(mu^2)*(x-2.4)^2 -2};
    \addplot[mark=*,only marks,mark size=0.8pt] coordinates {(0,0)};
    \node[anchor=center, align=left] at (axis cs:2.4,-3) {\small damped \\ oscillation}; 
    \addplot+[only marks,mark=*,mark size=2.2pt, mark options={fill=blue,draw=blue}]
      coordinates {(\xstart,{-1.2*(0.5*mu^2)*(\xstart+.1)^2 + 1.2*(1/24)*(\xstart+.1)^4+.5})} 
      node[left,align=left,yshift=10pt,black]{\small start \\[-3pt] $r=0$};
  \end{axis}
\end{tikzpicture}
        \caption{$V = \dfrac{m^2}{2}\phi^2 - \dfrac{1}{4!}\phi^4$}
        \label{fig:pot2}
    \end{subfigure}

\caption{
(Left) Particle-mechanics interpretation of the radial equation of motion in the inverted potential $-V(\phi)$. With initial conditions $\phi(0)$ arbitrary and $\phi'(0)=0$, the trajectory monotonically approaches $\phi\to 0$ as $r\to\infty$ in the massless theory, corresponding to the instanton solution.
(Right) Adding a mass term eliminates the finite-action instanton solution.
For any arbitrary initial value $\phi(0)$ with $\phi'(0)=0$, the motion results in damped oscillations.
}
    \label{fig:pot12}
\end{figure}
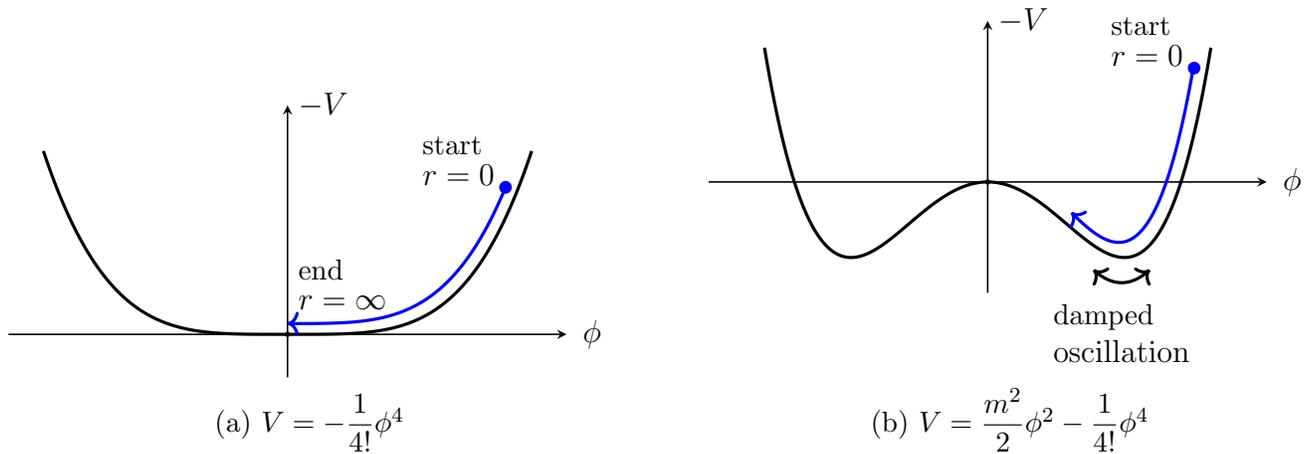

In the discussion below,
we rely on an observation that Eq.\,\eqref{eq: Euler--Lagrange eq in phi4}
can be interpreted as the motion of a point particle in the inverted potential $-V(\phi)$ with a time-dependent friction term $(3/r)\phi'$, upon identifying $r$ with time. 
In the massless case, the instanton corresponds to a trajectory that starts at an arbitrary initial height, $\phi(0)=4\sqrt{3}\rho^{-1}$, with $\phi'(0)=0$ at $r=0$. The friction term then drives the solution $\phi(r)$ to asymptotically approach zero as $r\to\infty$ 
(Fig.\,\ref{fig:pot1}).

In the massive case, by contrast, no finite-action solution with $\phi(0)\neq0$ and $\phi'(0)=0$ exists.
The existence of the instanton solution at $m^2=0$ also implies that, for $m^2>0$, no choice of initial height with $\phi'(0)=0$ can reach $\phi=0$ at infinity, since the mass term provides an outward force that prevents $\phi$ from reaching zero.
This solution asymptotically oscillates around the vacuum as $r\to\infty$,
which eventually leads to an infinite action.
This argument shows that no instanton solution exists in this case 
(Fig.\,\ref{fig:pot2}).

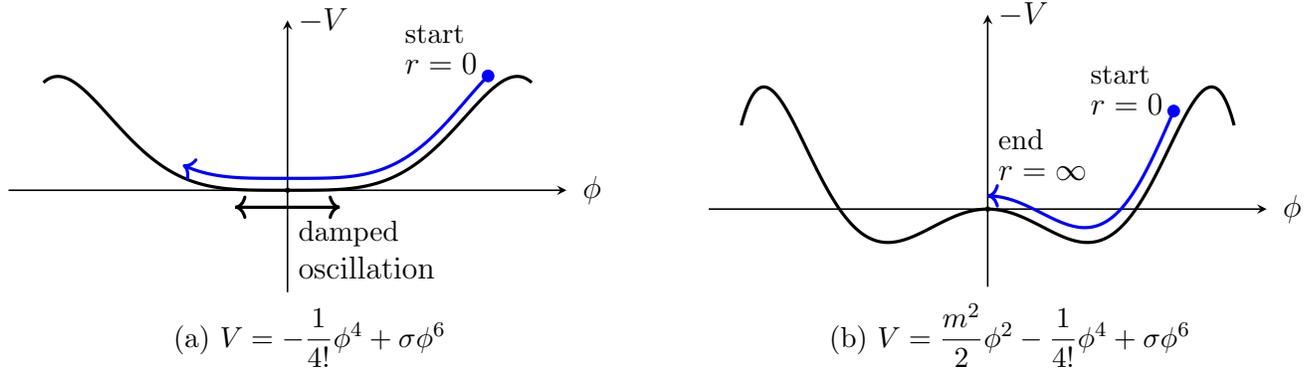
\begin{figure}[t]
    \centering
    \begin{subfigure}{0.48\textwidth}
        \centering
        \begin{tikzpicture}
  \begin{axis}[
    width=9cm,height=5.2cm,
    axis lines=middle,
    axis line style={semithick},
    xmin=-3.2,xmax=3.2,
    ymin=-0.6,ymax=1,
    ticks=none,
    xlabel={$\,\phi$},
    ylabel={$-V$},
    xlabel style={at={(xticklabel* cs:1)},anchor=west},
    ylabel style={at={(yticklabel* cs:1)},anchor=west},
    clip=false,
    declare function={sigma=0.004;}
  ]
    \pgfmathsetmacro{\xstart}{2.3}
    \addplot[domain=-2.8:2.8, samples=400, very thick]
      {(1/24)*x^4-sigma*x^6};
      \addplot[domain=-1.2:\xstart, samples=400, very thick,<-,blue]
      {1.05*(1/24)*x^4-1.05*sigma*x^6+.07};
      \addplot[domain=-.6:.6, samples=400, very thick,<->]
      {-.1};
    \addplot[mark=*,only marks,mark size=0.8pt] coordinates {(0,0)};
    \addplot+[only marks,mark=*,mark size=2.2pt, mark options={fill=blue,draw=blue}]
      coordinates {(\xstart,{1.05*(1/24)*(\xstart)^4-1.05*sigma*\xstart^6+.07})}
      node[left,align=left,yshift=10pt,black]{\small start \\[-3pt] $r=0$};
      \node[anchor=west,align=left,] at (axis cs:-0.0,-0.35) {\small damped\\ oscillation}; 
  \end{axis}
\end{tikzpicture}
        \caption{$V = -\dfrac{1}{4!}\phi^4+{\sigma}\phi^6$}
        \label{fig:pot1phi6}
    \end{subfigure}
    \hfill
    \begin{subfigure}{0.48\textwidth}
        \centering
        \begin{tikzpicture}
  \begin{axis}[
    width=9cm,height=5.2cm,
    axis lines=middle,
    axis line style={semithick},
    xmin=-3,xmax=3,
    ymin=-0.1,ymax=.25,
    ticks=none,
    xlabel={$\,\phi$},
    ylabel={$-V$},
    xlabel style={at={(xticklabel* cs:1)},anchor=west},
    ylabel style={at={(yticklabel* cs:1)},anchor=west},
    clip=false,
    declare function={sigma=0.004;},
    declare function={mu=0.4;}
  ]
    \pgfmathsetmacro{\xstart}{2.}
    \addplot[domain=-2.65:2.65, samples=400, very thick]
      {-(0.5*mu^2)*x^2 +(1/24)*x^4-sigma*x^6};
      \addplot[domain=0:\xstart, samples=400, very thick,<-,blue]
      {-(0.5*mu^2)*x^2 +1.045*(1/24)*x^4-1.045*sigma*x^6+.017};
    \addplot[mark=*,only marks,mark size=0.8pt] coordinates {(0,0)};
    \addplot+[only marks,mark=*,mark size=2.2pt, mark options={fill=blue,draw=blue}]
      coordinates {(\xstart,
      {-(0.5*mu^2)*\xstart^2 +1.045*(1/24)*\xstart^4-1.045*sigma*x^6+.017})} node[left,align=left,yshift=8pt]{\small start \\[-3pt] $r=0$};
      \node[anchor=west , align=left] at (axis cs:-0.0,0.07) {\small end \\[-3pt] $r=\infty$}; 
  \end{axis}
\end{tikzpicture}
        \caption{$V = \dfrac{m^2}{2}\phi^2 - \dfrac{1}{4!}\phi^4+\sigma\phi^6$}
        \label{fig:pot2phi6}
    \end{subfigure}

    \caption{
(Left) Effect of the positive $\phi^6$ constraint term in the particle-mechanics picture for $m^2=0$: the solution overshoots, and the finite-action instanton solution disappears.
(Right) Adding both a mass term and a positive $\phi^6$ term restores a finite-action constrained instanton solution.
}
    \label{fig:pot12phi6}
\end{figure}

We can also see from the particle-mechanics analogy that an additional contribution to the potential can restore the instanton solution. For example, let us add a positive $\phi^6$ term to the potential. 
Since the time-dependent friction decays during the time that the particle stays near the top of the hill, an appropriate choice of the initial height $\phi(0)$ can yield an instanton solution that asymptotically approaches $\phi(\infty)=0$ (see Fig.\,\ref{fig:pot12phi6}).

This instanton solution in the modified potential is referred to as a constrained instanton. This modification is equivalent to searching for a stationary point of the action subject to a constraint on the value of 
\begin{align}
    S_\mathrm{con} =
    \int \dd^4x \,\mathcal{O}_{\mathrm{con}} 
    = \int \dd^4x\, \phi^6
    \ .
\end{align}
The stationary point can be obtained using the Lagrange-multiplier method by extremizing the modified action,
\begin{align}
    S+\sigma (S_\mathrm{con}-\lambda)
    =
    \int\dd^4 x\,\qty[\mathcal{L}+\sigma\mathcal{O}_\mathrm{con}]-\sigma\lambda\, , 
    \label{eq: Stot in phi4}
\end{align}
where $\lambda$ represents the specific value to which $S_\mathrm{con}$ is constrained, and $\sigma$ is the Lagrange multiplier. Varying with respect to $\phi$ and $\sigma$ yields
\begin{align}
\phi''+\frac{3}{r}\phi'-\tilde{V}'(\phi)&=0\,,\qquad 
\tilde{V}(\phi)=\frac{m^2}{2}\phi^2-\frac{1}{4!}\phi^4+\sigma \phi^6\ ,
\label{eq: Constrained Euler--Lagrange for phi}
\\
S_\mathrm{con}-\lambda &=0\,,
\end{align}
which are equivalent to the equation of motion for a particle moving in the modified potential $\tilde{V}(\phi)$.

In the above discussion, we introduced the additional term to the potential by hand as a contribution from the constraint.
In fact, introducing the constraint as in Eq.\,\eqref{eq: Stot in phi4} can be understood as inserting a $\delta$-function into the path integral of the original $\phi^4$ theory to fix the instanton size $\rho$ (see Appendix\,\ref{sec: constraint method of inserting 1}).
Therefore, by appropriately integrating over $\lambda$ or equivalently over the size $\rho$, we can recover the path integral of the original $\phi^4$ theory.

We now turn to the construction of the constrained instanton solution in the regime $\rho m \ll 1$ by means of a perturbative expansion in $(\rho m)^2$. 
However, N\&N pointed out  that a perturbative analysis with $\mathcal{O}_{\mathrm{con}}=\phi^p \,(p\ge5)$ may exhibit problematic asymptotic behavior.
We therefore formulate the expansion with care and return to this issue later.

\subsection{Asymptotic Behaviors of Constrained Instanton}
In the following, we discuss the asymptotic behaviors of the constrained instanton solution at $r\to0$ and at $r\to\infty$.
The boundary conditions
for the constrained instanton solution are defined by
\begin{align}
    \phi(r=0)=\phi_\mathrm{ini} >0\ ,  \quad \phi'(r=0)  = 0 \ .
\end{align}
In addition, finiteness of the action requires that
\begin{align}
    \phi(r)\to0 \qquad (r\to\infty)\ .
\end{align}

We consider the situation where the mass is non-zero but small compared with $\phi_\mathrm{ini}$, i.e., $\phi_\mathrm{ini} \gg m$.
In this case, we expect that the solution becomes similar to the instanton solution around $r\to 0$, 
\begin{align}
\label{eq:app instanton}
    \phi(r) \simeq  \phi_0(r)=\frac{4\sqrt{3}\rho}{r^2 + \rho^2}\ ,\quad (r \ll \rho)\ ,
\end{align}
where $\rho$ is the instanton size given by 
\begin{align}
    \rho \simeq \frac{4\sqrt{3}}{\phi_\mathrm{ini}} \ .
\end{align}
In the situation $\rho m\ll1$, which corresponds to $m\ll \phi_\mathrm{ini}$, the resultant Lagrange multiplier $\sigma$ is expected to be small, that is,
\begin{align} 
\label{eq:small sigma}
|\sigma \phi_\mathrm{ini}^2| \ll 1 \ ,
\end{align}
since the deviation from the massless $\phi^4$ theory by the mass term is small.
In fact, as we will see shortly, $\sigma$ has a relation,
\begin{align}
    \sigma \sim  m^2 \rho^4\ ,
\end{align}
which satisfies Eq.\,\eqref{eq:small sigma} for $m\ll\phi_\mathrm{ini}$.

As for the asymptotic behavior at $r \to \infty$, the Euler--Lagrange equation is dominated by the mass term in the region where $r\gg m^{-1}$,
\begin{align}
    (\partial_\mu\partial_\mu - m^2)\phi = 0\ ,\quad (r\gg m^{-1})\ .
\end{align}
Thus, the asymptotic solution should be of the form
\begin{align}
\label{eq: K1}
    \phi(r) \simeq \kappa\,\frac{K_1(mr)}{r}\ ,\quad (r\gg m^{-1})\ ,
\end{align}
where $K_1$ denotes the modified Bessel function of order one.
The coefficient $\kappa$ is determined by the matching procedure discussed later.
At finite radius $r$, the solution deviates from Eq.\,\eqref{eq: K1}, but the deviation is further suppressed by a factor
$e^{-mr}$ relative to the leading term as $r\to\infty$.

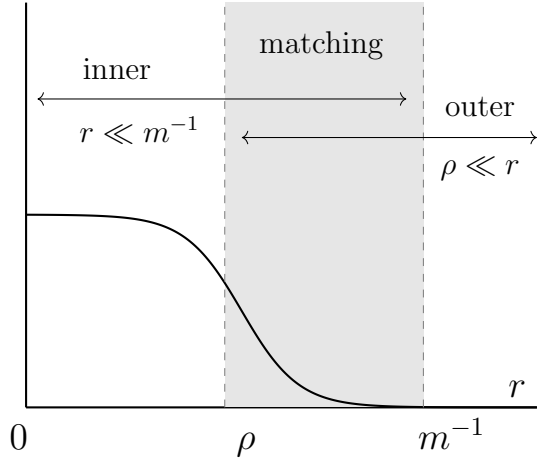
\begin{figure}[t]
    \centering
\begin{tikzpicture}
  \begin{axis}[
    domain=-1:8, 
    samples=200,
    axis lines=middle,
    xlabel={\large{$r$}},
    ymin=0, ymax=1.1,
    xmin=-1, xmax=8,
    xtick={0},
    ytick=\empty,
    enlargelimits=false,
    clip=false,
    x axis line style={-,thick},
    y axis line style={draw=none},
    tick style={black},
  ]

    \path[fill=gray!20,draw=none] 
      (axis cs:2.5,0) rectangle (axis cs:6,1.05);

    \addplot[thick, black] {0.5/(1 + exp(2*(x - 2.8)))};

    \draw[black,thick] (axis cs:-1,0) -- (axis cs:-1,1.05);
    \draw[dashed, gray] (axis cs:2.5,0) -- (axis cs:2.5,1.05);
    \draw[dashed, gray] (axis cs:6,0) -- (axis cs:6,1.05);

    \node[] at (axis cs:1.,0.72){$r \ll m^{-1}$};
    \node[] at (axis cs:0.6,0.88){inner};
    \draw[<->] (axis cs:-0.8,0.8)--(axis cs:5.7,0.8);
    \draw[<->] (axis cs:2.8,0.7)--(axis cs:8.0,0.7);

    \node[] at (axis cs:7.,0.62){$\rho \ll r$};
    \node[] at (axis cs:7,0.78){outer};
    \node[] at (axis cs:4.2,0.94){matching};
    \node[below] at (axis cs:3,0) {\large{$\rho^{\phantom{0}}$}};
    \node[below] at (axis cs:6.5,0) {\large{$m^{-1}$}};
    \node[below] at (axis cs:-1,0) {\large{$0^{\phantom{2}}$}};
  \end{axis}
\end{tikzpicture}
    \caption{
Schematic illustration of the matching procedure.
We extend the inner solution to $r\ll m^{-1}$ and the outer solution to $r\gg \rho$.
The inner and outer solutions are matched in the overlap region $\rho\ll r\ll m^{-1}$.}
    \label{fig: Regions}
\end{figure}
To construct the solution, we need to match the behavior in Eq.\,\eqref{eq:app instanton} for $r\ll \rho$ with that in Eq.\,\eqref{eq: K1} for $r\gg m^{-1}$. To this end, we extend the former solution to the region $r\ll m^{-1}$ and the latter solution to the region $r\gg\rho$, and refer to them as the inner and outer solutions, respectively.
These two solutions are then matched in the overlapping region $\rho \ll r \ll m^{-1}$ (see Fig.\,\ref{fig: Regions}). 
In what follows, we assume $\rho m \ll 1$ and construct both the inner and outer solutions as series
expansions in $(\rho m)^2$.

\subsection{Inner Solution}

The inner solution can be extended to $r>\rho$ by performing a perturbative expansion in the small parameter $(\rho m)^2 \ll 1$,
\begin{align}
    \phi(r) &= f_0(r/\rho)+(\rho m)^2 f_2(r/\rho)+(\rho m)^4 f_4(r/\rho)+\cdots\,,
    \label{eq: phi perturbative series}
    \\
    \rho f_0(r/\rho) &=
    \frac{4\sqrt{3}}{(r/\rho)^2+1}\,,
    \label{eq: f0}
\end{align}
where each $f_n(r/\rho)$ has mass dimension one and does not carry any power-dependence on $m$. Any remaining dependence on $m$ may appear only through logarithms such as $\log(\rho m)$. The first term corresponds to the instanton solution for the massless $\phi^4$ theory, which we refer to as the leading-order (LO) inner solution.

We also expand the Lagrange multiplier $\sigma$ by
\begin{align}
    \sigma = \sigma_0 + (\rho m)^2 \sigma_2 + (\rho m)^4 \sigma_4+\cdots \ ,
    \label{eq: lagrange multiplier expaneded in phi4}
\end{align}
where $\sigma_0 = 0$, and $\sigma_{2,4,\ldots}$ do not carry any power-dependence on $m$. As in the case of 
$f_n(r/\rho)$, any remaining dependence on $m$ may appear only through logarithms such as $\log(\rho m)$.

By substituting 
Eqs.\,\eqref{eq: phi perturbative series} and \eqref{eq: lagrange multiplier expaneded in phi4}
into the Euler--Lagrange equation
Eq.\,\eqref{eq: Constrained Euler--Lagrange for phi},
the next-to-leading order (NLO) contribution satisfies,
\begin{align}
    f_2''(\hat{r})+\frac{3}{\hat{r}}f_2'(\hat{r})
+\frac{\rho^2}{2}f_0^2 f_2(\hat{r})
    = f_0
    + 6\sigma_2\rho^2 f_0^5\ , 
    \quad\quad
    \hat{r}:=\frac{r}{\rho}\,,
\end{align}
where the right-hand side provides the source terms for $f_2$.
A general solution to this differential equation can be written as
\begin{align}
    f_2(\hat{r})=
    c_1
    f_2^{(\mathrm{hom},1)}(\hat{r})
    +c_2
    f_2^{(\mathrm{hom},2)}(\hat{r})
    +f_2^{(m^2)}(\hat{r})+f_2^{(\sigma_2)}(\hat{r}) 
    \,.
    \label{eq: NLO phi}
\end{align}
The last two terms, $f_2^{(m^2)}(\hat{r})$ and $f_2^{(\sigma_2)}(\hat{r})$ are the particular solutions originating from the mass term and the constraint term, respectively. 
They are given by
\begin{align}
    &\rho f_2^{(\mathrm{hom},1)}(\hat{r})
    =
    \frac{1-\hat{r}^2}{\left(\hat{r}^2+1\right)^2}\,,
    \label{eq: hom1}
    \\
    &\rho f_2^{(\mathrm{hom},2)}(\hat{r})=
    \frac{\hat{r}^6-9 \hat{r}^4-9 \hat{r}^2+12 \left(\hat{r}^2-1\right) \
    \hat{r}^2 \log \left(\hat{r}\right)+1}{\left(\hat{r}^3+\hat{r}\right)^2}
    \,,
    \label{eq: hom2}
    \\
&\rho f_2^{(m^2)}(\hat{r}) =
    \sqrt{3}\qty[\hat{r}\left(\hat{r}^2+1\right)]^{-2} \left[-6 \left(\hat{r}^2-1\right) \hat{r}^2 \mathrm{Li}_2\left(\frac{1}{\hat{r}^2+1}\right)
    -3 \left(\hat{r}^2-5\right) \hat{r}^4
    \right.\notag
    \\&\quad\left.
    +\left(\hat{r}^6-9 \hat{r}^4-9 \hat{r}^2+3 \left(\hat{r}^2-1\right) \hat{r}^2 \log \left(\frac{\hat{r}^4}{\hat{r}^2+1}\right)+1\right) \log \left(\hat{r}^2+1\right)\right]\,,
    \label{eq: m^2 particular solution in phi4 theory}
    \\
    &\rho f_2^{(\sigma_2)}(\hat{r}) =\frac{\sigma_2}{\rho^2}
    \times
    \frac{1152 \sqrt{3} \left(6 \hat{r}^4-3 \hat{r}^2+6 \left(\hat{r}^4-1\right) \hat{r}^2 \log \left(\frac{\hat{r}^2}{\hat{r}^2+1}\right)+1\right)}{5 \hat{r}^2 \left(\hat{r}^2+1\right)^3}
    \,.
    \label{eq: sigma2 particular solution in phi4 theory}
\end{align}
The particular solutions are not unique, since homogeneous solutions can be added arbitrarily as
\begin{align}
    f_2^{(m^2,\sigma_2)}
    \to
    f_2^{(m^2,\sigma_2)}
    +a
    f_2^{(\mathrm{hom},1)}
    +b
    f_2^{(\mathrm{hom},2)}\,,
\end{align}
where $a,b$ are arbitrary constants.
Using this freedom, we choose the particular solutions above so that they exhibit the useful asymptotic behaviors described below.

The resulting asymptotic behaviors are as follows.
The asymptotic behavior of \(f_2^{(m^2)}\) is
\begin{align}
\rho f_2^{(m^2)}
=
\begin{cases}
    \sqrt{3} \left(1+\pi ^2\right)-\dfrac{\sqrt{3}\left(5+6 \pi ^2\right)}{2}   \hat{r}^2+\cdots
    \quad
    &
    \text{at}~ \hat{r}\to0
    \,,
    \\[.5em]
    \sqrt{3} \left(2 \log\hat{r}-3\right)+2 \sqrt{3} \left(6 (\log\hat{r})^2-11 \log\hat{r}+11\right)\hat{r}^{-2}+\cdots
    &
    \text{at}~ \hat{r}\to\infty\,.
\end{cases}
\end{align}
We have used the above freedom so that \(\rho f_2^{(m^2)}\) is regular as \(r\to0\).
The asymptotic behavior of \(f_2^{(\sigma_2)}\) is
\begin{align}
    \rho f_2^{(\sigma_2)}
    =
    \dfrac{\sigma_2}{\rho^2}\times
    \begin{cases}
        \dfrac{1152 \sqrt{3}}{5 \hat{r}^2}-\dfrac{6912}{5} \sqrt{3} \left(2 \log \hat{r}+1\right)+\cdots
        &\text{at}~ \hat{r}\to0
        \,,
        \\[.7em]
        1152 \sqrt{3}\hat{r}^{-8}-\dfrac{19008 \sqrt{3}}{5}\hat{r}^{-10}
        +\cdots
        &\text{at}~ \hat{r}\to\infty
        \,.
    \end{cases}
\end{align}
We have again used the above freedom so that \(\rho f_2^{(\sigma_2)}\) is suppressed as \(r\to\infty\).
Accordingly, the NLO inner solution \eqref{eq: NLO phi} can be made finite as $r\to0$ by setting
\begin{align}
    \frac{\sigma_2}{\rho^2}=-\frac{5 c_2}{1152 \sqrt{3}}\,,
    \label{eq: phi NLO Lagrange multiplier}
\end{align}
which ensures that the $\hat{r}^{-2}$ and $\log\hat{r}$ dependences in $f_2^{(\mathrm{hom},2)}$ and $f_2^{(\sigma_2)}$ cancel each other out in the $\hat{r}\to0$ limit.

The first homogeneous solution  $f^{(\mathrm{hom},1)}_2$ can be written as
\begin{align}
    \rho f^{(\mathrm{hom},1)}_2
    =
    \frac{-\rho^2}{4\sqrt{3}}
    \frac{\partial}{\partial\rho}\qty(\frac{4\sqrt{3}\rho}{\rho^2+r^2})
    =\frac{-\rho^2}{4\sqrt{3}}
    \frac{\partial}{\partial\rho}f_0
    \,,
\end{align}
which shows that this term corresponds to the variation of $f_0$ in Eq.\,\eqref{eq: NLO phi} under a rescaling of $\rho$.
Because of this, for a small change of the size parameter of order $(\rho m)^2$, that is,
\begin{align}
    \tilde{\rho}=\rho+(\rho m)^2\delta\rho\,,
\end{align}
the LO instanton solution with size $\tilde{\rho}$ can be related to that with size $\rho$ as
\begin{align}
    \frac{4\sqrt{3}\tilde{\rho}}{\tilde{\rho}^2+r^2}
    &=
    \frac{4\sqrt{3}\rho}{\rho^2+r^2}+(\rho m)^2\delta\rho\frac{\partial}{\partial\rho}\qty(\frac{4\sqrt{3}\rho}{\rho^2+r^2})+\order{\rho^4 m^4}
    \\
    &=
    \frac{4\sqrt{3}\rho}{\rho^2+r^2}
    -4\sqrt{3}(\rho m)^2\frac{\delta\rho}{\rho} f_2^{(\mathrm{hom},1)}+\order{\rho^4 m^4}
    \ .
\end{align}
Thus, the NLO inner solution parametrized by $(\tilde{\rho},c_1)$ is equivalent to that parametrized by $(\rho, c_1 -4\sqrt{3} \delta \rho/\rho)$.
Using this redundancy, we may fix the value of $c_1$ to be
\begin{align}
    c_1=
    -\sqrt{3} \left(1+\pi ^2\right)+5 c_2\,.
    \label{eq: c_1 choice}
\end{align}
This choice is convenient because it implies $f_2(\hat{r}=0)=0$ (see also Eq.\,\eqref{eq: phi NLO Lagrange multiplier}), thereby fixing the definition of the size parameter as
\begin{align}
    \rho:=\frac{4\sqrt{3}}{|\phi(r=0)|}(1+\order{\rho^4m^4})\,.
\end{align}

\subsection{Outer Solution}

The outer solution can also be extended to $r<m^{-1}$ by a perturbative expansion in the small parameter $(\rho m)^2\ll1$ as
\begin{align}
    \phi(r)=\kappa_0\qty[\frac{K_1(mr)}{r}+(\rho m)^2g_2(mr)+(\rho m)^4g_4(mr)+\cdots]\,,
    \label{eq: outer solution perturbed}
\end{align}
where each $g_n(mr)$ has mass dimension one and does not carry any power of $\rho$.%
\footnote{In general, odd-order terms of the form \((\rho m)^{2i+1} g_{2i+1}(mr)\) \((i\in\mathbbm{N})\) may also appear in the outer expansion. However, these terms turn out to vanish by the matching with the inner solution, since the corresponding orders are absent in the inner expansion.}
Any remaining dependence on $\rho$ may appear only through logarithms such as $\log(\rho m)$. We refer to the first term inside the brackets as the LO outer solution, $g_0(mr)$.
Here, the coefficient $\kappa$ in Eq.\,\eqref{eq: K1} has been expanded in powers of $(\rho m)^2$ as
\begin{align}
    \kappa = \kappa_0 \qty(1+\order{\rho^2 m^2})\,.
    \label{eq: kappa and kappa0}
\end{align}
The coefficient $\kappa_0$ in Eq.\,\eqref{eq: outer solution perturbed} is fixed by matching to the $\order{(\rho m)^0}$ term of the inner solution in Eq.\,\eqref{eq: phi perturbative series}, which gives
\begin{align}
\label{eq: kappa value}
    \kappa_0 = 4\sqrt{3}\,\rho m\,.
\end{align}

By substituting the outer expansion into the Euler--Lagrange equation, Eq.\,\eqref{eq: Constrained Euler--Lagrange for phi}, we obtain
\begin{align}
    g_2''+\frac{3}{R}g_2'-g_2=8m\qty[\frac{K_1(R)}{R}]^3\times\qty(1+\order{\rho^2 m^2}+\frac{\sigma_2}{\rho^2}\order{\rho^6 m^6})\,,
    \quad\quad
    R:=mr\,,
    \label{eq: phi4 outer NLO equation}
\end{align}
where the prime denotes differentiation with respect to $R$.
The last term in the parentheses on the right-hand side shows that the contribution from the constraint does not enter at the present order.
We can then obtain the NLO outer solution for $r\ll m^{-1}$ ($R\ll1$),
\begin{align}
    m^{-1} g_2(R) = -\frac{1}{R^4}+\frac{3(\log R)^2}{R^2}
    +\frac{1+12\gamma_E-12\log2}{2}\frac{\log R}{R^2}
    +\frac{c^{(\mathrm{out})}}{R^2}+\order{R^0}\,,
    \label{eq: outer NLO phi4}
\end{align}
where $c^{(\mathrm{out})}$ is one of the two arbitrary coefficients of the homogeneous solutions, and 
$\gamma_E$ is the Euler--Mascheroni constant.
The other constant appears in the abbreviated $\order{R^0}$ contributions.

\subsection{Matching Conditions and Consistency}

To perform the matching, we rewrite the outer solution in terms of $\rho m$ and $\hat{r}=r/\rho$ via
\begin{align}
    R := mr = \rho m \hat{r}\,.
    \label{eq: rewriting R into harr}
\end{align}
The matching is carried out using a double expansion in $(\rho m)^2$ and $\hat{r}^2$ in the overlap region between the inner and outer solutions,
\begin{align}
    \rho \ll r \ll m^{-1} \ .
\end{align}
Within this region,
\begin{align}
    \hat{r} \gg 1 \,,\quad R\ll 1
\end{align}
hold simultaneously.

To make the notation explicit, we parameterize the asymptotic expansions of
$f_n$ and $g_n$ at large $\hat r$ and small $R$ as
\begin{align}
  \rho\, f_n(\hat r)
  &\;=\; \sum_{k} f_n^{(k)}\,\hat r^{-k}
      \,,
  \qquad (\hat r \gg 1), 
  \label{eq: fn-expansion}\\
  m^{-1} g_n(R)
  &\;=\; \sum_{k} g_n^{(k)}\,R^{-k}
  \qquad (R \ll 1) 
  \\
  &\;=\; \sum_{k} (\rho m)^{-k} g_n^{(k)}\,\hat{r}^{-k}
  \qquad (\hat{r} \gg 1)\,.
  \label{eq: gn-expansion}
\end{align}
Here $f_n^{(k)}$ and $g_n^{(k)}$ are dimensionless coefficients, which may
contain logarithmic dependences on $\hat r$ and on~$\rho m$.

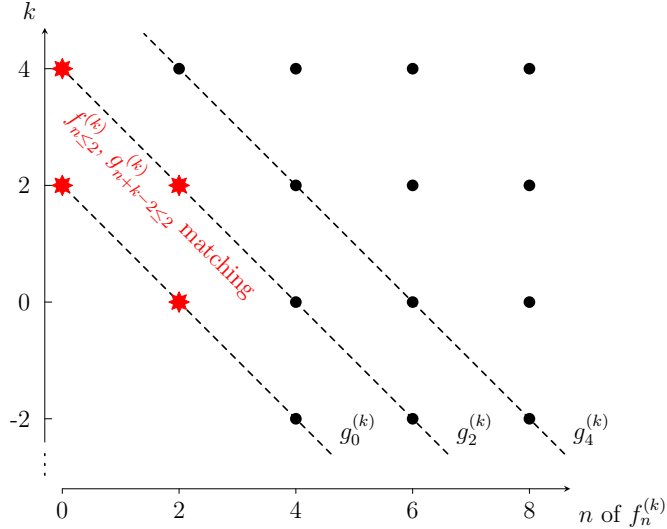
\begin{figure}
    \centering
  \begin{tikzpicture}[>=stealth, line cap=round, line join=round]

\draw[->] (-0,-3.2) -- (8.7,-3.2) node[below right] {$n$ of $f_n^{(k)}$};
\foreach \x in {0,2,4,6,8}{
  \draw (\x,-3.2) -- ++(0,-0.08) node[below] {\x};
}

\def\xaxisx{-0.3}
\draw[->] (\xaxisx,-2.4) -- (\xaxisx,4.7) node[above left] {$k$};
\foreach \y in {-2,0,2,4}{
  \draw (\xaxisx,\y) -- ++(0.12,0) node[left=3pt] at (\xaxisx,\y) {\y};
}
\draw[loosely dotted,thick]
  (\xaxisx,-2.6) -- (\xaxisx,-3.0);

\node[red,rotate=-45] at (1.7,1.7) {$f_{n\le2}^{(k)},\,g_{n+k-2\le2}^{(k)}$ matching};

\def\ptR{0.1}
\foreach \k in {-2,0,2,4}{ 
  \foreach \n in {0,2,4,6,8}{ 
    \pgfmathtruncatemacro{\keep}{\n >= -\k + 2 ? 1 : 0}
    \ifnum\keep=1
      \fill[black] (\n,\k) circle (\ptR);
    \fi
  }
}

\draw[thick,dashed] (4.6,-2.6) -- (0,2) node[pos=0,above right] {$g_0^{(k)}$};
\draw[thick,dashed] (6.6,-2.6) -- (0,4) node[pos=0,above right] {$g_2^{(k)}$};
\draw[thick,dashed] (8.6,-2.6) -- (2-.6,4+.6) node[pos=0,above right] {$g_4^{(k)}$};

\tikzset{
  mystar/.style={
    star, star points=8, star point ratio=1.75,
    minimum size=100*\ptR, inner sep=0pt,
    draw=red, fill=red
  }
}

\fill[white] (0,2) circle (\ptR);
\fill[white] (2,0) circle (\ptR);
\fill[white] (0,4) circle (\ptR);
\fill[white] (2,2) circle (\ptR);

\node[mystar] at (0,2) {};
\node[mystar] at (2,0) {};
\node[mystar] at (0,4) {};
\node[mystar] at (2,2) {};

\end{tikzpicture}
    \caption{
Illustration of the $(n,k)$ matching conditions between the coefficients $f_n^{(k)}$ and $g_{n+k-2}^{(k)}$.
The coefficient $f_n^{(k)}$ vanishes for $k<2-n$.
The dashed lines indicate the orders appearing in the outer expansion in Eq.\,\eqref{eq: outer solution perturbed}, while the horizontal axis represents the order $n$ in the inner expansion in Eq.\,\eqref{eq: phi perturbative series}.
The $(0,2)$ condition corresponds to the LO matching condition, while $(0,4)$, $(2,0)$, and $(2,2)$ correspond to the NLO matching conditions.
}

    \label{fig: matching double expansion}
\end{figure}

In the overlap region we can expand the same quantity $\rho\,\phi(r)$
either from the inner side or from the outer side. From the inner solution we
have
\begin{align}
    \rho\phi(r)
    =
    \rho\sum_n (\rho m)^n f_n
    =\sum_{n,k}(\rho m)^n f_n^{(k)}\hat{r}^{-k}\,,
\end{align}
where we have used Eqs.\,\eqref{eq: phi perturbative series} and \eqref{eq: fn-expansion}. On the other hand, from the outer solution we have
\begin{align}
    \rho\phi(r)
    =
    \rho\kappa_0\sum_n (\rho m)^n g_n(R)
    =
    \kappa_0\sum_{n,k}(\rho m)^{n+1-k}
    g_n^{(k)}\hat{r}^{-k}
    =
    4\sqrt{3}\sum_{n,k}(\rho m)^{n+2-k} g_n^{(k)}\hat{r}^{-k}\,,
\end{align}
where we have used Eqs.\,\eqref{eq: outer solution perturbed} and \eqref{eq: gn-expansion}. Requiring that the coefficients of $(\rho m)^n\hat{r}^{-k}$ agree between the two expansions, we obtain
\begin{align}
  4\sqrt{3}\,g_{n-2+k}^{(k)} = f_{n}^{(k)}\,,
  \label{eq: n,k condition}
\end{align}
which we refer to as the $(n,k)$ condition.
Since $g_{n-2+k}$ exists only for $n-2+k\ge0$,
the expansion coefficient 
$f_n^{(k)}$ must vanish for 
\begin{align}
    k < 2-n\ ,
\end{align}
as illustrated in Fig.\,\ref{fig: matching double expansion}.

Let us first consider the LO matching condition, which corresponds to 
the $(n,k)=(0,2)$ condition. In fact, we have already imposed this condition in Eq.\,\eqref{eq: kappa value} to determine $\kappa_0$, thereby matching the LO inner solution and the LO outer solution. This condition is consistent with 
the matching condition used in Ref.\,\cite{Affleck:1980mp} and N\&N.

Next, let us discuss the NLO matching conditions, which consist of the $(0,4)$, $(2,0)$, and $(2,2)$ conditions.
Details of some of the matching conditions are deferred to Appendix\,\ref{app: phi4 matching general}.
For the $(0,4)$ condition, 
we can read off directly from the large-$\hat{r}$ expansion of Eq.\,\eqref{eq: f0} and from the
first term of the NLO outer solution in Eq.\,\eqref{eq: outer NLO phi4} that
\begin{align}
f_0^{(4)}= 4\sqrt{3}\,g_2^{(4)}=-4\sqrt{3} \ .
\end{align}
The $(2,0)$ condition can be satisfied by taking
\begin{align}
 \label{eq: c_2}
    c_2&=2 \sqrt{3} \qty(1+\gamma_E+\log\frac{\rho m}{2})\ ,
\end{align}
where $c_2$ is the coefficient of the homogeneous solution in Eq.\,\eqref{eq: NLO phi}. 
Finally, the $(2,2)$ condition gives two independent relations through the $(\log \hat{r})^0$ and $(\log \hat{r})^1$ terms. The part proportional to $(\log \hat{r})^0$ determines $c^{(\mathrm{out})}$ in Eq.\,\eqref{eq: outer NLO phi4} as
\begin{align}
    c^{(\mathrm{out})}=-\frac{c_1+11 c_2}{4 \sqrt{3}}-3 \qty[\log (\rho m)]^2+\left(-\frac{1}{2}-6 \gamma_E
   +6\log2\right) \log (\rho m)+\frac{11}{2}\,.
   \label{eq: cout in phi4}
\end{align}
Note that, while $c_1$ has been chosen as in Eq.\,\eqref{eq: c_1 choice}, Eq.\,\eqref{eq: cout in phi4} holds for a general choice of $c_1$.
The remaining term proportional to $(\log \hat{r})^1$ in the $(2,2)$ condition
is found to be satisfied by Eq.\,\eqref{eq: c_2}, providing a nontrivial consistency check.
In this way, all the coefficients $\kappa_0$, $c_1$, $c_2$, $c^{(\mathrm{out})}$, $\sigma_0$, and $\sigma_2$ are fixed so that all matching conditions up to NLO are satisfied.

It is worth emphasizing that our successful matching appears to differ from the conclusion reached in N\&N.
In that work, it was argued that no solution starting from $\phi(0)=\phi_\mathrm{ini}$ with $\phi'(0)=0$ could be smoothly matched to the outer behavior $\kappa K_1(mr)/r$ for constraints of the form $\mathcal{O}_\mathrm{con}=\phi^p$ with $p\ge5$. 
In contrast, our analysis shows that a consistent matching can indeed be achieved with a $\phi^6$ constraint term. 
This apparent discrepancy can be traced to the fact that N\&N used the asymptotic form \(\kappa K_1(mr)/r\) even in the region \(mr\lesssim 1\), effectively disregarding the order-by-order distinction among the functions \(g_n\). By contrast, a consistent matching requires that the expansion in $(\rho m)^2$ be carried out systematically on both the inner and outer solutions.
We further show in Appendix~\ref{app: phi4 matching general} that the same matching strategy can also be implemented for $\mathcal{O}_\mathrm{con}=\phi^p$ with $p\ge5$.

\subsection{Numerical Verification of the Matching}

To verify the above matching procedure, 
we numerically solved the Euler--Lagrange equation given in Eq.\,\eqref{eq: Constrained Euler--Lagrange for phi}.
We imposed the boundary conditions $\phi'(0)=0$ and $\phi(\infty)=0$.
For each fixed value of $\sigma$, we determined $\phi_\mathrm{ini}:=\phi(0)$ using a shooting method and thereby obtained the configuration.%
\footnote{In the numerical analysis, 
instead of integrating from $r=0$, we started at a sufficiently small radius $r=\epsilon$ with the small-$r$ expansion of the Euler--Lagrange equation.
In the shooting method, we searched for the value of $\phi_\mathrm{ini}$
such that the solution smoothly matches onto the large-$r$ behavior proportional to $K_1(mr)/r$.}
The size is given by
\begin{align}
    \rho = \frac{4\sqrt{3}}{\phi_{\mathrm{ini}}}\ ,
\end{align}
which follows from the choice of $c_1$ in Eq.\,\eqref{eq: c_1 choice}.

\begin{figure}[t]
    \centering
    \includegraphics[width=0.5\linewidth]{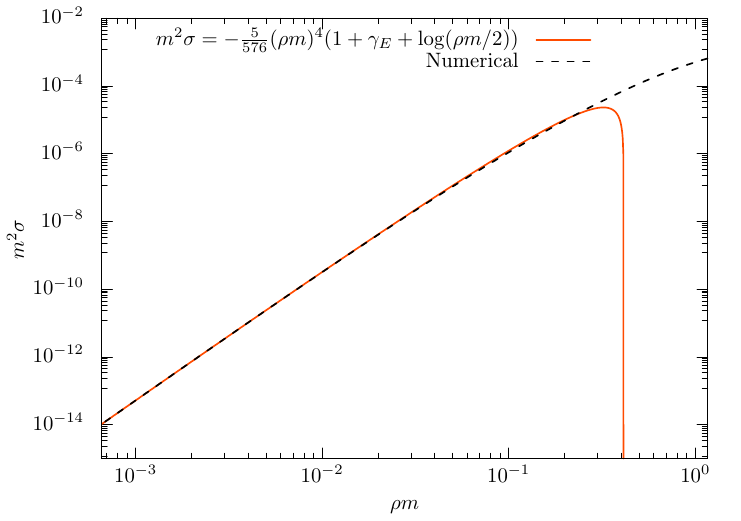}
    \caption{
Dimensionless Lagrange multiplier $m^2\sigma$ as a function of the instanton size $\rho m$.
The numerical result is compared with the analytic NLO prediction in Eq.\,\eqref{eq:sigma2}, showing agreement for sufficiently small $\rho m$.
}
    \label{fig:phi4 Lagrange multiplier value}
\end{figure}

Figure~\ref{fig:phi4 Lagrange multiplier value} shows the relation between the dimensionless Lagrange multiplier $m^2\sigma$ and the dimensionless size $\rho m$ for the numerically obtained solutions. 
We also plot the analytic prediction for the relation between the instanton size and the Lagrange multiplier,
obtained from Eq.\,\eqref{eq: phi NLO Lagrange multiplier} at order $(\rho m)^2$,
\begin{align}
\label{eq:sigma2}
    \frac{\sigma_2}{\rho^2}
    =
    -\frac{5}{576}\qty(1+\gamma_E+\log\frac{\rho m}{2})\ , 
\end{align}
where we have substituted Eq.\,\eqref{eq: c_2}.
The figure shows that the numerical results agree with Eq.\,\eqref{eq:sigma2} for sufficiently small $\rho m$, supporting the validity of the matching procedure described above.

\begin{figure}[t]
    \centering
    \includegraphics[width=0.7\linewidth]{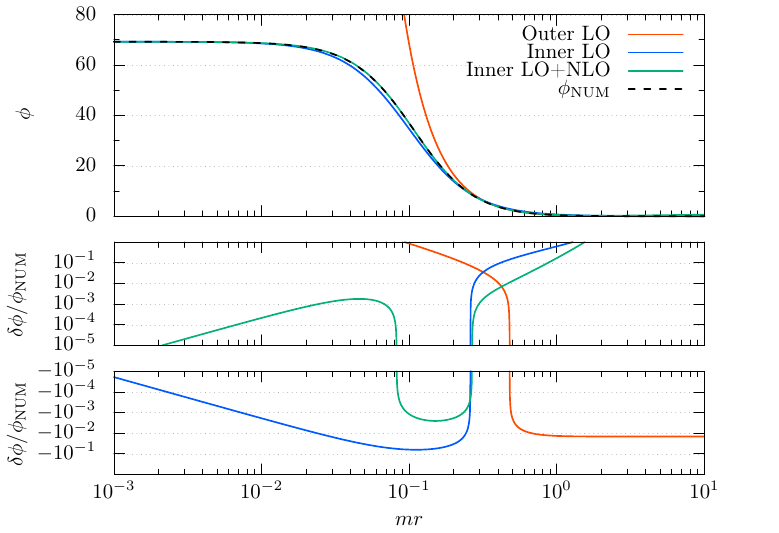}
    \caption{
Comparison between the numerical solution and the analytic expansions for $\rho m=10^{-1}$.
In the top panel, the dashed line represents the numerical configuration $\phi_\mathrm{NUM}$.
The blue line shows the LO inner solution $f_0$, while the green line shows the inner LO+NLO solution.
The orange line represents the LO outer solution $g_0$.
In the middle and bottom panels, the relative deviation from the numerical configuration,
$\delta\phi(r)/\phi_\mathrm{NUM}(r):=(\phi(r)-\phi_\mathrm{NUM}(r))/\phi_\mathrm{NUM}(r)$, is shown for the inner LO, inner LO+NLO, and outer LO contributions, respectively.
Including the NLO correction improves the agreement of the inner solution with the numerical result.
}
    \label{fig:phi4 numerical configurations}
\end{figure}

As a representative example, Fig.\,\ref{fig:phi4 numerical configurations} shows the numerical solution for the case $\rho m=10^{-1}$, together with 
the inner solution at LO and LO+NLO with the outer solution at LO.
In the middle and the bottom panels, we show the relative deviations of those contributions from the numerical solution.
For the inner solution, the LO+NLO contribution yields a smaller relative deviation than the LO contribution alone.
We also see that the relative deviation of the outer LO solution approaches a constant for large $mr$. This constant corresponds to $(\kappa-\kappa_0)/\kappa$ and is indeed of order $(\rho m)^2 \simeq 10^{-2}$, as expected from Eq.\,\eqref{eq: kappa and kappa0}.

\begin{figure}[ht]
    \begin{subfigure}{.32\textwidth}
        \centering
        \includegraphics[width=\linewidth]{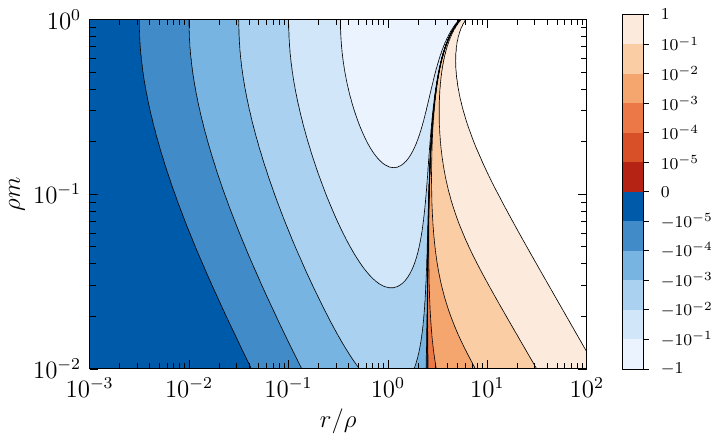}
        \caption{Inner LO solution}
    \end{subfigure}
    \begin{subfigure}{.32\textwidth}
        \centering
        \includegraphics[width=\linewidth]{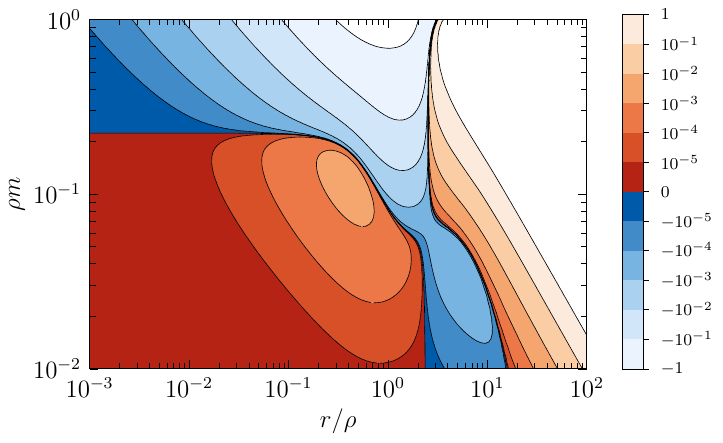}
        \caption{Inner LO+NLO solution}
    \end{subfigure}
    \begin{subfigure}{.32\textwidth}
        \centering
        \includegraphics[width=\linewidth]{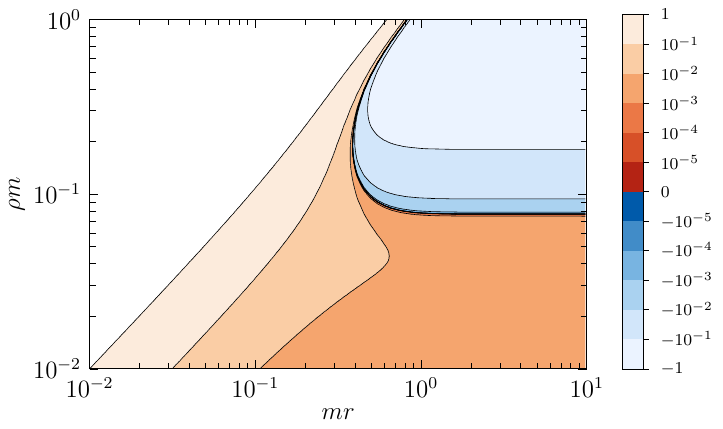}
        \caption{Outer LO solution}
    \end{subfigure}
    \caption{
Contour plots of the relative deviation from the numerical configuration,
$\delta\phi(r)/\phi_\mathrm{NUM}(r)$, for $\phi$ given by
(a) the inner LO solution,
(b) the inner LO+NLO solution, and
(c) the outer LO solution.
The deviations decrease as $\rho m$ becomes smaller, consistent with the validity of the matched asymptotic expansion.
}
    \label{fig:phi4 error contour}
\end{figure}

Figure~\ref{fig:phi4 error contour} 
shows the relative deviations of those contributions from the numerical solutions for various instanton sizes.
We find that the relative deviations decrease for smaller $\rho m$, as expected.


\section{Constrained Instanton in Yang--Mills Theory}
\label{sec: revisit YM constrained instanton}

As mentioned in Introduction,
instantons as exact local minima of the action are absent in the broken phase of the non-Abelian gauge theory, as in the massive $\phi^4$ theory. In this section, we construct a constrained instanton in the broken phase by imposing a size constraint, adopting the methodology established for $\phi^4$ theory (see also Refs.~\cite{Affleck:1980mp,Espinosa:1989qn}).

\subsection{Instanton-like Configurations in the Broken Phase}
\label{sec: No instanton-like strict minima in broken phase}

Let us first consider an $\SU(N)$ Yang--Mills theory without matter fields in Euclidean space.  
Using the identity
\begin{align}
(F_{\mu\nu} \pm \tilde{F}_{\mu\nu})^2 
= F_{\mu\nu}F_{\mu\nu} \pm F_{\mu\nu}\tilde{F}_{\mu\nu} \ge 0\ ,
\end{align}
for the field strength $F_{\mu\nu}$ and its dual $\tilde{F}_{\mu\nu}=\epsilon_{\mu\nu\rho\sigma}F_{\rho\sigma}/2$, we obtain the inequality
\begin{align}
 \int\tr F_{\mu\nu}F_{\mu\nu}\, d^4x 
\ge \qty|\int \tr F_{\mu\nu}\tilde{F}_{\mu\nu}\, d^4x| 
= 16\pi^2 n_w\ .
\label{eq: BPS bound}
\end{align}
See Appendix~\ref{app:notations} 
for our notation in Euclidean space.
Here, $n_w$ is the topological winding number characterizing the map from the boundary sphere $S^3$ at infinity to the $\SU(N)$ gauge group manifold. 

The instanton configuration is a stationary point of the Euclidean action for fixed $n_w$, satisfying the (anti-)self-dual condition $F_{\mu\nu} = \pm \tilde{F}_{\mu\nu}$. 
For simplicity, we hereafter restrict our attention to the $\SU(2)$ case.
For $n_w=1$,  $\SU(2)$ instanton configuration centered at the origin, expressed in the singular gauge (see, e.g., Ref.\,\cite{Shifman:2012zz}), is given by
\begin{align}
A_\mu = \dfrac{2\rho^2 x_\nu}{r^2(r^2+\rho^2)}\,\bar{\tau}_{\mu\nu}\ .
\label{eq: singular gauge pure YM instanton}
\end{align}
Here, $\rho$ denotes the instanton size, $r^2 = x_\mu x_\mu$, and $\bar{\tau}_{\mu\nu}$ represents the antisymmetric combination of the Pauli matrices.
By construction, this instanton saturates the inequality in Eq.\,\eqref{eq: BPS bound}, yielding an action that is independent of the size parameter $\rho$.

In the semiclassical treatment of Yang--Mills theory, such solutions play a central role in computing transition amplitudes between topologically distinct sectors. Because the action is independent of $\rho$, the path integral formulation requires treating the instanton size as a collective coordinate, integrating over $\rho$ as well as the quantum fluctuations around the instanton background.

Let us now consider an $\SU(2)$ gauge theory coupled to a scalar $\SU(2)$ doublet, described by the Euclidean action
\begin{align}
    S = 
    \frac{1}{g^2}\int d^4 x\qty[
        \dfrac{1}{2}\,\mathrm{Tr}(F_{\mu\nu}F_{\mu\nu})
        + |D_{\mu}H|^2
        + \dfrac{\lambda_H}{4}\bigl(|H|^2 - v^2\bigr)^2
    ]\,,
    \label{eq: action for SU(2) gauge theory with Higgs}
\end{align}
where the field strength and the covariant derivative are defined by
\begin{align*}
    &
    F_{\mu\nu} = \partial_{\mu}A_{\nu} - \partial_{\nu}A_{\mu} - i\left[A_{\mu},A_{\nu}\right]\ ,\\[2mm]
    &
    D_{\mu} = \partial_{\mu} - iA_{\mu}\ .
\end{align*}
Here, $g$ is the $\SU(2)$ gauge coupling, $\lambda_H$ is the quartic coupling, 
$v$ is a dimensionful parameter, and $A_\mu = A_\mu^a \tau^a/2$ ($a=1,2,3$), with $\tau^a$ denoting the Pauli matrices.
Note that we have rescaled the scalar field so that the gauge coupling $g$ appears as an overall factor in the action.
Assuming $\lambda_H > 0$, the $\SU(2)$ symmetry is spontaneously broken by the vacuum expectation value (VEV) of the scalar field, which can be chosen as $\expval{H}=(0,v)^T$ by a suitable gauge transformation.
Consequently, due to the aforementioned rescaling of the fields, the gauge bosons and the radial mode of the scalar field acquire masses $m_A=v/\sqrt{2}$ and $m_H=\sqrt{\lambda_H}\,v$, respectively.

In the symmetric phase, i.e., $\langle H\rangle=0$, instanton configurations saturating the equality in Eq.\,\eqref{eq: BPS bound} for $n_w=1$ exist as in the Yang--Mills theory without matter fields discussed earlier, taking the form
\begin{align}
    A_\mu = \dfrac{2\rho^2 x_\nu}{r^2(r^2+\rho^2)} \bartau_{\mu\nu}\ ,\ \ \ \ \ H= 0  \ .
    \label{eq: instanton w/o Higgs}
\end{align}

In contrast, in the broken phase, i.e., $\langle H\rangle \neq 0$, there are no local minima of the action for $n_w \neq 0$.
To demonstrate this, we consider a finite-action configuration $(A_\mu(x),H(x))$ with $n_w \neq 0$ and introduce a scale parameter $a$, defining the rescaled fields
\begin{align}
   x':=a^{-1}x\,,\quad
   A_\mu'(x')=a A_\mu(x)\,,\quad
   H'(x')=H(x)\,.
   \label{eq: rescaling in gauge theory}
\end{align}
For $a>1$, the profile is scaled to a smaller spatial size, while $n_w$ is unchanged. 
On the other hand, the action for these rescaled fields decreases as
\begin{align}
    S[A',H']
    &=
    \frac{1}{g^2}\int\mathrm{d}^4x'
    \qty[\dfrac{1}{2 }\mathrm{Tr}(F_{\mu\nu}'(x')F_{\mu\nu}'(x')) + |D_{\mu}H'(x')|^2 + \dfrac{\lambda_H}{4}(|H'(x')|^2 - v^2)^2]
    \\
    &=
    \frac{1}{g^2}\int\mathrm{d}^4 x
    \qty[\dfrac{1}{2}\mathrm{Tr}(F_{\mu\nu}(x)F_{\mu\nu}(x)) + a^{-2}|D_{\mu}H(x)|^2 + a^{-4}\dfrac{\lambda_H}{4}(|H(x)|^2 - v^2)^2]
    \\
    &< S[A,H]\, ,
    \label{eq: action decreases as the size shrinks}
\end{align}
where the inequality is strict for $a>1$ because the covariant derivative and potential terms for the scalar field cannot vanish simultaneously.
This confirms the absence of local minima for $n_w \neq 0$: any such profile can be continuously scaled to a smaller spatial size to further lower its action.

\subsection{Semiclassical Approximation with a Constraint}
Because the preceding analysis establishes the absence of exact local minima for $n_w \neq 0$, the conventional saddle-point analysis using instantons cannot be directly applied.
To overcome this, we employ the same methodology discussed for the $\phi^4$ theory in Sec.\,\ref{sec: phi4 case}. Specifically, as proposed in Ref.\,\cite{Affleck:1980mp}, the semiclassical approximation can be implemented by imposing a constraint that prevents the action from decreasing under the scale transformation in Eq.\,\eqref{eq: action decreases as the size shrinks}, thereby yielding a constrained minimum.
The path integral then requires integrating over both the quantum fluctuations and the parameter that fixes the value of the constraint functional.

To illustrate how such a constraint can be introduced, we consider the path integral
\begin{align}
    Z
    =
    \int\mathcal{D}\Phi\,
    e^{-S[\Phi]}\,,
\end{align}
where $\Phi$ collectively denotes the fields, $A_\mu$ and $H$.
We insert the identity,
\begin{align}
    1 &= \int \mathrm{d}\lambda\, \delta(\lambda - S_\mathrm{con}[\Phi])\,,
    \label{eq: 1}
\end{align}
into the path integral where
$\lambda$ is a real parameter, and $S_\mathrm{con}$ is a constraint functional
chosen so as to forbid the rescaling, defined by the integral of a local operator $\mathcal{O}_\mathrm{con}$:
\begin{align}
    S_\mathrm{con}
    =
    \int\dd^4x\,\mathcal{O}_\mathrm{con}\,.
    \label{eq: constraint operator FFtilde}
\end{align}
Consequently, the path integral takes the form
\begin{align}
    Z &= \int \mathcal{D}\Phi\, \dd\lambda\, 
        \delta(\lambda - S_\mathrm{con}[\Phi])\, e^{-S[\Phi]}\,.
    \label{eq: path integral with delta inserted}
\end{align}

By exchanging the order of integration over $\Phi$ and $\lambda$, we can proceed with the semiclassical evaluation of Eq.\,\eqref{eq: path integral with delta inserted} by finding the stationary points of the modified action~\cite{Gervais:1977me} (see Appendix~\ref{sec: constraint method of inserting 1}),
\begin{align}
    S_\mathrm{tot}^{(\lambda)}[\Phi,\sigma]
    :=
    S[\Phi]+\frac{\sigma}{g^2}\bigl(S_\mathrm{con}[\Phi]-\lambda\bigr)\,.
    \label{eq: modified action in gauge theory}
\end{align}
Here, $\sigma$ acts as a Lagrange multiplier enforcing the constraint $S_\mathrm{con} = \lambda$.
As shown below, this modified action admits a stationary point for a fixed $\lambda$, analogous to the situation in $\phi^4$ theory.

As a concrete example, we choose the constraint operator
\begin{align}
    \mathcal{O}_\mathrm{con}=\qty(\frac{1}{2}\Tr F_{\mu\nu}\tilde{F}_{\mu\nu})^2\,.
    \label{eq: Ocon = FFtilde}
\end{align}
Under the scale transformation in Eq.\,\eqref{eq: rescaling in gauge theory}, the corresponding functional scales as
\begin{align}
    S_\mathrm{con}[A']=a^4S_\mathrm{con}[A]\,.
\end{align}
Consequently, for a fixed value of $\lambda$, the delta function constraint fixes the scale of the configuration, thereby preventing the action from decreasing under the rescaling with $a>1$.
In what follows, we restrict our attention to this specific choice.

\section{Explicit Construction of Constrained Instanton}
\label{sec: explicit construction in YM}

With the size of the configuration fixed by the constraint, the modified action $S_\mathrm{tot}^{(\lambda)}$ admits a stationary point, namely the constrained instanton.
Conventionally, higher-dimensional operators such as $\tr(F_{\mu\nu}F_{\nu\rho}F_{\rho\mu})$ and $(\tr{F_{\mu\nu}\tilde{F}_{\mu\nu}})^2$ have been adopted for $\mathcal{O}_\mathrm{con}$ \cite{Affleck:1980mp,Ringwald:1989ee,Klinkhamer:1993kn,Aoyama:1995ca}.
However, N\&N argued that these conventional gauge-invariant choices exhibit problematic asymptotic behaviors.
Specifically, as discussed for the $\phi^4$ theory, they reported an inconsistency when matching the inner and outer solutions at the NLO in the $(\rho m_{A,H})^2$ expansion, suggesting an obstruction to constructing a valid constrained instanton.

In this section, we demonstrate that a consistent constrained instanton can nevertheless be constructed at both the LO and NLO, entirely free from such asymptotic issues.
Analogous to our resolution in the $\phi^4$ case, refining the analysis by N\&N reveals that a careful treatment of the higher-order terms in the outer region ($r\gg\rho$) is crucial for consistently matching the inner and outer solutions.
Consequently, our analysis shows that the obstruction discussed by N\&N does not arise when using conventional gauge-invariant constraint operators.
Finally, we evaluate the configuration numerically and confirm agreement with the perturbative results.

\subsection{Profile Functions of Constrained Instanton}

The constrained instanton configuration $\bar\Phi^{(\lambda)}$
is defined as the stationary point of the modified action \eqref{eq: modified action in gauge theory}, satisfying
\begin{align}
    \frac{\delta S_\mathrm{tot}^{(\lambda)}}{\delta \Phi}=0\,.
\end{align}
To solve this variational problem for $\Phi=(A_\mu,H)$, we adopt the ansatz
\begin{align}
    A_\mu(x)
    =
    \bar\tau_{\mu\nu}
    \frac{x_\nu}{r^2}
    \mathcal{A}(r)\,,\quad
    H(x)
    =
    \mqty(0 \\ v) \mathcal{H}(r)\,,
    \label{eq: ansatz}
\end{align}
where $\mathcal{A}(r)$ and $\mathcal{H}(r)$ are scalar radial profile functions subject to the boundary conditions
\begin{align}
    \mathcal{A}(r) = 
    \begin{cases}
    2 &  r=0\,,
    \\
    0 &  r=\infty\,,
    \end{cases}
    \quad\quad\quad
    \mathcal{H}(r) = 
    \begin{cases}
    0 &  r=0\,,
    \\
    1 &  r=\infty\,.
    \label{eq: boundary conditions for A, H}
    \end{cases}
\end{align}
This gauge field ansatz is motivated by the instanton solution in the singular gauge, Eq.\,\eqref{eq: singular gauge pure YM instanton}, and ensures a unit topological charge,
\begin{align}
    \frac{1}{16\pi^2}\int\mathrm{d^4}x\,\tr F_{\mu\nu}\tilde{F}_{\mu\nu}
    =
    1\,.
\end{align}
For the scalar field, the ansatz restricts $H$ to a constant direction within the $\mathrm{SU}(2)$ doublet space. This is motivated by the observation that fluctuations away from this direction strictly increase the action.
Furthermore, the condition $\mathcal{H}(\infty)=1$ is necessary for the action to be finite.
Finally, the condition $\mathcal{H}(0)=0$ is required 
to ensure that the scalar field configuration in the non-singular gauge is regular at $r=0$.%
\footnote{
The gauge transformation from the singular gauge to the non-singular gauge, $H\to UH$, is given by
\begin{align}
    U=\frac{i\bar\sigma_\mu x_\mu}{\sqrt{x_\nu x_\nu}}\,.
\end{align}
}

Substituting the ansatz into the action yields
\begin{align}
S
&=\int_0^\infty
\dd r\,
    2 \pi ^2 r^3 L(\mathcal{A}(r),\mathcal{A}'(r),\mathcal{H}(r),\mathcal{H}'(r))\,,
\end{align}
where $L$ is given by
\begin{align}
L=
\frac{3 \left(r^2 \mathcal{A}'^2+\mathcal{A}^4-4 \mathcal{A}^3+4
   \mathcal{A}^2\right)}{2 r^4}
   +\frac{3 v^2 \mathcal{A}^2 \mathcal{H}^2}{4 r^2}+
   \frac{1}{4} \lambda_H  v ^4 \left(\mathcal{H}^2-1\right)^2+v^2 \mathcal{H}'^2
   \,.
   \label{eq: YMH lagrangian in terms of profiles}
\end{align}
For the specific constraint operator chosen in Eq.\,\eqref{eq: Ocon = FFtilde},
the constraint functional takes the form,
\begin{align}
    S_\mathrm{con}
    &=\int_0^\infty\dd r\,
    2\pi^2r^3
    L_\mathrm{con}\,,
    \\
    L_\mathrm{con}&=
    \frac{9 (\mathcal{A}-2)^2 \mathcal{A}^2 \
    \mathcal{A}^{\prime2}}{r^6}\,.
\end{align}
The Euler--Lagrange equations for the profile functions derived from $S_\mathrm{tot}$ \eqref{eq: modified action in gauge theory} are
\begin{align}
    \mathcal{A}''(r)
    +\frac{1}{r}\mathcal{A}'(r)
    -\frac{v^2}{2}\mathcal{A}(r)\mathcal{H}(r)^2
    +\frac{2}{r^2}\qty(-\mathcal{A}(r)^3
    +3\mathcal{A}(r)^2
    -2\mathcal{A}(r)
    )
    -\frac{\sigma}{6\pi^2r}\frac{\delta S_\mathrm{con}}{\delta \mathcal{A}(r)}&=0\,,
    \label{eq: gauge field ansatz EOM}
    \\
    \mathcal{H}''(r)+\frac{3}{r}\mathcal{H}'(r)
    -\frac{3}{4r^2}\mathcal{A}(r)^2\mathcal{H}(r)
    +\frac{\lambda_H v^2}{2}(\mathcal{H}(r)-\mathcal{H}(r)^3)-\frac{\sigma}{4\pi^2 r^3 v^2}\frac{\delta S_\mathrm{con}}{\delta \mathcal{H}(r)}&=0\,.
    \label{eq: scalar field ansatz EOM}
\end{align}
Instead of determining the Lagrange multiplier $\sigma$ for a given constraint value $\lambda$, their one-to-one correspondence (see Appendix\,\ref{sec: constraint method of inserting 1}) justifies the reverse procedure in practice: for a given $\sigma$, the corresponding $\lambda$ is obtained by substituting the resulting configuration into the spatial integral of $\mathcal{O}_\mathrm{con}$.

\subsection{Matching Procedure of Constrained Instanton}
\label{sec: strategy for construction}

In the following, we construct the constrained instanton solution, which behaves as
\begin{align}
    \mathcal{A}=\frac{2\rho^2}{r^2+\rho^2}+\order{r^4/\rho^4}\,,
    \label{eq: size definition for A}
\end{align}
for $r\ll\rho$.
As we will see later, this asymptotic behavior up to quadratic order in $r/\rho$ serves as the definition of the size parameter $\rho$; equivalently, $\rho$ is defined through the second derivative of the profile $\mathcal{A}$ at the origin as
\begin{align}
\rho = \frac{2}{\sqrt{-\mathcal{A}''(0)}}\,.
\label{eq: size definition for A explicit}
\end{align}

Analogous to our treatment of the $\phi^4$ theory, both
 $\mathcal{A}$ and $\mathcal{H}$ admit perturbative expansions in powers of $(\rho m_{A})^2$ and $(\rho m_{H})^2$, respectively.
Defining the mass scale $m=\max\{m_A,m_H\}$, the inner ($r \ll m^{-1}$) and outer ($r \gg \rho$) solutions can be expanded as
\begin{align}
    &\mathcal{A}(r)=
    \sum_{n=0,2,4,\ldots} \begin{cases}
    \displaystyle{
    (\rho m_A)^n \mathcal{A}^\mathrm{(in)}_n(r/\rho)
    }
    &r\ll m^{-1}
    \\[.4em]
    \displaystyle{
    (\rho m_A)^n \mathcal{A}^\mathrm{(out)}_n(m_A r)
    }
    &r\gg\rho
    \end{cases}
    ,
    \label{eq: Expansion of inner/outer solutions A}
    \\[.6em]
   &\mathcal{H}(r)= \begin{cases}
    \displaystyle{\sum_{n=0,2,4,\ldots}
     (\rho m_H)^n \mathcal{H}^{(\mathrm{in})}_n(r/\rho)
    }
    &r\ll m^{-1}
    \\[.4em]    \displaystyle{1+\sum_{n=0,2,4,\ldots}
    (\rho m_H)^n \mathcal{H}^{(\mathrm{out})}_n(m_Hr)
    }
    &r\gg\rho
    \end{cases}
    ,
    \label{eq: Expansion of inner/outer solutions H}
\end{align}
similarly to Eqs.\,\eqref{eq: phi perturbative series} and \eqref{eq: outer solution perturbed} in the $\phi^4$ case.
Although we label the two series by the corresponding mass scales for convenience, they are in fact expansions in $(\rho v)^2$. In general, because the field equations are coupled, the dimensionless combinations of $m_A$ and $m_H$ may appear in the coefficient functions of either sector, and hence they may depend on $\lambda_H$.

Imposing the boundary conditions in Eq.\,\eqref{eq: boundary conditions for A, H} yields 
\begin{align}
    \mathcal{A}_n^{(\mathrm{in})}(0)
    =
    \begin{cases}
        2
        &n=0
        \\
        0
        &n>0
    \end{cases}\,,
    \quad
    \mathcal{H}_n^{(\mathrm{in})}(0)=0\,,
    \quad
    \mathcal{A}_n^{(\mathrm{out})}(\infty)
    =0\,,
    \quad
    \mathcal{H}_n^{(\mathrm{out})}(\infty)=0\,.
    \label{eq: boundary condition at each order}
\end{align}
As we will see in Sec.\,\ref{sec: matching at LO}, matching conditions dictate that $\mathcal{A}_0^{(\mathrm{out})}=\mathcal{H}_0^{(\mathrm{out})}=0$.

Since the specific constraint operator chosen in Eq.\,\eqref{eq: Ocon = FFtilde} depends only on the gauge field, it is convenient to expand the Lagrange multiplier  $\sigma$ in powers of $(\rho m_A)^2$ as
\begin{align}
    \sigma 
    &=
    \sigma_0 + (\rho m_A)^2\sigma_2+(\rho m_A)^4\sigma_4\,+\cdots
    .
    \label{eq: Expansion of Lagrange multiplier}
\end{align}

Two comments are in order regarding the expansions in Eqs.\,\eqref{eq: Expansion of inner/outer solutions A}, \eqref{eq: Expansion of inner/outer solutions H}, and \eqref{eq: Expansion of Lagrange multiplier}. First, the expansion coefficients, namely $\mathcal{A}^{(\mathrm{in},\mathrm{out})}_n$, $\mathcal{H}^{(\mathrm{in},\mathrm{out})}_n$, and $\sigma_n$, contain no power-law dependence on $v$, depending on it solely through the logarithmic factor $\log(\rho v)$. Second, the zeroth-order Lagrange multiplier vanishes as
\begin{align}
    \sigma_0 = 0\,,
    \label{eq: sigma0}
\end{align}
since the unperturbed solution in Eq.\,\eqref{eq: instanton w/o Higgs} with $\sigma=0$
satisfies the Euler--Lagrange equations \eqref{eq: gauge field ansatz EOM} and \eqref{eq: scalar field ansatz EOM} in the limit $v\to0$.

Similarly to the $\phi^4$ theory, the matching in Yang--Mills theory with spontaneous symmetry breaking is performed in the overlap region $\rho \ll r \ll m^{-1}$ using a double expansion in $(\rho m_{A,H})^2$ and $\hat r^{-2}$ (see Fig.~\ref{fig: Regions}),
where $\hat{r}$ is again defined by $\hat{r}:=r/\rho$.
For the gauge and scalar fields, collectively denoted by $\Phi = \mathcal{A}, \mathcal{H}$, the inner and outer solutions are expanded as
\begin{align}
  \Phi(r)
  &= \sum_{n=0}^\infty (\rho m_\Phi)^n\,
     \Phi_n^{(\mathrm{in})}(\hat r)
   \;=\; \sum_{n,k} (\rho m_\Phi)^n\,
     \Phi_{n}^{(\mathrm{in})(k)}\,\hat r^{-k}\,,
  \label{eq:Phi-inner-series}
\\
  \Phi(r)
  &= \sum_{n=0}^\infty (\rho m_\Phi)^n\,
     \Phi_n^{(\mathrm{out})}(m_\Phi r)
   \;=\; \sum_{n,k} (\rho m_\Phi)^{n-k}\,
     \Phi_{n}^{(\mathrm{out})(k)}\,\hat r^{-k}\,.
  \label{eq:Phi-outer-series}
\end{align}
Since these two expressions must describe the same function in the overlap region, the coefficients of each monomial $(\rho m_\Phi)^n \hat r^{-k}$ must identically agree. Comparing Eq.\,\eqref{eq:Phi-inner-series} with Eq.\,\eqref{eq:Phi-outer-series} yields the $(n,k)$ condition for Yang--Mills theory with spontaneous symmetry breaking,
\begin{align}
  \Phi_{n+k}^{(\mathrm{out})(k)}
  = \Phi_{n}^{(\mathrm{in})(k)}
  \qquad (n\ge 0,\;k\ge -n)\,.
  \label{eq: n,k condition in YM}
\end{align}
For the gauge field $\Phi=\mathcal{A}$, this relation holds universally for all
$n\ge0$ and $k\ge-n$.
For the scalar field $\mathcal{H}$, however, due to the outer-region definition in Eq.\,\eqref{eq: Expansion of inner/outer solutions H}, the $(0,0)$ condition for $\mathcal{H}$ is given by
\begin{align}
  \mathcal{H}_0^{(\mathrm{in})(k=0)}
  = 1 + \mathcal{H}_0^{(\mathrm{out})(k=0)}\,,
\end{align}
while the other conditions are given by Eq.\,\eqref{eq: n,k condition in YM}.

Note that the constraint operator $\mathcal{O}_{\mathrm{con}}$ in Eq.\,\eqref{eq: Ocon = FFtilde}
has a higher mass dimension than the mass term.
Accordingly, the outer solutions $\mathcal{A}^{(\mathrm{out})}_n$ and $\mathcal{H}^{(\mathrm{out})}_n$ are not affected by $\mathcal{O}_\mathrm{con}$ and are
governed by the free-field equations of motion
\begin{align}
    \partial^2 A_\mu-\partial_\mu\partial_\nu A_\nu-m_A^2A_\mu &= 0\qquad (r\to\infty)\,,
    \label{eq: free EOM for A}
    \\
    (\partial^2-m_H^2)\qty[H-\mqty(0 \\ v)] &=0\qquad (r\to\infty)\,.
    \label{eq: free EOM for H}
\end{align}
With the ansatz \eqref{eq: ansatz}, the solutions to these equations are given by
\begin{align}
    \mathcal{A}(r)
    &= 
    \kappa_A\,
    K_2(m_A r)
    \phantom{1+\kappa_H\,\frac{K_1(m_H r)}{m_H r}} 
    \!\!\!\!\!\!\!\!\!\!\!\!\!\!\!\!\!\!\!\!(r\to\infty)\,,
    \label{eq: asymptotic solution of A}
    \\    
    \mathcal{H}(r)
    &= 1+\kappa_H\,\frac{K_1(m_H r)}{m_H r}
    \phantom{\kappa_A\,
    K_2(m_A r)}
    \!\!\!\!\!\!\!\!\!\!\!\!\!\!\!\!\!\!\!\!(r\to\infty)\,.
    \label{eq: asymptotic solution of H}
\end{align}
While the full solutions at finite radii deviate from these forms, such corrections are further suppressed by a factor of $e^{-m_{A,H}r}$ relative to the leading terms as $r \to \infty$, just as in the $\phi^4$ theory.

\subsection{Leading-Order Solutions}
\label{sec: LO}

Let us first determine the LO solutions
$\mathcal{A}_0^{(\mathrm{in})}(r)$, 
$\mathcal{H}_0^{(\mathrm{in})}(r)$,
$\mathcal{A}_2^{(\mathrm{out})}(r)$,
and 
$\mathcal{H}_2^{(\mathrm{out})}(r)$.
As will be shown later, the $(0,0)$ condition implies that
$\mathcal{A}_0^{(\mathrm{out})}=\mathcal{H}_0^{(\mathrm{out})}=0$.
For this reason, the first non-vanishing terms in outer solutions are
$\mathcal{A}_2^{(\mathrm{out})}(r)$ and $\mathcal{H}_2^{(\mathrm{out})}(r)$,
which we therefore regard as LO in the outer region.
Accordingly, the inner and outer LO solutions are matched through
the $(0,2)$ condition, which we call the LO matching condition.

\subsubsection*{Inner solution}

The solution in Eq.\,\eqref{eq: instanton w/o Higgs}  suggests that the inner expansion in Eq.\,\eqref{eq: Expansion of inner/outer solutions A} should begin with
\begin{align}
    \mathcal{A}(r)
    &= \sum_{n=0,2,4,\ldots}(\rho m_A)^n \mathcal{A}^{(\mathrm{in})}_n(\hat{r})
    \qquad\text{with}\qquad
    \mathcal{A}^{(\mathrm{in})}_0(\hat{r})=\frac{2}{\hat{r}^2+1}\,.
    \label{eq: leading order inner solution of A}
\end{align}
As for the scalar field, Eq.\,\eqref{eq: scalar field ansatz EOM} at LO in $\rho m_{A,H}$ reduces to
\begin{align}
    \mathcal{H}_0^{(\mathrm{in})\prime\prime}(\hat r)
    +\frac{3}{\hat r}\,\mathcal{H}_0^{(\mathrm{in})\prime}(\hat r)
    -\frac{3}{4\hat r^{2}}\mathcal{A}_0^{(\mathrm{in})}(\hat r)^2
      \mathcal{H}_0^{(\mathrm{in})}(\hat r)
    =0\,,
\end{align}
where primes ($'$) denote derivatives with respect to $\hat r$.
The general solution to this equation is given by
\begin{align}
    \mathcal{H}_0^{(\mathrm{in})}(\hat r)
    = C_1\qty(\frac{\hat{r}^2}{\hat{r}^2+1})^{1/2}
      + C_2\qty(\frac{\hat{r}^2+1}{\hat{r}^2})^{3/2}\,.
    \label{eq: leading order inner solution of H, general solution}
\end{align}
The coefficient $C_1$ will be fixed by matching to the outer solution, whereas the boundary condition $\mathcal{H}_0^{(\mathrm{in})}(0)=0$ in Eq.\,\eqref{eq: boundary condition at each order} requires $C_2=0$.
Consequently, the LO inner expansion of $\mathcal{H}$ takes the form
\begin{align}
    \mathcal{H}(r)
    &= \sum_{n=0,2,4,\ldots}(\rho m_H)^n \mathcal{H}^{(\mathrm{in})}_n(\hat{r})
    \quad\text{with}\quad
    \mathcal{H}^{(\mathrm{in})}_0(\hat{r})
    = C_1\qty(\frac{\hat{r}^2}{\hat{r}^2+1})^{1/2}\,.
    \label{eq: leading order inner solution of H}
\end{align}

\subsubsection*{Outer solution}

The LO outer solutions satisfy the free-field equations of motion and are given by
\begin{align}
    (\rho m_A)^2\,\mathcal{A}_2^{(\mathrm{out})}(m_A r) &= \kappa_{A0}\,K_2(m_A r)\,,
    \\
    (\rho m_H)^2\,\mathcal{H}_2^{(\mathrm{out})}(m_H r) &= \kappa_{H0}\,\frac{K_1(m_H r)}{m_H r}\,.
\end{align}
Here, the constant $\kappa_{\Phi0}$ $(\Phi = A,H)$ denotes the LO term in the expansion of $\kappa_\Phi$, the coefficients of the asymptotic solutions at spatial infinity introduced in Eqs.\,\eqref{eq: asymptotic solution of A} and \eqref{eq: asymptotic solution of H}:
\begin{align}
    \kappa_{\Phi}
    = \kappa_{\Phi0}\Bigl(1+\order{\rho^2 m_\Phi^2}\Bigr)\,.
\end{align}
Note that the constraint operator does not contribute at LO.

\subsubsection*{Matching}
\label{sec: matching at LO}

To perform the LO matching, 
we first expand the outer solutions for small $m_{A,H} r$ as
\begin{align}
K_2(m_Ar) &= 
\dfrac{2}{m_A^2r^2}-\dfrac{1}{2}
-\dfrac{1}{8}m_A^2r^2\log{(m_Ar)}+\frac{1}{32}
    \qty(3-4\gamma_E+4\log 2)m_A^2 r^2+\order{m_A^4 r^4}\ ,
    \label{eq: asymptotic expanding Bessel of A}\\
    \dfrac{K_1(m_Hr)}{m_H r} &= \dfrac{1}{m_H^2 r^2} + \dfrac{1}{2}\log{(m_Hr)} 
-\frac{1}{4}\qty(1-2\gamma_E + 2 \log2)
    + \order{m_H^2 r^2}
   \ ,
    \label{eq: asymptotic expanding Bessel of H}
\end{align}
which can be rewritten in terms of the inner variable $\hat r = r/\rho$ as
\begin{align}
    (\rho m_A)^2\mathcal{A}_2^{(\mathrm{out})} &\simeq \overbrace{2\kappa_{A0}(\rho m_A)^{-2}\hat{r}^{-2}}^{(0,2)}
    \overbrace{
    -\frac{\kappa_{A0}}{2}}^{(2,0)}
    +\order{\kappa_{A0}(\rho m_A)^2}\,,
    \label{eq: expanding the leading order outer solurion of A around the origin}
\\
    (\rho m_H)^2\mathcal{H}_2^{(\mathrm{out})}
    &\simeq
    \underbrace{\kappa_{H0}
    (\rho m_H)^{-2}
    \hat{r}^{-2}}_{(0,2)}
    +
    \underbrace{
    \frac{\kappa_{H0}}{2}\qty(\log\frac{\rho m_H}{2}
    +\log\hat{r}
    -\frac{1}{2}+\gamma_E
    )
    }_{(2,0)}
    +\order{\kappa_{H0}(\rho m_H)^2}
    \,.
    \label{eq: expanding the leading order outer solurion of H around the origin}
\end{align}
From the $(0,0)$ condition we obtain
\begin{align}
    C_1=1\,,\qquad
    \mathcal{H}^{(\mathrm{out})}_0(m_Hr)
    =0\,,
    \qquad
    \mathcal{A}^{(\mathrm{out})}_0(m_Ar)
    =0\,,
\end{align}
and from the $(0,2)$ condition we obtain
\begin{align}
\kappa_{A0} &= (\rho m_A)^2 \ ,
\label{eq: matching of A at LO}
\\
\kappa_{H0}   
    &=-\frac{1}{2} (\rho m_H)^2
    \ .
    \label{eq: matching of H at LO}
\end{align}

With these constants fixed by the LO matching,
the LO inner and outer solutions turn out to be
\begin{align}
    \mathcal{A}^{(\mathrm{out})}_2(m_Ar)&=K_2(m_Ar)
    \,,
    \label{eq: leading order outer solution of A}\\
    \mathcal{H}^{(\mathrm{out})}_2(m_Hr)
    &=
    -\frac{1}{2}\frac{K_1(m_Hr)}{m_Hr}
    \,,
    \label{eq: leading order outer solution of H}
    \\
    \mathcal{A}^{(\mathrm{in})}_0(\hat{r})
    &=\frac{2}{\hat{r}^2+1}
    \,,
    \\
    \mathcal{H}^{(\mathrm{in})}_0(\hat{r})
    &=\qty(\frac{\hat{r}^2}{\hat{r}^2+1})^{1/2}
    \,.
\end{align}
These LO inner and outer solutions coincide with those obtained by
N\&N.

\subsection{Next-to-Leading Order}
\label{sec: NLO}

We next determine the NLO inner and outer solutions similarly to above.

\subsubsection*{Inner solution of the gauge field}

The NLO correction to the inner solution, 
$\mathcal{A}_2^{(\mathrm{in})}$, satisfies
\begin{align}
    \mathcal{A}_2''(\hat{r})
    +\frac{1}{\hat{r}}\,\mathcal{A}_2'(\hat{r})    
    -\frac{4 \bigl(\hat{r}^4-4 \hat{r}^2+1\bigr)}{\hat{r}^2\bigl(\hat{r}^2+1\bigr)^2}\,
     \mathcal{A}_2(\hat{r})
    -\frac{2\hat{r}^2}{\bigl(\hat{r}^2+1\bigr)^2}
    -\frac{\sigma_2}{6\pi^2\hat{r}}\,\rho
      \left.\frac{\delta S_\mathrm{con}}{\delta\mathcal{A}(r)}\right|_{\mathcal{A}=\mathcal{A}_0^{(\mathrm{in})}}
    =0\,.
    \label{eq: NLO EOM for A}
\end{align}
For the specific choice of the constraint operator $\mathcal{O}_\mathrm{con}=(\tr F_{\mu\nu}\tilde{F}_{\mu\nu}/2)^2$, we have
\begin{align}
    \rho^{5}\left.\frac{\delta S_\mathrm{con}}{\delta\mathcal{A}(r)}\right|_{\mathcal{A}=\mathcal{A}_0^{(\mathrm{in})}}
    =\frac{-2^{11}3^2\pi^2\,\hat{r}^3}{\bigl(1+\hat{r}^2\bigr)^7}\,.
\end{align}
The general solution to Eq.\,\eqref{eq: NLO EOM for A} is
\begin{align}
    \mathcal{A}^{(\mathrm{in})}_2(\hat{r})
    =a_2^{(m^2)}(\hat{r})+a_2^{(\sigma_2)}(\hat{r})
    +c_1
    \frac{\hat{r}^2}{\bigl(\hat{r}^2+1\bigr)^2}
    +c_2
    \frac{-\hat{r}^8-8 \hat{r}^6-24 \hat{r}^4 \log (\hat{r})
          +8 \hat{r}^2+1}{\hat{r}^2\bigl(\hat{r}^2+1\bigr)^2}\,,
    \label{eq: NLO inner general solution of A}
\end{align}
where the particular solutions corresponding to the last two terms in Eq.\,\eqref{eq: NLO EOM for A} are given by
\begin{align}
    a_2^{(m^2)}(\hat{r})
    &=\frac{\hat{r}^4 \bigl(\hat{r}^2+2\bigr)}{12 \bigl(\hat{r}^2+1\bigr)^2}\,,
    \\
    a_2^{(\sigma_2)}(\hat{r})
    &=
    -\rho^{-4}\sigma_2\,
    \frac{16 \Bigl(-12 \hat{r}^8-30 \hat{r}^6-22 \hat{r}^4+11 \hat{r}^2
      +12 \bigl(\hat{r}^2+1\bigr)^3
      \hat{r}^4 \log \bigl(\hat{r}^{-2}+1\bigr)+1\Bigr)}
         {7 \hat{r}^2 \bigl(\hat{r}^2+1\bigr)^5}\,.
     \label{eq: NLO constraint particular solution of A}
\end{align}
Here, as in the $\phi^4$ theory, $a_2^{(m^2)}(\hat{r})$ is chosen to be suppressed as $\hat{r}\to0$, while $a_2^{(\sigma_2)}(\hat{r})$ is chosen to be suppressed as $\hat{r}\to\infty$, by adding appropriate homogeneous solutions.
Namely, the asymptotic behavior of these particular solutions are given by 
\begin{align}
    a_2^{(m^2)}(\hat{r})
    &=
    \begin{cases}
        \dfrac{\hat{r}^4}{6}-\dfrac{\hat{r}^6}{4}
        +\order{\hat{r}^8}
        &
        \text{at}~ \hat{r}\to0\,,
        \\[.7em]
        \dfrac{\hat{r}^2}{12}-\dfrac{1}{12 \hat{r}^2}+\order{\hat{r}^{-4}}
        &
        \text{at}~ \hat{r}\to\infty\,,
    \end{cases}
\\
    a_2^{(\sigma_2)}(\hat{r})
    &=
    \begin{cases}
        -\dfrac{16}{7 \hat{r}^2}-\dfrac{96}{7}+\order{\hat{r}^2}
        &
        \text{at}~ \hat{r}\to0\,,
        \\[.7em]
        -\dfrac{32}{\hat{r}^{10}}+\dfrac{5568}{35 \hat{r}^{12}}+\order{\hat{r}^{-14}}
        &
        \text{at}~ \hat{r}\to\infty\,.
    \end{cases}
    \label{eq: constraint particular solution asymptotics of A}
\end{align}

The boundary condition $\mathcal{A}_2^{(\mathrm{in})}(0)=0$ in Eq.\,\eqref{eq: boundary condition at each order} is satisfied, provided that
\begin{align}
    \frac{\sigma_2}{\rho^4}=\frac{7}{16}\,c_2\,.
    \label{eq: sigma2 from c2}
\end{align}
Note that if this condition is not met, the action diverges due to the contribution near the spatial origin.
As in the $\phi^4$ theory,
the redundancy associated with the size parameter $\rho$ is fixed by imposing Eq.\,\eqref{eq: size definition for A}, which leads to
\begin{align}
    c_1=7\,\qty(7c_2-\frac{32\sigma_2}{\rho^4})\,.
    \label{eq: c_1 choice in YMH}
\end{align}

\subsubsection*{Outer solution of the gauge field}

The NLO correction to the outer solution,
$\mathcal{A}^{(\mathrm{out})}_4$,
satisfies
\begin{align}
    &\mathcal{A}_4^{(\mathrm{out})\prime\prime}(R)+\frac{\mathcal{A}_4^{(\mathrm{out})\prime}(R)}{R}-\frac{(4+R^2) \mathcal{A}_4^{(\mathrm{out})}(R)}{R^2}
    +\frac{6 K_2(R){}^2}{R^2}+\frac{\sqrt{2\lambda_H}K_2(R) K_1\left(R_H\right)}{R}
    =0\,,
    \label{eq: NLO outer equation for A}
\end{align}
where $R:=m_Ar$ and $R_H:=m_Hr=\sqrt{2\lambda}R$.
In the outer region, the contribution from the constraint operator is suppressed by the additional powers of $(\rho m_A)^2$
and thus does not enter at NLO (see Appendix \,\ref{app: constraint order counting}).

For $R := m_A r \ll 1$, the general solution is expanded as
\begin{align}
    \mathcal{A}^{(\mathrm{out})}_4(R)
    =
    \overbrace{-\frac{2}{R^4}}^{(0,4)}
    \underbrace{-\frac{\log R^2}{R^2}
    +\frac{c^{(\mathrm{out})}}{R^2}}_{(2,2)}
    +\overbrace{\order{R^{0}}}^{(4,0),\ldots}\,.
\end{align}
As in the $\phi^4$ theory,
$c^{(\mathrm{out})}$
is one of the arbitrary coefficients of the two homogeneous solutions, while the other is contained in the $\order{R^0}$ terms.

\subsubsection*{Matching of the gauge field}

In the matching region $\rho m\ll\hat{r}^{-1}\ll1$, the LO and NLO contributions to the inner solution can be expanded for $\hat{r}\gg1$ as
\begin{align}
    \mathcal{A}_0^{(\mathrm{in})}(\hat{r})+&(\rho m)^2\mathcal{A}^{(\mathrm{in})}_2(\hat{r})=
    \overbrace{2\hat{r}^{-2}}^{(0,2)}\overbrace{-2\hat{r}^{-4}}^{(0,4)}+\overbrace{\order{\hat{r}^{-6}}}^{(n=0,k\ge6)}
    \notag\\
    &-(\rho m_A)^2
    \qty(\underbrace{\qty(c_2-\frac{1}{12})\hat{r}^2}_{(2,-2)}
        +\underbrace{6c_2}_{(2,0)} + \underbrace{\qty(24c_2\log \hat{r}+36c_2+\frac{1}{12})\hat{r}^{-2}}_{(2,2)}
        +\underbrace{\order{\hat{r}^{-4}}}_{(n=2,k\ge4)}
    )
    \,,
    \label{eq : NLO Gauge inner}
\end{align}
while the LO and NLO contributions to the outer solution can be expanded at $R(=\rho m\hat{r})\ll1$ as
\begin{align}
    (\rho m_A)^2\mathcal{A}_2^{(\mathrm{out})}(R)
    +
    (\rho &m_A)^4\mathcal{A}_4^{(\mathrm{out})}(R)
    =
    \overbrace{2\hat{r}^{-2}}^{(0,2)}\overbrace{-2\hat{r}^{-4}}^{(0,4)}
    \notag\\
    &-(\rho m_A)^2\qty(\underbrace{\frac{1}{2}}_{(2,0)}+\underbrace{\qty(2\log\hat{r}+2\log{\rho m_A}-c^{(\mathrm{out})})\hat{r}^{-2}}_{(2,2)})+\underbrace{\mathcal{O}(\rho^4m_A^4)}_{(n\ge4,{}^\forall k)}
    \,.
    \label{eq : NLO Gauge outer}
\end{align}

A few remarks about the above expressions are in order.
For the outer solution in Eq.\,\eqref{eq : NLO Gauge outer}, note that the right-hand side is expressed as a double expansion in $(\rho m_A)^2$ and $\hat{r}^{-2}$, whereas the left-hand side is written in terms of $R$. Consequently, the powers of $\rho m_A$ appear to be organized differently on the two sides, although they are consistent as $R=\rho m_A\hat{r}$. 
Specifically, the terms corresponding to the $(0,2)$ and $(2,0)$ conditions originate from $(\rho m_A)^2\mathcal{A}^{(\mathrm{out})}_2(R)$, while those corresponding to the $(0,4)$ and $(2,2)$ conditions stem from $(\rho m_A)^4\mathcal{A}^{(\mathrm{out})}_4(R)$.

The NLO matching involves the $(n,k)=(0,4), (2,-2), (2,0)$, and $(2,2)$ conditions.
The $(0,4)$ condition is satisfied without any adjustment of the integration constants, as in the $(0,4)$ condition in the $\phi^4$ theory.

The $(n,k)=(2,0)$ and $(2,2)$ conditions are simultaneously satisfied by setting
\begin{align}
    c_2=\frac{1}{12}\,,\quad c^{(\mathrm{out})}=2\log{\rho m_A}-\frac{37}{12}\,.
    \label{eq: c2 and cout by matching}
\end{align}
Note that the $(2,2)$ condition splits into two independent equations corresponding to the coefficients of $(\log \hat r)^0$ and $(\log \hat r)^1$; both are satisfied by the above choice of constants.
We also emphasize that the value of $c_2$ in Eq.\,\eqref{eq: c2 and cout by matching} does not depend on how the size parameter $\rho$ is defined, namely, on the choice of $c_1$.

With the constant $c_2$ fixed as above, the term in the inner solution corresponding to the $(2,-2)$ condition is found to vanish. If such a term were present, it implied $\mathcal{A}_0^{(\mathrm{out})}\neq0$. Its absence is therefore consistent with the LO matching, which dictated that $\mathcal{A}_0^{(\mathrm{out})}=0$, implying that the outer solution begins at $\order{\rho^2m_A^2}$ with $\mathcal{A}_2^{(\mathrm{out})}$

Note that the $\order{\hat{r}^{-6}}$ and $(\rho m_A)^2\order{\hat{r}^{-4}}$ terms appearing in the LO and NLO inner expansions match with higher-order outer solutions beyond NLO. However, they do not give the independent NLO matching conditions, while the beyond-NLO matching provides consistency checks for the NLO matching.

Combining Eq.\,\eqref{eq: sigma2 from c2} and the NLO matching in Eq.\,\eqref{eq: c2 and cout by matching}, we obtain
\begin{align}
    \frac{\sigma_2}{\rho^4}=\frac{7}{192}\,.
    \label{eq: sigma2 value for YMH}
\end{align}
This value of $\sigma_2$ can also be derived without invoking the matching procedure.
Indeed, as discussed in Appendix \ref{app: Lagrange multiplier value}, the Lagrange multiplier satisfies
\begin{align}
    \frac{\sigma}{g^2}=-\frac{\dd S}{\dd \rho}\qty[\frac{\dd S_\mathrm{con}}{\dd \rho}]^{-1}\,,
    \label{eq: sigma dSdlambda}
\end{align}
for the constrained instanton solution parametrized by $\rho$ (or $\lambda$).
Expanding Eq.\,\eqref{eq: sigma dSdlambda} in powers of $(\rho m_A)^2$ reproduces the identical value of $\sigma_2$ as in Eq.\,\eqref{eq: sigma2 value for YMH}.
The agreement between these two independent derivations provides a nontrivial consistency check of the matching.

\subsubsection*{Scalar Field}
In the scalar-field case as well, the matching can be carried out consistently up to NLO.
The NLO correction to the inner solution $\mathcal{H}_2(\hat{r})$ satisfies
\begin{align}
    \mathcal{H}_2^{(\mathrm{in})\prime\prime}(\hat{r})
    +
    \frac{3}{\hat{r}}\mathcal{H}_2^{(\mathrm{in})\prime}(\hat{r})
    -
    \frac{3 \mathcal{H}_2^{(\mathrm{in})}(\hat{r})}{\hat{r}^2\left(\hat{r}^2+1\right)^2}
    +\frac{\hat{r}}{2(\hat{r}^2+1)^{3/2}}
    -\frac{3\mathcal{A}_2^{(\mathrm{in})}(\hat{r})}{2\lambda_H \hat{r}(\hat{r}^2+1)^{3/2}}
    =0\,.
\end{align}
Note that the constraint does not appear for the choice of $\mathcal{O}_\mathrm{con}$ in Eq.\,\eqref{eq: Ocon = FFtilde}.
The general solution is
\begin{align}
    \mathcal{H}_2^{(\mathrm{in})}(\hat{r})=
    h_2^{(m^2)}(\hat{r})+h_2^{(\mathcal{A}_2)}(\hat{r})
    +d_1\qty(\frac{\hat{r}^2}{1+\hat{r}^2})^{1/2}
    +d_2\qty(\frac{\hat{r}^2+1}{\hat{r}^2})^{3/2}.
\end{align}
For the explicit forms of the particular solutions $h_2^{(m^2)}$ and $h_2^{(\mathcal{A}_2)}$,
see Appendix\,\ref{app: higgs NLO detail}.
The boundary condition $\mathcal{H}_2^{(\mathrm{in})}(0)=0$ in Eq.\,\eqref{eq: boundary condition at each order} fixes $d_2=0$.

The NLO correction to the outer solution $\mathcal{H}_2(\hat{r})$ satisfies
\begin{align}
\mathcal{H}^{(\mathrm{out})\prime\prime}\left(R_H\right)+\frac{3 \mathcal{H}^{(\mathrm{out})\prime}\left(R_H\right)}{R_H}
-\mathcal{H}\left(R_H\right)
-\frac{3 \left(2 \
\lambda_H ^2 K_1\left(R_H\right){}^2+K_2\left(R\right){}^2\right)}{16 \lambda_H ^2 R_H^2}=0\,,
\end{align}
where $R:=m_Ar$ and $R_H:=m_Hr$\,.
The general solution takes the form
\begin{align}
    \mathcal{H}^{(\mathrm{out})}_2(R_H) =\frac{3}{8R_H^4}+\frac{3(1-\lambda_H)\log R_H^2}{16\lambda_H R_H^2}
    +\frac{d^{(\mathrm{out})}}{R_H^2}+\order{R_H^0}
    \,,
\end{align}
where $d^{(\mathrm{out})}$ represents the coefficient of one of the two linearly independent homogeneous solutions, while the contribution from the second linearly independent solution is contained in the higher-order $\order{R_H^0}$ terms.

The NLO corrections to the inner and outer solutions are matched similarly to the gauge field. The $(2,0)$ condition fixes $d_1$ as
\begin{align}
    d_1
    &=-\frac{1}{4} \left( \log \frac{\rho  m_H}{2}+ \gamma_E +\frac{1}{4}\right)\,,
    \label{eq: d1}
\end{align}
while the $(\log\hat{r})^0$-dependent part of the $(2,2)$ condition determines $d^{(\mathrm{out})}$ as 
\begin{align}
    d^{(\mathrm{out})}
    &=
    \frac{-15 \left(16 d_1+3\right) \lambda_H +180 (\lambda_H -1) \log \left(\rho  m_H\right)+244}{480 \lambda_H }
   \notag\\
    &=
    \frac{30 (4 \lambda_H -3) \log \left(\rho  m_H\right)+15 \lambda_H  (-1+2 \gamma_E -\log4)+122}{240 \lambda_H
   }\,.
   \label{eq: dout}
\end{align}
Here,
we have substituted Eq.\,\eqref{eq: d1} in the last equality.
The remaining $(0,4)$ condition and the $(\log\hat{r})^1$-dependent part of the $(2,2)$ condition are satisfied without adjusting the constants, providing consistency checks.
We thus obtain
\begin{align}
    \mathcal{H}_2^{(\mathrm{in})}(\hat{r})
    =&
    \frac{60 \left(6 \hat{r}^4+3 \hat{r}^2+1\right) \left(\hat{r}^2+1\right)^3
   \log \left(\hat{r}^2+1\right)+\hat{r}^2 \left(976 \hat{r}^8+2652
   \hat{r}^6+2228 \hat{r}^4+457 \hat{r}^2-60\right)}{1920 \hat{r}^3
   \left(\hat{r}^2+1\right)^{9/2}\lambda_H}
   \notag
   \\
   ~~
   &+
   \frac{-2 \hat{r}^4 \log \left(\frac{\rho m_H}{2}\right)-2 \gamma_E
   \hat{r}^4+\hat{r}^4+\hat{r}^2-\left(\hat{r}^2+1\right)^2 \log
   \left(\hat{r}^2+1\right)}{8 \hat{r}^3 \sqrt{\hat{r}^2+1}}
    \,,
   \\
    \mathcal{H}^{(\mathrm{out})}_2(R_H) =&
    \frac{3}{8R_H^4}+\frac{3(1-\lambda_H)\log R_H^2}{16\lambda_H R_H^2}
    +\frac{30 (4 \lambda_H -3) \log \left(\rho  m_H\right)+15 \lambda_H 
   (-1+2 \gamma_E -\log 4)+122}{240 \lambda_H R_H^2 }
   \notag\\&+\order{R_H^0}\,.
\end{align}

In this way, we have achieved a consistent matching between the inner and outer solutions, supporting the existence of a constrained instanton solution within our perturbative framework.
Using the configurations determined by the matching, we estimate the action as
\begin{align}
    g^2S&=8\pi^2+4\pi^2(\rho m_A)^2
    +(\rho m_A)^4\pi^2
    \qty[3\log{\frac{\rho m_A}{2}}+3\gamma_E-\frac{188333}{27720}]
    \notag\\&\quad
    -\rho^4m_A^2m_H^2\pi^2
    \qty[\log\frac{\rho m_H}{2}+\gamma_E+\frac{1}{2}]
    +\order{(\rho m_{A,H})^6}\,.
    \label{eq: action value from matching, 4th order}
\end{align}
Here, we additionally used a partial next-to-NLO (NNLO) inner solution to obtain $(\rho m_{A,H})^4$ order contribution. 
See Appendix\,\ref{app: action estimation}
for details.
Note that this expression depends on the definition of the size parameter in Eq.\,\eqref{eq: size definition for A}
and on the choice of the constraint operator in Eq.\,\eqref{eq: Ocon = FFtilde}.
In the following subsection, we cross-check this conclusion against an explicit numerical minimization of the constrained action, which validates our inner-outer matching procedure and supports the existence of the solution beyond fixed-order perturbation theory.

\subsection{Numerical Verification of the Matching}

To verify our perturbative analysis,
we numerically obtained constrained instanton configurations by minimizing $S_\mathrm{tot}$ \eqref{eq: modified action in gauge theory} for given values of $\sigma$,
using the ansatz in Eq.\,\eqref{eq: ansatz} and the boundary conditions in Eq.\,\eqref{eq: boundary conditions for A, H}.
Note that, for a class of constraint operators $\mathcal{O}_{\mathrm{con}}$ including the one in Eq.\,\eqref{eq: Ocon = FFtilde}, the constrained instanton configuration satisfying Eqs.\,\eqref{eq: gauge field ansatz EOM} and \eqref{eq: scalar field ansatz EOM} can be obtained by minimizing $S_{\mathrm{tot}}$, as discussed in Appendix\,\ref{app: remarks on constraint}.
The size parameter $\rho $ of the numerical configuration is determined from the second derivative of $\mathcal{A}$ near the origin  by Eq.\,\eqref{eq: size definition for A explicit}.

\begin{figure}[t]
    \centering
    \includegraphics[width=0.7\linewidth]{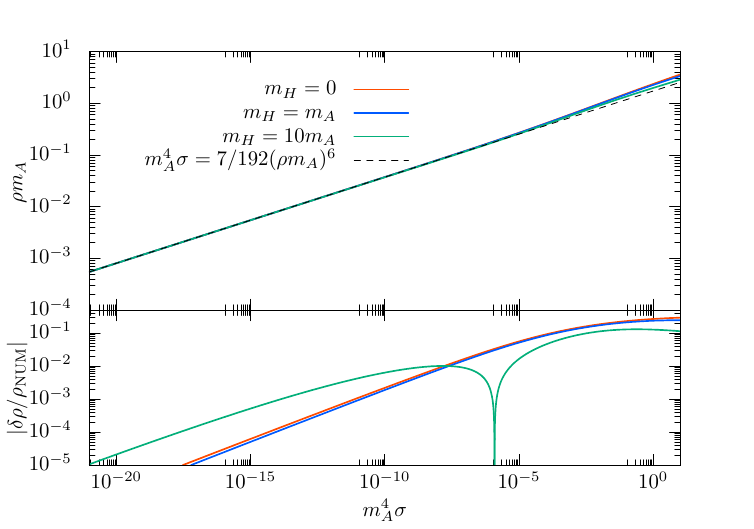}
    \caption{
Relation between the dimensionless Lagrange multiplier $m_A^4\sigma$ and the instanton size $\rho m_A$ for several values of $m_H/m_A$.
The solid orange, blue, and green curves denote the numerical results for $m_H=0$, $m_H=m_A$, and $m_H=10m_A$, respectively.
The dashed black curve shows the analytic NLO prediction, $m_A^4\sigma=\frac{7}{192}(\rho m_A)^6$.
In the lower panel, the relative deviation of the analytic NLO prediction from the numerical results is shown for the same three choices of $m_H/m_A$. Here, we define $\delta\rho:=\rho-\rho_\mathrm{NUM}$.
The numerical results approach the analytic prediction in the small-$\rho m_A$ region.
}
    \label{fig: size and sigma for YM}
\end{figure}

Figure \,\ref{fig: size and sigma for YM}
shows the numerically obtained relation between the dimensionless Lagrange multiplier $m_A^4\sigma$ and the size $\rho m_A$, alongside the analytic prediction given in Eq.\,\eqref{eq: sigma2 value for YMH}.
For sufficiently small $\rho m_A$, the numerical results agree with Eq.\,\eqref{eq: sigma2 value for YMH} for $m_H/m_A=0,1$ and $10$, thereby supporting the validity of both the analytic expansion and the matching procedure.

\begin{figure}[t]
        \centering
        \includegraphics[width=.6\linewidth]{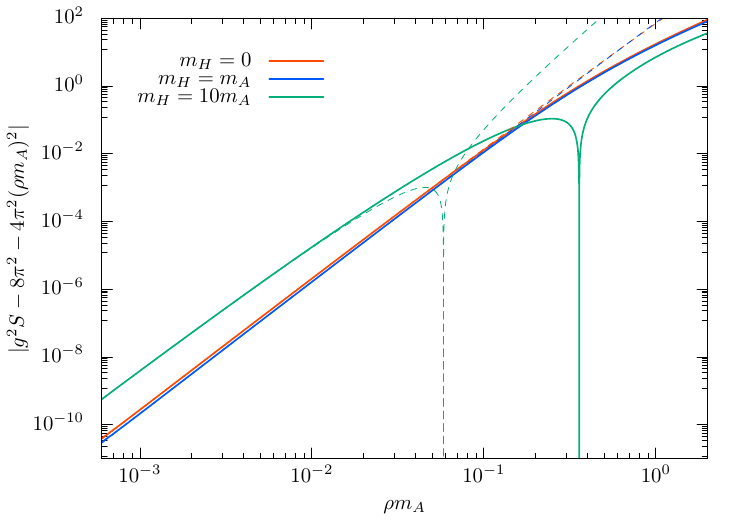}
\caption{
Numerically obtained $\rho m_A$ dependence of the action difference $g^2S-8\pi^2-4\pi^2(\rho m_A)^2$, which is expected to be of order $\order{\rho^4 m_{A,H}^4}$.
The solid orange, blue, and green curves denote the numerical results for $m_H=0$, $m_H=m_A$, and $m_H=10m_A$, respectively.
The dashed curves of the same colors show the corresponding analytic $\order{\rho^4 m_{A,H}^4}$ corrections given in Eq.\,\eqref{eq: action value from matching, 4th order}.
In the small-$\rho m_A$ region, the dashed and solid curves nearly overlap and are difficult to distinguish visually.
This agreement in the small-$\rho m_A$ region supports the perturbative matching analysis.
}
    \label{fig: size dependence of action}
\end{figure}

Figure \ref{fig: size dependence of action} shows the numerically computed values of $g^2S-8\pi^2-4\pi^2(\rho m_A)^2$ as a function of $\rho m_A$.
This quantity represents the $\order{(\rho m_{A,H})^4}$ correction to the action, and we plot it alongside its corresponding analytic prediction derived in Eq.\,\eqref{eq: action value from matching, 4th order}.
As expected, the analytic curve coincides with the numerically obtained results at small sizes ($\rho m_A \ll 1$).
Altogether, these numerical findings confirm that the obtained solutions are consistent with our inner--outer matching procedure.

\section{Conclusions}
\label{sec: conclusion}

In this work, we revisited the construction of constrained instantons, investigating the asymptotic structure of their profiles near the origin and at large distances, with particular emphasis on the obstruction claimed by N\&N (Ref.\,\cite{Nielsen:1999vq}). 
We carefully examined the expansion parameters required to construct the asymptotic profiles in Sec.\,\ref{sec: phi4 case} for the massive $\phi^4$ theory and in Sec.\,\ref{sec: explicit construction in YM} for Yang--Mills theory with spontaneous symmetry breaking. 
Our analysis demonstrates that, once the expansion parameters controlling the inner and outer regions are treated systematically, the inner and outer solutions can be consistently matched using conventional gauge-invariant constraint operators $\mathcal{O}_\mathrm{con}$, without encountering the purported obstruction.

We first reconsidered the constrained instanton in the massive $\phi^4$ theory, constructing inner and outer solutions for $r\to 0$ and for $r\to\infty$, respectively. 
Both solutions were written as double expansions in $(\rho m)^2$ and $(\rho/r)^2$, where $\rho$ denotes the instanton size, and were matched in an intermediate overlap region. 
Working consistently up to NLO in $(\rho m)^2$, we showed that the constrained instanton can be constructed consistently with conventional choices of $\mathcal{O}_\mathrm{con}=\phi^p (p=5,6,\ldots)$, in contrast to the conclusion by N\&N. The key point is that consistency is restored when higher-order terms in $(\rho m)^2$ are properly retained in the outer solution. We further validated this analytic expansion by comparing it against numerical solutions of the Euler--Lagrange equation with the constraint.

We then applied the same strategy to the constrained instanton in spontaneously symmetry-broken $\SU(2)$ Yang--Mills theory with a scalar doublet.  
Taking $\mathcal{O}_\mathrm{con} = (\frac{1}{2}\tr F_{\mu\nu}\tilde{F}_{\mu\nu})^2$, we derived the profile equations, constructed inner and outer profile functions for the gauge and scalar fields, and expanded them in powers of $(\rho m_{A,H})^2$.
After establishing the corresponding matching conditions and determining the LO and NLO solutions, we showed that the matching procedure is again consistent through NLO.  
To corroborate the analytic construction, we numerically obtained constrained instanton solutions and tested the analytic predictions against them, confirming the overall validity of the expansion and matching procedure.

Taken together, we confirmed that constrained instanton solutions can be consistently constructed up to NLO with conventional choices of $\mathcal{O}_\mathrm{con}$, both in the massive $\phi^4$ theory and in Yang--Mills theory with spontaneous symmetry breaking. This demonstrates that the NLO matching procedure can be successfully carried out, resolving the potential difficulties previously raised by N\&N.

We conclude with a few remarks.
First, while our explicit analytic construction is carried out only up to NLO and does not address higher-order corrections, the agreement with the numerical solutions supports the overall consistency of the matching procedure.
Second, a more precise evaluation of the path integral requires computing the prefactor in the partition function $Z$, i.e., the functional determinant around the constrained instanton background~\cite{tHooft:1976snw}.
Although our focus here has been on the classical action, the profile functions obtained in this work provide the necessary groundwork for computing this determinant.
In future applications,
such precise evaluations
will be valuable for computing high-energy scattering amplitudes with baryon-number violation in the electroweak theory~\cite{Espinosa:1989qn} and the small-instanton contributions to the axion mass in certain axion models \cite{Agrawal:2017evu,Agrawal:2017ksf,Gaillard:2018xgk}.
We anticipate that the systematic framework developed here will serve as a robust foundation for exploring these non-perturbative phenomena in theories with spontaneous symmetry breaking.

\section*{Acknowledgments}
This work is supported by Grant-in-Aid for Scientific Research from the Ministry of Education, Culture, Sports, Science, and Technology (MEXT), Japan, 21H04471, 22K03615, and 26K07102 (M.I.) and by World Premier International Research Center Initiative (WPI), MEXT, Japan. 
This work is also supported by Grantin-Aid for JSPS Research Fellow 25KJ0851 (T.A.). This work is supported by FoPM,
WINGS Program, the University of Tokyo (T.A.). 

\appendix

\section{Notation in Euclidean Space}
\label{app:notations}
Throughout this paper, quantities in Minkowski spacetime are labeled with the subscript $\mathrm{M}$, while Euclidean quantities are written without such a label. The coordinates and derivatives in Euclidean and Minkowski spaces are related via
\begin{align}
    x^{i} = x_\mathrm{M}^i \ , 
    \quad
    \partial^i = \partial_{\mathrm{M},i}\ , \quad
    x^4 = i x^0_\mathrm{M} \ , 
    \quad 
    \partial^4 = -i \partial_{\mathrm{M},0}\ ,
\end{align}
for $i=1,2,3$.
The metric tensors are defined as
\begin{align}
    g_{\mathrm{M},\mu\nu} = (+1,-1,-1,-1) \ ,
\end{align}
in Minkowski spacetime and 
\begin{align}
    g_{\mu\nu} = (+1,+1,+1,+1) \ ,
\end{align}
in Euclidean space.
The antisymmetric tensor in Euclidean space is denoted by $\epsilon_{\mu\nu\rho\sigma}$ with the convention $\epsilon_{1234}=+1$\,.
The Euclidean and Minkowski gauge potentials are related by
\begin{align}
    A^i  = A_{\mathrm{M},i}\ ,\quad A^4 = - i A_{\mathrm{M},0} \ .
\end{align}

For the generators of the $\SU(2)$ gauge group,
we use the Pauli matrices $\boldtau$.
To describe the instanton solution, it is convenient to introduce the matrices
\begin{align}
    (\tau^\mu)
    =(\boldtau,i\mathbb{1}_2)\ , \quad  (\bartau^\mu)\ 
    =(\boldtau,-i\mathbb{1}_2)\ , \quad (\mu = 1,2,3,4)\ ,
\end{align}
and 
\begin{align}
\label{eq:generators}
    \tau_{\mu\nu} 
    = \frac{1}{4i} (\bartau_\mu \tau_\nu - \bartau_\nu \tau_\mu)\ ,\quad
    \bartau_{\mu\nu} 
    = \frac{1}{4i} (\tau_\mu \bartau_\nu - \tau_\nu \bartau_\mu)\ .
\end{align}

\section{\texorpdfstring{Inner and Outer Matching in $\phi^4$ Theory}{}}
\label{app: phi4 matching general}
In this appendix, 
we further show that the matching strategy developed  in Sec.\,\ref{sec: phi4 case} can be applied to the constraint operator $\mathcal{O}_\mathrm{con}=\phi^p$ with $p\ge5$.
Although the procedure below does not directly apply to the case $p=3$, we have verified that the matching can still be achieved through a separate analysis.
Note that the case $p=4$ merely corresponds to a shift in the quartic coupling and therefore does not provide a constraint on the instanton size.

\subsubsection*{Relations between the Constants in the Inner Solution}
The NLO contribution to the inner solution satisfies
\begin{align}
    f_2''(\hat{r})+\frac{3}{\hat{r}}f_2'(\hat{r})
+\frac{\rho^2}{2}\phi_0^2 f_2(\hat{r})
    = \phi_0
    + p\sigma_2\rho^2 \phi_0^{p-1}\ , 
    \quad\quad
    \hat{r}:=\frac{r}{\rho}\,.
\end{align}
The general solution can be written as
\begin{align}
    f_2(\hat{r})=
    c_1
    f_2^{(\mathrm{hom},1)}
    +c_2
    f_2^{(\mathrm{hom},2)}
    +f_2^{(m^2)}+f_2^{(\sigma_2;p)} 
    \,.
    \label{eq: NLO phi, general p}
\end{align}
where the explicit dependence on $p$ appears only in the particular solution $f_2^{(\sigma_2;p)}$.
Here, $f_2^{(\sigma_2;p)}$ is chosen to be as suppressed as possible as $r\to\infty$,
by exploiting the freedom to add homogeneous solutions, as in Sec.\,\ref{sec: phi4 case}.
The other terms $f_2^{(\mathrm{hom},1)}$, $f_2^{(\mathrm{hom},2)}$ and $f_2^{(m^2)}$ are given by Eqs.\,\eqref{eq: hom1}--\eqref{eq: m^2 particular solution in phi4 theory} in Sec.\,\ref{sec: phi4 case}.

As discussed in Sec.\,\ref{sec: phi4 case},
the integration constants $c_1$ and $c_2$, as well as the expansion coefficient $\sigma_2$ of the Lagrange multiplier, are restricted by the boundary conditions at the origin. Requiring $f_2(r=0)$ to be finite imposes a relation between $\sigma_2$ and $c_2$, because the particular solution $f_2^{(\sigma_2;p)}$ is singular at $r=0$ (see Eq.\,\eqref{eq: sigma2 particular solution in phi4 theory} for $p=6$). We further fix $c_1$ by imposing $f_2(r=0)=0$, which eliminates the redundancy associated with the definition of the size parameter $\rho$. These conditions yield
\begin{align}
    \frac{\sigma_2}{\rho^{p-4}}
    &=
    -\frac{3^{\frac{1}{2}-\frac{p}{2}} 4^{2-p} \left(p^2-3 p+2\right)}{p-4}c_2 \,,
    \label{eq: phi4 sigma_2 in terms of c_2, general constraint}
    \\
    c_1
    &=
    -\sqrt{3} \left(1+\pi ^2\right)+c_2 \left(-6 H_{p-2}-p+\frac{12}{p-4}+\frac{6}{p-3}+\frac{6}{p-2}+14\right)\,,
    \label{eq: phi4 c_1, general constraint}
\end{align}
which are valid for $p\ge5$\,. Here, $H_{p-2}$ denotes the $(p-2)$-th harmonic number, defined by
\begin{align}
    H_N:=1+\frac{1}{2}+\frac{1}{3}+\cdots+\frac{1}{N}\,.
\end{align}
Equations~\eqref{eq: phi4 sigma_2 in terms of c_2, general constraint} and~\eqref{eq: phi4 c_1, general constraint}
reduce to Eqs.\,\eqref{eq: phi NLO Lagrange multiplier} and~\eqref{eq: c_1 choice}, respectively, for $p=6$.

\subsubsection*{List of Matching Coefficients}
\label{app: phi4 coefficients list}

The outer solution is not affected by $\mathcal{O}_\mathrm{con}$ at both LO and NLO for $p\ge5$. Regarding the inner solution, the expansion coefficient $\sigma_2$ of the Lagrange multiplier does not directly enter the NLO matching conditions for $p\ge 5$ (see Sec.\,\ref{sec: phi4 case}),
while the value of $\sigma_2$ is determined through Eq.\,\eqref{eq: phi4 sigma_2 in terms of c_2, general constraint}
to avoid the singularity of $f_2(r=0)$.
Consequently, the matching relations between $f_n^{(k)}$ and $g_{n+2-k}^{(k)}$ are independent of $p$.
The coefficients $f_n^{(k)}$ and $g_{n+2-k}^{(k)}$ for $p=5,6,\ldots$ are summarized in Tab.\,\ref{tab: f and g coefficients}.

Using the expressions in Tab.\,\ref{tab: f and g coefficients}, we can systematically evaluate the $(n,k)$ matching conditions defined in Eq.\,\eqref{eq: n,k condition}.
First, the LO matching condition corresponding to $(n,k)=(0,2)$ leads directly to Eq.\,\eqref{eq: kappa value}.
At NLO, the $(2,0)$ condition reproduces Eq.\,\eqref{eq: c_2}.
Furthermore, the $(2,2)$ condition yields Eq.\,\eqref{eq: cout in phi4} from the $\mathcal{O}((\log\hat{r})^0)$ term, while its $\mathcal{O}((\log\hat{r})^1)$ term provides a nontrivial consistency check for Eq.\,\eqref{eq: c_2}.
Finally, the $(0,4)$ condition, as well as the $\mathcal{O}((\log\hat{r})^2)$ term in the $(2,2)$ condition, are automatically satisfied without the need to adjust the constants $c_1$, $c_2$, or $c^{(\mathrm{out})}$, serving as additional consistency checks of our matching procedure.

\begin{table}[ht]
\centering
\caption{Matching coefficients $f_n^{(k)}$ and $g_{n+k-2}^{(k)}$ for $p\ge5$, labeled by the indices $(n,k)$\, (see Fig.\,\ref{fig: matching double expansion}, Eqs.\,\eqref{eq: fn-expansion} and \eqref{eq: gn-expansion}).
The integer $n$ denotes the order in $\rho m$ of the inner solution, corresponding to the vertical entries in Fig.\,\ref{fig: matching double expansion}.
The order in $\rho m$ for the outer solution is $n+k-2$, represented by the diagonal entries in Fig.\,\ref{fig: matching double expansion}.
The LO matching condition corresponds to $(n,k)=(0,2)$, while the NLO matching conditions correspond to $(0,4)$, $(2,0)$ and $(2,2)$.
}
\label{tab: f and g coefficients}

\begin{minipage}{\linewidth}
\centering
\subcaption{Table of $\rho f_n^{(k)}/(4\sqrt{3})$}

\begin{tabular}{c|c|c|c}
 & $n=0$ & $n=2$ & $n=4$\\
 \hline
 $\rotatebox{90}{\ldots}$ & $\rotatebox{90}{\ldots}$ & $\rotatebox{90}{\ldots}$ & $\rotatebox{90}{\ldots}$\\ 
\hline
$k=4$ & $-1$ & $\rotatebox{90}{\ldots}$ & $\rotatebox{90}{\ldots}$ \\ \hline
$k=2$ & $1$ & $3(\log\hat{r})^2+ \left(\sqrt{3} c_2-11/2\right)\log\hat{r}+11/2-(c_1+11c_2)/(4 \sqrt{3})$ & $\rotatebox{90}{\ldots}$ \\ \hline
$k=0$ & 0 & $1/2\,\log\hat{r}+(c_2/\sqrt{3}-3)/4$ & $\rotatebox{90}{\ldots}$ \\ \hline
$\rotatebox{90}{\ldots}$ & 0 & 0 & $\rotatebox{90}{\ldots}$
\end{tabular}
\hspace{6mm}$\ldots$

\end{minipage}

\begin{minipage}{\linewidth}
\centering
\subcaption{Table of $m^{-1}g_{n+k-2}^{(k)}$}

\begin{tabular}{c|c|c|c}
 & $n=0$ & $n=2$ & $n=4$\\
 \hline
 $\rotatebox[origin=c]{90}{\ldots}$ & $\rotatebox[origin=c]{-30}{\ldots}$& $\rotatebox[origin=c]{-30}{\ldots}$ & $\rotatebox[origin=c]{-30}{\ldots}$\\ 
\hline
$k=4$ & $-1$ & $\rotatebox[origin=c]{-30}{\ldots}$ & $\rotatebox[origin=c]{-30}{\ldots}$ \\ \hline
$k=2$ & $1$ & $\begin{aligned}3(\log\hat{r})^2+(1/2+6\gamma_E+6\log(\rho m/2))\log\hat{r}\\+c_\mathrm{out}+(1/2+6\gamma_E-6\log2)\log(\rho m)+3(\log(\rho m))^2
\end{aligned}$& $\rotatebox[origin=c]{-30}{\ldots}$ \\ \hline
$k=0$ & $0$ & $1/2\, \log\hat{r}+(-1+2\gamma_E+2\log(\rho m/2))/4$ & $\rotatebox[origin=c]{-30}{\ldots}$ \\ \hline
$\rotatebox[origin=c]{90}{\ldots}$ & $0$ & $0$ & $\rotatebox[origin=c]{-30}{\ldots}$
\end{tabular}
\hspace{6mm}$\ldots$
\end{minipage}

\end{table}

\clearpage
\section{Semiclassical Approximation with Constraint Operator}
\label{sec: constraint method of inserting 1}

Our method for evaluating the path integral by introducing constraints is based on Ref.\,\cite{Gervais:1977me}.
In contrast to the main text, we work with canonically normalized fields in this appendix.
When working with rescaled fields, the normalization must be adjusted accordingly.
By expressing all integration variables by fields $\Phi_i\,(i=1,2,\ldots)$, we can write the path integral as
\begin{align}
    Z
    =
    \int\prod_i\mathcal{D}\Phi_i\,
    e^{-S[\Phi]}\,.
\end{align}
In the following, the index $i$ and the associated products and sums are suppressed for notational simplicity.
The core idea is to insert unity (``$1$") into the path integral. That is, we introduce the identity
\begin{align}
    1 = \int\mathrm{d}\lambda\, \delta(\lambda-S_\mathrm{con}[\Phi])\,,
    \label{eq: 1 app}
\end{align}
where $\lambda$ is a real variable representing the constrained value of $S_\mathrm{con}$, the constraint functional given by the spatial integral of the local operator $\mathcal{O}_\mathrm{con}$. The functional $S_\mathrm{con}$ must depend monotonically on the instanton size $\rho$, and thereby provides a one-to-one mapping between the integration variable $\lambda$ and the size $\rho$.

Inserting Eq.\,\eqref{eq: 1 app} into the path integral, we obtain
\begin{align}
    Z &=
    \int\mathcal{D}\Phi\,
    \mathrm{d}\lambda\,
    \delta(\lambda-S_\mathrm{con}[\Phi])\,
    \exp(-S[\Phi])
    \\
    &=
    \int\mathcal{D}\Phi\,
    \mathrm{d}\lambda\,
    \delta(\lambda-S_\mathrm{con}[\Phi])\,
    \exp(-S[\Phi]-\sigma(S_\mathrm{con}[\Phi]-\lambda))\,.
    \label{eq: path integral with delta insertion}
\end{align}
In the second equality, we have used the fact that the term $\sigma(S_\mathrm{con}-\lambda)$ identically vanishes for any arbitrary parameter $\sigma$ on the constraint surface enforced by the delta function $\delta(\lambda-S_\mathrm{con})$\,.
We then define the constrained action $S_\mathrm{tot}^{(\lambda)}$ as 
\begin{align}
    S_\mathrm{tot}^{(\lambda)}[\Phi,\sigma]:=
    S[\Phi]+\sigma(S_\mathrm{con}[\Phi]-\lambda)\ .
    \label{eq: total action in app}
\end{align}

In what follows, we perform the semiclassical approximation of the path integral in Eq.\,\eqref{eq: path integral with delta insertion}.
As a first step, for a given value of $\lambda$, we seek the stationary points of $S_\mathrm{tot}^{(\lambda)}$ with respect to both the field $\Phi$ and the Lagrange multiplier $\sigma$. These points satisfy the extremization conditions:
\begin{align}  &\frac{\delta S^{(\lambda)}_\mathrm{tot}}{\delta \Phi}= 
\frac{\delta S}{\delta \Phi} + \sigma \frac{\delta S_\mathrm{con}}{\delta \Phi}=0\,,
    \label{eq: constrained Euler Lagrange equation}
    \\
&\frac{\partial S^{(\lambda)}_\mathrm{tot}}{\partial \sigma}=
    S_\mathrm{con}[\Phi]-\lambda=0\,.
    \label{eq: Euler Lagrange equation for multiplier: constraint}
\end{align}
We denote this stationary point of $S_\mathrm{tot}^{(\lambda)}$ as $(\bar\Phi^{(\lambda)},\bar\sigma^{(\lambda)})$\,.
Furthermore, we define the stationary value of the action as $\bar{S}^{(\lambda)}:=S_\mathrm{tot}^{(\lambda)}[\bar\Phi^{(\lambda)},\bar\sigma^{(\lambda)}]=S[\bar\Phi^{(\lambda)}]$,
where the second equality follows from Eq.\,\eqref{eq: Euler Lagrange equation for multiplier: constraint}.
Hereafter, we adopt a shorthand dot ($\odot$) product notation for the fields, which represents local multiplication at each point and integration over the coordinates:
\begin{align}
    f(x)\odot_x g(x) = \int\mathrm{d}^4x\,f(x)g(x)\,.
\end{align}

Expanding the total action $S_\mathrm{tot}^{(\lambda)}$ around this stationary point, we obtain
\begin{align}
S_\mathrm{tot}^{(\lambda)}[\Phi,\sigma]
    =
\bar{S}^{(\lambda)}+\frac{1}{2}\varphi(x)\odot_x\left.\frac{\delta^2 S_\mathrm{tot}^{(\lambda)}}{\delta \Phi(x)\delta \Phi(y)}\right|_{\Phi=\bar{\Phi}^{(\lambda)}}\odot_y\varphi(y)+\order{\varphi^3,(\mathit{\Delta}\sigma)^2}\,,
    \label{eq: quadratic expansion of Stot}
\end{align}
where we have introduced the fluctuations $\varphi := \Phi -\bar\Phi^{(\lambda)}$
and 
$\mathit{\Delta}\sigma=\sigma-\bar\sigma^{(\lambda)}$\,.
Similarly, the argument of the delta function can be expanded as
\begin{align}
    \lambda-S_\mathrm{con}[\Phi]
    =-
   \left. \frac{\delta S_\mathrm{con}}{\delta \Phi(x)}\right|_{\Phi=\bar{\Phi}^{(\lambda)}}\odot_x\varphi(x) + \order{\varphi^2}
    \,,
    \label{eq: linear expansion of delta function}
\end{align}
where the zeroth-order term in $\varphi$ vanishes due to Eq.\,\eqref{eq: Euler Lagrange equation for multiplier: constraint}.
Hereafter, we omit the evaluation symbol ($|_{\Phi=\bar\Phi^{(\lambda)}}$) for brevity, with the understanding that all functional derivatives are evaluated at the stationary point.

Substituting Eqs.\,\eqref{eq: quadratic expansion of Stot} and \eqref{eq: linear expansion of delta function}
into
Eq.\,\eqref{eq: path integral with delta insertion},
we obtain
\begin{align}
    Z|_{\varphi\sim0}
    =
    \int &
    \mathcal{D}\varphi\,
    \mathrm{d}\lambda\,
    \delta\qty(-\frac{\delta S_\mathrm{con}}{\delta \Phi(x)}\odot_x\varphi(x)+\order{\varphi^2})
    \notag
    \\
    &
    \times
    \exp(-\bar{S}^{(\lambda)}-\frac{1}{2}\varphi(x)\odot_x\frac{\delta^2 S_\mathrm{tot}^{(\lambda)}}{\delta \Phi(x)\delta \Phi(y)}\odot_y\varphi(y)+\order{\varphi^3})\,.
\label{eq: semiclassical contributions from neighborhood of constrained minima}
\end{align}
Using the Fourier representation of the delta function,
\begin{align}
    \delta(X) = \int_{-\infty}^{\infty} \frac{\mathrm{d}\mu}{2\pi}\exp(i\mu X)\,,
\end{align}
we can rewrite the path integral as
\begin{align}
    Z|_{\varphi\sim0} &\simeq
    \int\mathrm{d}\lambda \exp(-\bar{S}^{(\lambda)})
    \notag\\
    &\times\frac{\mathrm{d}\mu}{2\pi}
    \mathcal{D}\varphi\,
    \exp(-i\mu\frac{\delta S_\mathrm{con}}{\delta \Phi(x)}\odot_x\varphi(x)-\frac{1}{2}\varphi(x)\odot_x\frac{\delta^2 S_\mathrm{tot}^{(\lambda)}}{\delta \Phi(x)\delta \Phi(y)}\odot_y\varphi(y))\,.
    \label{eq: partition function under semiclassical approximation}
\end{align}
Here, the $\order{\varphi^3}$ terms in $S_\mathrm{tot}$ are neglected in accordance with the semiclassical approximation.
Furthermore, the $\order{\varphi^2}$ terms in the argument of the delta function is also neglected~\cite{Gervais:1977me}. This can be explicitly verified by restoring and counting the powers of $\hbar$.

The linear term in $\varphi$ in the exponent, originating from the delta function, can be eliminated by shifting the integration variables as
\begin{align}
    \tilde\varphi^{(\lambda)}(x):=\varphi(x)+\mu
    \qty[-i\frac{\delta^2 S_\mathrm{tot}^{(\lambda)}}{\delta \Phi\delta \Phi}]^{-1}_{(x,y)}\odot_y
    \frac{\delta S_\mathrm{con}}{\delta \Phi(y)}\,.
\end{align}
Here, the Green's function $[-i{\delta^2 S_\mathrm{tot}^{(\lambda)}}/{(\delta \Phi\delta \Phi)}]^{-1}_{(x,y)}$ satisfies,
\begin{align}
 \frac{\delta^2 S_\mathrm{tot}^{(\lambda)}}{\delta \Phi(x)\delta \Phi(y)}\odot_y\qty[-i\frac{\delta^2 S_\mathrm{tot}^{(\lambda)}}{\delta \Phi\delta \Phi}]^{-1}_{(y,z)}
    =
    i\delta^{(4)}(x-z)\,.
\end{align}
Performing this shift, we obtain
\begin{align}
    &Z|_{\varphi\sim0} \simeq
    \int\mathrm{d}\lambda \exp(-\bar{S}^{(\lambda)})
    ~~
    \frac{\mathrm{d}\mu}{2\pi}
    \mathcal{D}\tilde\varphi
    \exp(-\mathit{\Delta} S)\,,
\end{align}
where 
\begin{align}
    \mathit{\Delta} S=
    \frac{1}{2}
    \tilde\varphi(x)\odot_x\frac{\delta^2 S_\mathrm{tot}^{(\lambda)}}{\delta \Phi(x)\delta \Phi(y)}\odot_y\tilde\varphi(y)
    -
    \frac{i\mu^2}{2} 
    \frac{\delta S_\mathrm{con}}{\delta \Phi(x)}
    \odot_x\qty[-i\frac{\delta^2 S_\mathrm{tot}^{(\lambda)}}{\delta \Phi\delta \Phi}]^{-1}_{(x,y)}\odot_y
    \frac{\delta S_\mathrm{con}}{\delta \Phi(y)}
    \,.
    \label{eq: constrained gaussian path integral}
\end{align}

To perform the integration over $\mu$,
let us consider the
total derivative of the
Euler--Lagrange equation 
Eq.\,\eqref{eq: constrained Euler Lagrange equation}
with respect to $\lambda$.
The Euler--Lagrange equation implicitly depends on $\lambda$ through the stationary-point solutions $(\bar\Phi^{(\lambda)},\bar\sigma^{(\lambda)})$. Taking the derivative with respect to $\lambda$ yields
\begin{align}
\frac{\delta^2 S^{(\lambda)}_\mathrm{tot}}{\delta\Phi(x')\delta \Phi(y)}\odot_y\frac{\mathrm{d}\bar\Phi^{(\lambda)}(y)}{\mathrm{d}\lambda} + \frac{\delta S_\mathrm{con}}{\delta \Phi(x')}
\frac{\mathrm{d}\bar\sigma^{(\lambda)}}{\mathrm{d}\lambda} =0\,,
\end{align}
evaluated at $(\Phi,\sigma)=(\bar\Phi^{(\lambda)},\bar\sigma^{(\lambda)})$.
Multiplying by $[-i\delta S^{(\lambda)}/(\delta\Phi\delta\Phi)]^{-1}_{(x,x')}$ and integrating over the coordinates $x'$ and $y$, we obtain
\begin{align}
    i\frac{\mathrm{d}\bar\Phi^{(\lambda)}(x)}{\mathrm{d}\lambda}
    +
    \qty[-i\frac{\delta^2 S_\mathrm{tot}^{(\lambda)}}{\delta \Phi\delta \Phi}]^{-1}_{(x,x')}\odot_{x'}
    \frac{\delta S_\mathrm{con}}{\delta \Phi(x')}
    \frac{\mathrm{d}\bar\sigma^{(\lambda)}}{\mathrm{d}\lambda}
    =0\,.
\end{align}
Further multiplying by $[\mathrm{d}\bar\sigma^{(\lambda)}/\mathrm{d}\lambda]^{-1}$, we find
\begin{align}
    i\frac{\mathrm{d}\bar\Phi^{(\lambda)}(x')}{\mathrm{d}\lambda}
    \qty[\frac{\mathrm{d}\bar\sigma^{(\lambda)}}{\mathrm{d}\lambda}]^{-1}
    +
    \qty[-i\frac{\delta^2 S_\mathrm{tot}^{(\lambda)}}{\delta \Phi\delta \Phi}]^{-1}_{(x',x)}\odot_x
    \frac{\delta S_\mathrm{con}}{\delta \Phi(x)}
    =0\,,
\end{align}
which leads to
\begin{align}
    \qty[-i\frac{\delta^2 S_\mathrm{tot}^{(\lambda)}}{\delta \Phi\delta \Phi}]^{-1}_{(x',x)}\odot_x
    \frac{\delta S_\mathrm{con}}{\delta \Phi(x)}=-i\frac{\mathrm{d}\bar\Phi^{(\lambda)}(x)}{\mathrm{d}\bar\sigma^{(\lambda)}}\,.
    \label{eq: Green's function formula for mu integration}
\end{align}
Substituting this relation into Eq.\,\eqref{eq: constrained gaussian path integral}, it follows that
\begin{align}
    -\frac{i\mu^2}{2}\frac{\delta S_\mathrm{con}}{\delta \Phi(x)}\odot_x
    \qty[-i\frac{\delta^2 S_\mathrm{tot}^{(\lambda)}}{\delta \Phi\delta \Phi}]^{-1}_{(x,y)}\odot_y
    \frac{\delta S_\mathrm{con}}{\delta \Phi(y)}
    &=
    -\frac{\mu^2}{2}
    \frac{\delta S_\mathrm{con}}{\delta \Phi(x)}\odot_x\frac{\mathrm{d}\bar\Phi^{(\lambda)}(x)}{\mathrm{d}\bar\sigma^{(\lambda)}}
    \\
    &=
    -\frac{\mu^2}{2}\frac{\mathrm{d} \lambda}{\mathrm{d}\sigma}\ .
\end{align}
Finally, performing the Gaussian integration over $\mu$, we obtain
\begin{align}
    Z|_{\varphi\sim0} &\simeq
    \int\mathrm{d}\lambda \exp(-\bar{S}^{(\lambda)})
    ~~\frac{\mathrm{d}\mu}{2\pi}
    \exp(-
    \frac{\mu^2}{2} 
    \left.\qty[-\frac{\mathrm{d}\lambda}{\mathrm{d}\sigma}]\right|_{\sigma=\bar\sigma^{(\lambda)}})
    \notag\\
    &\phantom{\simeq}~~~~~~~~~~
    \times
    \mathcal{D}\tilde\varphi
    \exp(-\frac{1}{2}
    \tilde\varphi_i(x)\odot_x\frac{\delta^2 S_\mathrm{tot}^{(\lambda)}}{\delta \Phi(x)\delta \Phi(y)}\odot_y\tilde\varphi(y))
    \\&=
    \int\frac{\mathrm{d}\lambda}{(2\pi)^{1/2}}\,
    \qty(-\frac{\mathrm{d}\bar\sigma^{(\lambda)}}{\mathrm{d}\lambda})^{1/2}
    e^{-\bar{S}^{(\lambda)}}
    \mathcal{D}\tilde\varphi
    \exp(-\frac{1}{2}
    \tilde\varphi(x)\odot_x\frac{\delta^2 S_\mathrm{tot}^{(\lambda)}}{\delta \Phi(x)\delta \Phi(y)}\odot_y\tilde\varphi(y))
    \,.
    \label{eq: constrained gaussian path integral simplified}
\end{align}
Here,  $\delta^2S_\mathrm{tot}^{(\lambda)}/(\delta\Phi\delta\Phi)$
is evaluated at the stationary point solutions $(\Phi,\sigma)=(\bar\Phi^{(\lambda)},\bar\sigma^{(\lambda)})$\,.
In the above derivation, we have assumed that 
$\bar\sigma^{(\lambda)}$ and $\lambda$ are in one-to-one correspondence and also that $-\mathrm{d}\bar\sigma^{(\lambda)}/\mathrm{d}\lambda$ is positive.

 We now verify these assumptions in the case of Yang--Mills theory with spontaneous symmetry breaking, where the normalization is given in Sec.\,\ref{sec: revisit YM constrained instanton}.
 We use the following identity, derived from
Euler--Lagrange equations Eqs.\,\eqref{eq: constrained Euler Lagrange equation} and \eqref{eq: Euler Lagrange equation for multiplier: constraint}:
\begin{align}
    \frac{\mathrm{d}\bar{S}^{(\lambda)}}{\mathrm{d}\lambda}
    =
    \frac{\mathrm{d}\bar\Phi^{(\lambda)}(x)}{\mathrm{d}\lambda}\odot_x\frac{\delta\bar{S}^{(\lambda)}}{\delta\Phi(x)}+\frac{1}{g^2}\frac{\mathrm{d}\bar\sigma^{(\lambda)}}{\mathrm{d}\lambda}\frac{\partial\bar{S}^{(\lambda)}}{\partial \sigma}+\frac{\partial\bar{S}^{(\lambda)}}{\partial \lambda}
    =
    \frac{\partial \bar{S}^{(\lambda)}}{\partial\lambda}
    =-\frac{\bar{\sigma}^{(\lambda)}}{g^2}\,.
    \label{eq: constraint value dependence of action}
\end{align}
Differentiating the relation once more with respect to $\lambda$, we obtain
\begin{align}
    -\frac{1}{g^2}\frac{\dd\bar\sigma^{(\lambda)}}{\dd\lambda}
    =
    \frac{\dd^2\bar{S}^{(\lambda)}}{\dd\lambda^2}\,.
    \label{eq: lagrange multiplier derivative in app}
\end{align}
We now focus on a typical situation in which the constraint functional scales as a power of the instanton size $\rho$,
\begin{align}
    S_\mathrm{con}[\bar{\mathcal{A}}]
    =C\rho^{-N}[1+\order{\rho^2v^2}]\,,
    \qquad
    N\neq0\,,
    \qquad
    C>0\,.
    \label{eq: typical behavior of constraint functional in app, constraint method}
\end{align} 
Specifically, the operator $\mathcal{O}_\mathrm{con}$ defined in Eq.\,\eqref{eq: Ocon = FFtilde} corresponds to the case $N=4$.
Substituting the behavior of the action $\bar{S}^{(\lambda)}-8\pi^2/g^2=4\pi^2\rho^2m_A^2/g^2+\order{\rho^4 v^4}$ into Eq.\,\eqref{eq: lagrange multiplier derivative in app} and using the constraint condition $S_\mathrm{con}[\bar{\mathcal{A}}]=\lambda$,
we obtain
\begin{align}
    -\frac{\dd\bar\sigma^{(\lambda)}}{\dd\lambda}
    &=
    g^2\frac{\dd^2\bar{S}^{(\lambda)}}{\dd\lambda^2}
    \\
    &=
    g^2\qty(\frac{\dd S_\mathrm{con}}{\dd\rho})^{-1}\frac{\dd}{\dd\rho}\qty[\qty(\frac{\dd S_\mathrm{con}}{\dd\rho})^{-1}\frac{\dd\bar{S}^{(\lambda)}}{\dd\rho}]
    \\
    &=
    8\pi^2C^{-2}\times(\rho^2 m_A^2)\rho^{2N}\qty[\frac{N+2}{N^2}+\order{\rho^2 v^2}]
    >0\qquad\text{for}\quad N>-2\,.
\end{align}
Therefore, for constraint operators obeying Eq.\,\eqref{eq: typical behavior of constraint functional in app, constraint method} with $N>-2$, 
the map between $\bar{\sigma}^{(\lambda)}$ and $\lambda$ is one-to-one, 
and $-\dd \bar\sigma^{(\lambda)}/\dd \lambda$ is guaranteed to be positive, provided that $\rho v\ll1$.

\section{Asymptotic Behavior of the Constraint Operator}
\label{app: constraint order counting}
While we adopted a specific constraint operator in the main text, this framework is applicable to a broader class of operators. In this appendix, we discuss the impact of such operators on the constrained instanton at NLO. For a certain class of constraint operators, we analyze their effect on the inner solution in the overlap region $\rho \ll r \ll m^{-1}$ and also show that the outer solution remains unchanged.

\subsubsection*{Asymptotic Behavior at \texorpdfstring{$\rho\ll r\ll m^{-1}$}{rho << r << 1/m}}

Let us first examine the asymptotic behavior of $\mathcal{O}_\mathrm{con}$ at $\rho\ll r\ll m^{-1}$. 
For simplicity, we restrict our attention to $\mathcal{O}_\mathrm{con}$ constructed as a higher-dimensional product of the field strength,
\begin{align}
    \mathcal{O}_\mathrm{con}
    \sim
    \qty(F_{\mu\nu}^a)^k\,,
    \qquad
    k=3,4,5,\ldots,
    \label{eq: constraint made of F and tildeF in app}
\end{align}
where the Lorentz and gauge indices are contracted in a gauge- and Lorentz-invariant manner; throughout this discussion, we do not distinguish between different contraction patterns (including possible substitutions of the dual field strength $\tilde{F}_{\mu\nu}^a$ instead of $F_{\mu\nu}^a$).
The asymptotic behavior of $\mathcal{O}_\mathrm{con}$ is obtained as
\begin{align}
    \mathcal{O}_\mathrm{con}
    \sim
    \qty(\frac{\rho}{r})^{4k}\rho^{-2k}
    \qty(1+\order{\rho^2/r^2})
    \quad\text{at}
    \quad
    r\gg\rho\,.
\end{align}
Here, we have used the scaling $F\sim\rho^2/r^4(1+\order{\rho^2/r^2})$
which follows from 
Eq.\,\eqref{eq: size definition for A}.

\subsubsection*{Inner Solution at \texorpdfstring{$\rho\ll r\ll m^{-1}$}{rho << r << 1/m}}
For a constraint operator of the form $\mathcal{O}_\mathrm{con}
\sim (F_{\mu\nu}^a)^k$,
the scaling of the source term appearing in the NLO Euler--Lagrange equation 
 Eq.\,\eqref{eq: NLO EOM for A} at $r\ll m^{-1}$,
can be estimated as
\begin{align}
    \left.\frac{\sigma_2}{\hat{r}}\rho\frac{\delta S_\mathrm{con}}{\delta \mathcal{A}(r)}\right|_{\mathcal{A}=\mathcal{A}_0^{(\mathrm{in})}}
    &\sim
    \sigma_2
    \frac{\rho^2}{r}\qty[\frac{r^3\mathcal{O}_\mathrm{con}}{\mathcal{A}_0^{(\mathrm{in})}}]
    \\
    &\sim\sigma_2
    \frac{\rho^2}{r}
    \qty[r^3\rho^{-2k}\qty(\frac{\rho}{r})^{4k}
    \qty(\frac{\rho^2}{r^2})^{-1}]
    \\
    &\sim(\rho^{4-2k}\sigma_2)\qty(\frac{\rho}{r})^{4(k-1)}\,.
    \label{eq: Ocon effect inner NLO in app}
\end{align}

Next, let us examine the asymptotic behavior of the particular solution $a_2^{(\sigma_2)}$
in the regime $\rho\ll r\ll m^{-1}$.
Substituting Eq.\,\eqref{eq: Ocon effect inner NLO in app},
the Euler--Lagrange equation at NLO for $a_2^{(\sigma_2)}$
scales as
\begin{align}
    \qty[\qty(\frac{\dd}{\dd\hat{r}})^2
    +\frac{1}{\hat{r}}\,\frac{\dd}{\dd\hat{r}}    
    -\frac{4 \bigl(\hat{r}^4-4 \hat{r}^2+1\bigr)}{\hat{r}^2\bigl(\hat{r}^2+1\bigr)^2}]a_2^{(\sigma_2)}
    \sim(\rho^{4-2k}\sigma_2)\hat{r}^{-4(k-1)}\,.
    \label{eq: NLO particular solution order counting}
\end{align}
Substituting the ansatz for the particular solution,
\begin{align}
    a_2^{(\sigma_2)}
    \sim
    (\mathrm{const}.)\times\hat{r}^{-n}\,,
\end{align}
into the left-hand side of Eq.\,\eqref{eq: NLO particular solution order counting}, we obtain
\begin{align}
    \qty[\qty(\frac{\dd}{\dd\hat{r}})^2
    +\frac{1}{\hat{r}}\,\frac{\dd}{\dd\hat{r}}    
    -\frac{4 \bigl(\hat{r}^4-4 \hat{r}^2+1\bigr)}{\hat{r}^2\bigl(\hat{r}^2+1\bigr)^2}]a_2^{(\sigma_2)}
    \sim(\mathrm{const}.)\times(n^2-4)\times\hat{r}^{-(n+2)}\,.
\end{align}
The cases $n=\pm2$ correspond to the homogeneous solutions and the genuine particular solution must have $|n|\neq2$.
Matching the power of $\hat{r}$ with Eq.\,\eqref{eq: NLO particular solution order counting},
we obtain $n=4k-6$, and hence,
\begin{align}
    a_2^{(\sigma_2)}
    \sim
    (\mathrm{const}.)\times\hat{r}^{-4k+6}\,.
    \label{eq: NLO particular solution suppression for general N}
\end{align}
This result is consistent with Eq.\,\eqref{eq: constraint particular solution asymptotics of A} for the specific case of $k=4$.

From the above argument, we conclude that for the higher-dimensional constraint operator taking the form of $\mathcal{O}_\mathrm{con}
\sim (F_{\mu\nu}^a)^k$ with $k\ge 3$, the particular solution is suppressed at large $\hat{r}$ at least as fast as
\begin{align}
a_2^{(\sigma_2)}=\order{\hat{r}^{-6}}\,,\qquad\hat{r}\to\infty\,.
\label{eq: NLO particular solution minimal suppression}
\end{align}
Since the NLO matching procedure requires the NLO inner solution only up to the order of $\hat{r}^{-2}$, as seen in Eq.\,\eqref{eq : NLO Gauge inner}, the contribution induced by $\mathcal{O}_\mathrm{con}$ is irrelevant in the region $\rho\ll r\ll m^{-1}$ for the NLO matching procedure.

\subsubsection*{Outer Solution at \texorpdfstring{$\rho\ll r\ll m^{-1}$}{rho << r << 1/m}}
 In the NLO Euler--Lagrange equation Eq.\,\eqref{eq: NLO outer equation for A} at $r\gg\rho$,
 the contribution from $\mathcal{O}_\mathrm{con}$ is restored as
\begin{align}
    &\mathcal{A}_4^{(\mathrm{out})\prime\prime}(R)+\frac{\mathcal{A}_4^{(\mathrm{out})\prime}(R)}{R}-\frac{(4+R^2) \mathcal{A}_4^{(\mathrm{out})}(R)}{R^2}
    \notag
    \\
    &~~~
    +\frac{6 K_2(R){}^2}{R^2}+\frac{\sqrt{2\lambda}K_2(R) K_1\left(\sqrt{2\lambda} R\right)}{R}
    -\frac{\sigma_2(\rho m_A)^{-2}}{6\pi^2m_AR}\left.\frac{\delta S_\mathrm{con}}{\delta\mathcal{A}(r)}\right|_{\mathcal{A}=(\rho m_A)^2\mathcal{A}_2^{(\mathrm{out})}}
    =0\,.
    \label{eq: NLO outer equation for A, constraint retained}
\end{align}
The last term represents the contribution from the constraint.
The factor $(\rho m_A)^{-2}$ originates from the difference in the powers of $\rho m_A$ between
$\mathcal{A}-(\rho m_A)^2\mathcal{A}_2^{(\mathrm{out})}=(\rho m_A)^4\mathcal{A}_4^{(\mathrm{out})}+\cdots$ and $\sigma=(\rho m_A)^2\sigma_2+\cdots$\,.

By substituting the LO outer solution $\mathcal{A}=(\rho m_A)^2\mathcal{A}_2^{(\mathrm{out})}$,
the asymptotic behavior 
of $\mathcal{O}_\mathrm{con}$ at $R\ll1$ is obtained as 
\begin{align}
    \mathcal{O}_\mathrm{con}
    \propto
    \frac{\qty(\rho m_A)^{2k}m_A^{2k}}{R^{4k}}
    \qty(1+\order{R^2})
    \quad\text{at}
    \quad
    R\ll 1\,.
\end{align}
We can then estimate the scaling of the constraint contribution with respect to $R$ as
\begin{align}
\frac{\sigma_2(\rho m_A)^{-2}}{m_AR}\left.\frac{\delta S_\mathrm{con}}{\delta\mathcal{A}(r)}\right|_{\mathcal{A}=(\rho m_A)^2\mathcal{A}_2^{(\mathrm{out})}}
    &\sim
    \sigma_2(\rho m_A)^{-2}\frac{m_A^{-1}}{R}
    \qty[\frac{r^3\mathcal{O}_\mathrm{con}}{(\rho m_A)^2\mathcal{A}_2^{(\mathrm{out})}}]
    \\
    &\sim
    \sigma_2(\rho m_A)^{-2}\frac{m_A^{-1}}{R}
    \qty[\frac{R^3m_A^{-3}}{(\rho m_A)^2}\frac{\qty(\rho m_A)^{2k}m_A^{2k}}{R^{4k}}
    \qty(\frac{(\rho m_A)^2}{R^{2}})^{-1}]
    \\
    &\sim
    (\rho^{4-2k}\sigma_2)(\rho m_A)^{4k-10}R^{-4(k-1)}\,.
\end{align}
Since the factor $\rho^{4-2k}\sigma_2$ does not include any powers of $\rho$ and $v$, we find that the constraint contribution in Eq.\,\eqref{eq: NLO outer equation for A, constraint retained}
is $\order{(\rho m_A)^{4k-10}}$. 
Therefore,
for $\mathcal{O}_\mathrm{con}$ given
in Eq.\,\eqref{eq: constraint made of F and tildeF in app},
the constraint contribution to the NLO Euler--Lagrange equation is further suppressed by the positive power of $(\rho m_A)^2$ since $k\ge3$,
and thus should not be included at NLO.

Note that our discussion above applies to a broad class of $\mathcal{O}_\mathrm{con}$ in Eq.\,\eqref{eq: constraint made of F and tildeF in app}, including the specific case of $(\tr F_{\mu\nu}\tilde{F}_{\mu\nu}/2)^2$.
From the observations above, we conclude that such constraint operators do not contribute to either the inner or the outer solution at the orders of $\hat{r}$ relevant for the NLO matching.
Therefore, the NLO matching between Eq.\,\eqref{eq : NLO Gauge inner} and Eq.\,\eqref{eq : NLO Gauge outer}
is achieved independently of $\mathcal{O}_\mathrm{con}$ in Eq.\,\eqref{eq: constraint made of F and tildeF in app}.

\section{Computation of the Leading Coefficient of the Lagrange Multiplier}
\label{app: Lagrange multiplier value}

In Sec.\,\ref{sec: NLO}, we have determined the value of the expansion coefficient $\sigma_2$ of the Lagrange multiplier through the matching \eqref{eq: c2 and cout by matching} and the finiteness of the profile at the origin \eqref{eq: sigma2 from c2}.
This value can also be obtained independently of that procedure by exploiting a general property of the Lagrange multiplier method.

Equation \,\eqref{eq: constraint value dependence of action} implies that
the Lagrange multiplier satisfies
\begin{align}
    \frac{\bar{\sigma}^{(\lambda)}}{g^2}=-\frac{\dd S}{\dd\rho}\qty[\frac{\dd S_\mathrm{con}}{\dd\rho}]^{-1}\,,
    \label{eq: sigma dSdlambda appendix}
\end{align}
where $S$ and $S_\mathrm{con}$ are evaluated at the constrained instanton solution parametrized by $\lambda$ (or equivalently $\rho$).
This relation allows us to determine the coefficient  $\sigma_2$.

For the specific choice of the constraint operator $\mathcal{O}_\mathrm{con}$ in Eq.\,\eqref{eq: Ocon = FFtilde}, we obtain
\begin{align}
    \bar{\sigma}^{(\lambda)} &= -g^2\frac{\dd S}{\dd\rho}\qty[\frac{\dd S_\mathrm{con}}{\dd\rho}]^{-1}
    \\
    &=
    -g^2\frac{\dd}{\dd \rho}\qty(\frac{8\pi^2+4\pi^2\rho^2m_A^2}{g^2}+\order{\rho^4v^4})
    \qty[\frac{\dd}{\dd\rho}\qty(\frac{384\pi^2}{7\rho^4}\qty(1+\order{\rho^2v^2}))]^{-1}
    \\
    &=
    \rho^4\frac{7}{192}(\rho m_A)^2
    \qty(1+\order{\rho^2v^2})\,.
\end{align}
Here, we have used the value of the constrained functional
\begin{align}
    S_\mathrm{con}[\bar{\mathcal{A}}]
    =
    S_\mathrm{con}[\mathcal{A}^{(\mathrm{in})}_0](1+\order{\rho^2v^2})
    =
    \frac{384\pi^2}{7\rho^4}(1+\order{\rho^2v^2})\,.
\end{align}

Note that the relation in Eq.\,\eqref{eq: sigma dSdlambda appendix} is applicable to both $\phi^4$ theory and gauge theory.
For $\phi^4$ theory,
the relationship in Eq.\,\eqref{eq: sigma dSdlambda appendix} is equivalent to Eq.\,(2.15) in Ref.\,\cite{Affleck:1980mp}.

\section{Scalar Field Profile \texorpdfstring{$\mathcal{H}$}{H} at NLO}
\label{app: higgs NLO detail}

In this appendix, we demonstrate the explicit form of the NLO correction to the scalar field profile $\mathcal{H}$.
The NLO correction to the inner solution $\mathcal{H}_2^{(\mathrm{in})}(\hat{r})$ satisfies

\begin{align}
    \mathcal{H}_2^{(\mathrm{in})\prime\prime}(\hat{r})
    +
    \frac{3}{\hat{r}}\mathcal{H}_2^{(\mathrm{in})\prime}(\hat{r})
    -
    \frac{3 \mathcal{H}_2^{(\mathrm{in})}(\hat{r})}{\hat{r}^2\left(\hat{r}^2+1\right)^2}
    +\frac{\hat{r}}{2(\hat{r}^2+1)^{3/2}}
    -\frac{3\mathcal{A}_2^{(\mathrm{in})}(\hat{r})}{2\lambda \hat{r}(\hat{r}^2+1)^{3/2}}
    =0\,.
\end{align}
The general solution to this equation is given by
\begin{align}
    \mathcal{H}_2^{(\mathrm{in})}(\hat{r})=
    h_2^{(m^2)}(\hat{r})+h_2^{(\mathcal{A}_2)}(\hat{r})
    +d_1\qty(\frac{\hat{r}^2}{1+\hat{r}^2})^{1/2}
    +d_2\qty(\frac{\hat{r}^2+1}{\hat{r}^2})^{3/2}\,,
\end{align}
where the particular solution $h_2^{(m^2)}(\hat{r})$ reads explicitly
\begin{align}
    h_2^{(m^2)}(\hat{r})=
    \frac{\hat{r}^2 \left(3 \hat{r}^2+2\right)-2 \left(\hat{r}^2+1\right)^2 \log \left(\hat{r}^2+1\right)}{16 \hat{r}^3 \sqrt{\hat{r}^2+1}}
    \quad\qty(\to\order{\hat{r}^3}
    ~\text{as}~\hat{r}\to0)\,.
\end{align}
The term $h_2^{(\mathcal{A}_2)}(\hat{r})$ is further decomposed as,
\begin{align}
    h_2^{(\mathcal{A}_2)}(\hat{r})
    &=
     h_2^{(\mathcal{A}_2,m_A^2)}(\hat{r})
    +h_2^{(\mathcal{A}_2,c_1)}(\hat{r})
    +h_2^{(\mathcal{A}_2,c_2)}(\hat{r})
    +h_2^{(\mathcal{A}_2,\sigma_2)}(\hat{r})\,,
    \\
    h_2^{(\mathcal{A}_2,m_A^2)}(\hat{r})&=
    \frac{3 \left(\hat{r}^2+1\right)^3 \log \left(\hat{r}^2+1\right)-\hat{r}^2 \left(2 \hat{r}^2+3\right)}{96 \hat{r}^3 \
    \left(\hat{r}^2+1\right)^{3/2}\lambda_H }\,,
    \\
    h_2^{(\mathcal{A}_2,c_1)}(\hat{r})  &=c_1
    \frac{-\hat{r}}{8 \left(\hat{r}^2+1\right)^{3/2}\lambda_H }
    \quad\qty(\to\order{\hat{r}^{-2}}~\text{as}~\hat{r}\to\infty)\,,
    \\
    h_2^{(\mathcal{A}_2,c_2)}(\hat{r})  &=c_2
    \frac{3 \left(-4 \hat{r}^4-\hat{r}^2-20 \left(\hat{r}^2-3\right) \hat{r}^4 \log \left(\hat{r}\right)+3\right)}{80 \hat{r}^3 \left(\hat{r}^2+1\right)^{3/2}\lambda_H }\,,
    \\
    h_2^{(\mathcal{A}_2,\sigma_2)}(\hat{r})&=\frac{\sigma_2}{\rho^4}\frac{5 \left(12 \hat{r}^6-6 \hat{r}^4-68 \hat{r}^2-63\right) \hat{r}^4-48 \hat{r}^2+60 \left(\hat{r}^2-3\right) \left(\hat{r}^2+1\right)^3 \hat{r}^4 \log \left(\frac{\hat{r}^2}{\hat{r}^2+1}\right)-18}{70\hat{r}^3 \left(\hat{r}^2+1\right)^{9/2}\lambda_H}
    \notag\\
    &\phantom{h_2^{(\mathcal{A}_2,\sigma_2)}(\hat{r})=\frac{\sigma_2}{\rho^4}5 \left(12 \hat{r}^6-6 \hat{r}^4-68 \hat{r}^2-63\right) \hat{r}^4-48 \hat{r}^2}
    (\to\order{\hat{r}^{-12}}~\text{as}~\hat{r}\to\infty)\,.
\end{align}
Here, the explicit dependence on the quartic coupling $\lambda_H$ appears through $\mathcal{A}-\mathcal{A}_0=\lambda_H^{-1}(\rho m_H)^2\mathcal{A}_2$.
Note that, in the limit $\lambda_H\to0$, the factor of $\lambda_H^{-1}$ in the particular solution cancels against the $\lambda_H$ in the expansion:
\begin{align}
\mathcal{H}&=\mathcal{H}_0+(\rho m_H)^2\mathcal{H}_2+\cdots
\\
&=\mathcal{H}_0+(\rho v)^2(\lambda_H\mathcal{H}_2)+\cdots\,,
\end{align}
where the combination $\lambda_H \mathcal{H}_2$ remains finite.

Note that there is a freedom of adding the homogeneous solutions when choosing these particular solutions.
While not essential for the matching procedure and the entire construction of the solution, let us comment on our specific choice here.
All these particular solutions are chosen to vanish at $\hat{r}=0$ 
when Eq.\,\eqref{eq: sigma2 from c2} is satisfied. 
The remaining freedom is used to make the particular solutions convenient for the matching procedure.

Substituting Eqs.\,\eqref{eq: sigma2 from c2}, \eqref{eq: c_1 choice in YMH} and \eqref{eq: c2 and cout by matching},
we obtain
\begin{align}
    &h_2^{(\mathcal{A}_2)}(\hat{r})=
    \notag\\&~~~~\frac{\hat{r}^2 \left(976 \hat{r}^8+2652 \hat{r}^6+2228 \hat{r}^4+457 \
\hat{r}^2-60\right)+60 \left(6 \hat{r}^4+3 \hat{r}^2+1\right) \
\left(\hat{r}^2+1\right)^3 \log \left(\hat{r}^2+1\right)}{1920 \
\hat{r}^3 \left(\hat{r}^2+1\right)^{9/2}\lambda_H}\,.
\end{align}
The boundary condition $\mathcal{H}_2^{(\mathrm{in})}(0)=0$ in Eq.\,\eqref{eq: boundary condition at each order} is then satisfied for
\begin{align}
    d_2=0\,.
\end{align}

\section{Yang--Mills Action at Order \texorpdfstring{$(\rho m_{A,H})^4$}{(rho m)\^4}}
\label{app: action estimation}
In this appendix, we estimate the action to $(\rho m_{A,H})^4$ order, using LO and NLO solutions together with partial next-to-NLO (NNLO) solutions.
Since the profile functions are constructed separately in the inner and outer regions, we split the radial integral accordingly. For a generic radial interval $r_\mathrm{low}\le r\le r_\mathrm{max}$, it is convenient to introduce
\begin{align}
    S[\mathcal{A},\mathcal{H};r_\mathrm{low},r_\mathrm{up}]
    :=
    \int_{r_\mathrm{low}}^{r_\mathrm{up}}\dd r\,
    2 \pi ^2 r^3 
    L(\mathcal{A}(r),\mathcal{A}'(r),\mathcal{H}(r),\mathcal{H}'(r),r)\,,
    \label{eq: action with cutoff}
\end{align} 
where $L$ is given in Eq.\,\eqref{eq: YMH lagrangian in terms of profiles}.

In the inner and outer regions,
we truncate the profile functions
using a perturbative expansion in $(\rho m_{A,H})^2$.
Hereafter, we use the following shorthand notation:
\begin{align}
\Phi^{(\mathrm{in,out})}_{\le 2n}:=\sum_{i=0}^n(\rho m_\Phi)^{2i}\Phi^{(\mathrm{in,out})}_{2i}\,,
\qquad
\Phi=\mathcal{A},\mathcal{H}\,.
\end{align}
For the scalar field $\mathcal{H}$ in the outer region, we include the constant shift appearing in Eq.\,\eqref{eq: Expansion of inner/outer solutions H} in $\mathcal{H}_0^{(\mathrm{out})}$, throughout this appendix.
Suppose that we have evaluated $S[\Phi_{\le2n}]$, where $S[\Phi]$ collectively denotes
the integral in Eq.\,\eqref{eq: action with cutoff}.
The effect of the neglected higher-order terms can be estimated by considering
    $\Phi= \Phi_{\le2n}+\epsilon\delta\Phi$
which yields
\begin{align}
    S[\Phi_{\le2n}+\epsilon\delta\Phi]
    &=
    S[\Phi_{\le 2n}]
    +
    \epsilon\int_{r_\mathrm{low}}^{r_\mathrm{{up}}}
    \dd r\,
    \qty(\left.\frac{\partial L}{\partial \Phi(r)}\right|_{\Phi=\Phi_{\le2n}}
    \!\!\!\!\!\!\!\!\!\!\!
    \delta\Phi(r)
    +
    \left.\frac{\partial L}{\partial \Phi'(r)}\right|_{\Phi=\Phi_{\le2n}}
    \!\!\!\!\!\!\!\!\!\!\!
    \delta\Phi'(r))
    +\order{\epsilon^2}
    \\
    &=
    S[\Phi_{\le 2n}]
    +
    \epsilon\left.\left.\left(\frac{\partial L}{\partial \Phi'(r)}\right|_{\Phi=\Phi_{\le 2n}}
    \!\!\!\!\!\!\!\!\!\!\!
    \delta\Phi(r)\right)\right|_{r=r_\mathrm{low}}^{r=r_{\mathrm{up}}}
    \!\!\!\!
    +
    \epsilon\int_{r_\mathrm{low}}^{r_\mathrm{{up}}}
    \dd r\,
    \left.\frac{\delta S}{\delta \Phi(r)}\right|_{\Phi=\Phi_{\le 2n}}
    \!\!\!\!\!\!\!\!\!\!\!
    \delta\Phi(r)
    +\order{\epsilon^2}\,,
    \label{eq: perturbation in general functional}
\end{align}
where $\epsilon=(\rho m_\Phi)^{2n+2}$.
Note that, although $\delta\Phi$ vanishes at $r=0$ and $r=\infty$ due to the boundary conditions in Eq.\,\eqref{eq: boundary condition at each order},
it remains non-vanishing at finite $r$, namely at $r=r_{\mathrm{up},\mathrm{low}}$.
Hence, the second term, the boundary contribution, does not vanish.

We evaluate $S[\Phi_{\le2}]$, which uses the LO and NLO solutions in the inner region and the LO solution in the outer region.
Note that
$S[\Phi_{\le2}]$ exhibits a logarithmic divergence as $r_\mathrm{up}\to\infty$ in Eq.\,\eqref{eq: action with cutoff}, when evaluated with the inner solution. Introducing
a cutoff $r_\mathrm{max}$ and setting
$r_\mathrm{low}=0$ and $r_\mathrm{up}=r_\mathrm{max}$ in Eq.\,\eqref{eq: action with cutoff},
we obtain
\begin{align}
    &g^2S[
    \mathcal{A}^{(\mathrm{in})}_{\le2}
    , 
    \mathcal{H}^{(\mathrm{in})}_{\le2}
    ;
    0,r_\mathrm{max}
    ]
    =
    8 \pi ^2
    +
    4 \pi ^2 (\rho m_A)^2
    \notag\\&\qquad
    +
    \pi ^2 (\rho m_A)^4 \left(\frac{m_H^2}{m_A^2} \left(-2\log \frac{\rho  m_H}{2}
    -\log \frac{r_{\max }}{\rho}
    -2\gamma_E+\frac{1}{2}\right)-3 \log \frac{r_{\max }}{\rho }-\frac{146753}{27720}\right)
    \notag\\
    &\qquad+f_\mathrm{in}(r_\mathrm{max})+\order{\rho^6m_{A,H}^6}\,.
    \label{eq: S inner in app}
\end{align}
Here, terms vanishing in the limit $r_\mathrm{max}/\rho\to\infty$ are denoted by $f_\mathrm{in}(r_\mathrm{max})$.

For the outer solution,
$S[\Phi_{\le2}]$ exhibits power divergences as $r_\mathrm{min}\to0$.
Evaluating the integral with the LO outer solution, we find
\begin{align}
    &g^2S[
    \mathcal{A}^{(\mathrm{out})}_{\le2}
    , 
    \mathcal{H}^{(\mathrm{out})}_{\le2}
    ;
    r_\mathrm{min},\infty
    ]
   =
   \pi ^2 (\rho m_A) ^4  \left(
   \frac{24}{m_A^4 r_{\min }^4}
   -
   \frac{4}{m_A^2r_{\min }^2}
   +
   \frac{m_H^2}{m_A^2} \left(\log \frac{m_H r_{\min
   }}{2}+\gamma_E-1\right)+\frac{3}{4}
   \right)
    \notag\\
    &\qquad\quad+f_\mathrm{out}(r_\mathrm{min})+\order{\rho^6m_{A,H}^6}\,,
    \label{eq: S outer in app}
\end{align}
where terms vanishing in the limit $m_{A,H}r_\mathrm{min}\to0$ are denoted by $f_\mathrm{out}(r_\mathrm{min})$.
Here, we have introduced a cutoff $r_\mathrm{min}$ by setting $r_\mathrm{low}=r_\mathrm{min}$ and $r_\mathrm{up}=\infty$ in Eq.\,\eqref{eq: action with cutoff}.

Before discussing how to treat the divergences encountered above, let us note that $S[\Phi_{\le2}]$ is not sufficient for evaluating the action to $(\rho m_{A,H})^4$-order.
To estimate the effect of higher-order profile corrections in the inner region, we take 
$n=1$ in Eq.\,\eqref{eq: perturbation in general functional} and consider
\begin{align}
    (\rho m_\Phi)^4\delta\Phi
    =
    \begin{cases}
    (\rho m_A)^4\mathcal{A}^{(\mathrm{in})}_4+\order{\rho^6m_{A,H}^6}\,,
    \\[1em]
    (\rho m_H)^4\mathcal{H}^{(\mathrm{in})}_4+\order{\rho^6m_{A,H}^6}\,.
    \end{cases}
\end{align}
Since the constrained instanton solutions are constructed order-by-order in $(\rho m_{A,H})^2$ so as to satisfy the Euler--Lagrange equation for $S_\mathrm{tot}$, we have
\begin{align}
    \left.\frac{\delta S}{\delta \Phi(r)}\right|_{\mathcal{A}=\mathcal{A}^{(\mathrm{in})}_{\le2},\mathcal{H}=\mathcal{H}^{(\mathrm{in})}_{\le2}}
    &=
    \left.\frac{\delta S_\mathrm{tot}}{\delta \Phi(r)}\right|_{\mathcal{A}=\mathcal{A}^{(\mathrm{in})}_{\le2},\mathcal{H}=\mathcal{H}^{(\mathrm{in})}_{\le2}}
    -
    \frac{\sigma}{g^2}
    \left.\frac{\delta S_\mathrm{con}}{\delta \Phi(r)}\right|_{\mathcal{A}=\mathcal{A}^{(\mathrm{in})}_{\le2},\mathcal{H}=\mathcal{H}^{(\mathrm{in})}_{\le2}}
    \\&
    =
    -
    (\rho m_A)^2\frac{\sigma_2}{g^2}
    \left.\frac{\delta S_\mathrm{con}}{\delta \Phi(r)}\right|_{\mathcal{A}=\mathcal{A}^{(\mathrm{in})}_{\le2},\mathcal{H}=\mathcal{H}^{(\mathrm{in})}_{\le2}}
    +\order{\rho^4 m_{A,H}^4}
    \\&
    =\order{\rho^2m_{A,H}^2}\,.
\end{align}
It then follows that the third term in Eq.\,\eqref{eq: perturbation in general functional}
contributes only at
$\order{\rho^6m_{A,H}^6}$, and is therefore negligible for computing the action to $(\rho m_{A,H})^4$-order.

As for the boundary contribution, which is non-vanishing at $r=r_\mathrm{max}$, we obtain
\begin{align}
    \left.\frac{\partial L}{\partial\mathcal{A}'(r)}\right|_{\mathcal{A}=\mathcal{A}^{(\mathrm{in})}_0,\mathcal{H}=\mathcal{H}^{(\mathrm{in})}_0}
    &=6 \pi ^2 r \mathcal{A}^{(\mathrm{in})\prime}_0(r)
    =-\frac{24 \pi ^2 \rho ^2 r^2}{\left(\rho ^2+r^2\right)^2}
    \,,
    \label{eq: dL/dA in app}
    \\
    \left.\frac{\partial L}{\partial \mathcal{H}'(r)}\right|_{\mathcal{A}=\mathcal{A}^{(\mathrm{in})}_0,\mathcal{H}=\mathcal{H}^{(\mathrm{in})}_0}
    &=4 \pi ^2 v ^2 r^3 \mathcal{H}^{(\mathrm{in})\prime}_0(r)=\order{\rho^2 m_A^2}\,,
    \label{eq: dL/dH in app}
\end{align}
and we see that the correction from
$\mathcal{A}^{(\mathrm{in})}_4$
contributes at
$(\rho m_{A,H})^4$-order through the boundary term in Eq.\,\eqref{eq: perturbation in general functional},
whereas the contribution from $\mathcal{H}^{(\mathrm{in})}_4$ is
$\order{\rho^6m_{A,H}^6}$ and can be neglected in the present evaluation.

By the $(4,-2)$ matching condition, which corresponds to the matching of the NNLO inner solution $\mathcal{A}^{(\mathrm{in})}_4$ to the LO outer solution, we find
\begin{align}
    \mathcal{A}^{(\mathrm{in})}_4 
    =
    \frac{\hat{r}^2}{8}  \left(-\log \frac{\rho m_A\hat{r}}{2} -\gamma_E +\frac{3}{4}\right)
    +\order{\hat{r}^0}\qquad\hat{r}\gg1\,.
    \label{eq: A4 relevant term}
\end{align}
Here, we used the LO outer solution \eqref{eq: leading order inner solution of A} with the expansion \eqref{eq: asymptotic expanding Bessel of A}.
The terms of $\order{\hat{r}^0}$ are unnecessary for evaluating the action.
Substituting Eq.\,\eqref{eq: dL/dA in app} and Eq.\,\eqref{eq: A4 relevant term} into the second term in Eq.\,\eqref{eq: perturbation in general functional},
we obtain an additional contribution to the action at the inner region in Eq.\,\eqref{eq: S inner in app},
\begin{align}
    &g^2S[
    \mathcal{A}^{(\mathrm{in})}_{\le4}, 
    \mathcal{H}^{(\mathrm{in})}_{\le4};
    0,r_\mathrm{max}
    ]
    -g^2S[
    \mathcal{A}^{(\mathrm{in})}_{\le2}, 
    \mathcal{H}^{(\mathrm{in})}_{\le2};
    0,r_\mathrm{max}
    ]
    \notag\\
    &\qquad\qquad
    =
    3 \pi ^2 \rho ^4 m_A^4 \left(\log \frac{m_A r_\mathrm{max}}{2}+ \gamma_E-\frac{3}{4}\right)
    +f_\mathrm{in}^{(\mathrm{NNLO})}(r_\mathrm{max})+\order{\rho^6m_{A,H}^6}
    \,.
    \label{eq: S inner NNLO in app}
\end{align}
Here, terms vanishing in the limit $r_\mathrm{max}/\rho\to\infty$
are denoted by
$f_\mathrm{in}^{(\mathrm{NNLO})}(r_\mathrm{max})$.

In the outer region, the higher-order profile corrections that are added to Eq.\,\eqref{eq: S outer in app} correspond to the case,
\begin{align}
    (\rho m_\Phi)^4\delta\Phi
    =
    \begin{cases}
    (\rho m_A)^4\mathcal{A}^{(\mathrm{out})}_4+\order{\rho^6m_{A,H}^6}\,,
    \\[1em]
    (\rho m_H)^4\mathcal{H}^{(\mathrm{out})}_4+\order{\rho^6m_{A,H}^6}\,.
    \end{cases}
\end{align}
Since every term in the action in Eq.\,\eqref{eq: YMH lagrangian in terms of profiles} is at least quadratic in either $\mathcal{A}$ or $1-\mathcal{H}$, and since $\mathcal{A}$ and $1-\mathcal{H}$ are  already $\order{\rho^2m_{A,H}^2}$  at LO in the outer region,
any contribution from $\mathcal{A}^{(\mathrm{out})}_4$ and $\mathcal{H}^{(\mathrm{out})}_4$  in Eq.\,\eqref{eq: perturbation in general functional} enters $S$ only at $\order{\rho^6m_{A,H}^6}$.
Therefore, these NNLO corrections are irrelevant for computing the action up to $(\rho m_{A,H})^4$-order.

Combining the inner and outer contributions 
$S[
\mathcal{A}^{(\mathrm{in})}_{\le4}, 
\mathcal{H}^{(\mathrm{in})}_{\le4};
0,r_\mathrm{max}
]$
and
$S[
\mathcal{A}^{(\mathrm{out})}_{\le2}, 
\mathcal{H}^{(\mathrm{out})}_{\le2};
r_\mathrm{min},\infty
]$,
we can determine the action to $(\rho m_{A,H})^4$-order.
To do this, we also need the integral in the overlap region $r_\mathrm{min}\le r\le r_\mathrm{max}$. 
In this region, we use the same double expansion in $(\rho m_{A,H})^2$ and $\hat{r}$
as in the matching procedure of Sec.\,\ref{sec: explicit construction in YM}, namely
\begin{align}
    &\mathcal{A}(r)
    =
    \frac{2}{\hat{r}^{2}}-\frac{2}{\hat{r}^{4}}
    -(\rho m_A)^2
    \qty[
        \frac{1}{2}+\frac{1}{\hat{r}^{2}}\qty(2\log\hat{r}+\frac{37}{12})
    ]
    \notag\\
    &\qquad\quad
    +(\rho m_A)^4
    \frac{\hat{r}^2}{8}  \left(-\log \frac{\rho m_A\hat{r}}{2} -\gamma_E +\frac{3}{4}\right)
    +\order{(\rho m_{A,H})^n\hat{r}^{-k}}
    \,,
    \\
    &\mathcal{H}(r)
    =
    1-\frac{1}{2 \hat{r}^2}+\frac{3}{8 \hat{r}^4}
    +
    (\rho m_H)^2 \left[
    \frac{1}{8} \left(-2 \log \frac{\rho   m_H\hat{r}}{2}-2 \gamma_E +1\right)
   \right.\notag\\
   &\qquad\qquad+
   \left.
    \frac{1}{8\hat{r}^{2}}\qty( \log \frac{\rho  m_H}{2}+\frac{61}{15 \lambda_H }+\frac{3 \log \hat{r}}{\lambda_H }-3 \log \hat{r}+\gamma_E -\frac{1}{2})
   \right]
   \notag\\
   &\qquad\quad
   +\order{(\rho m_{A,H})^n\hat{r}^{-k}}
    \,,
\end{align}
where the suppressed terms satisfy either $n\ge6$ or $n+k\ge4$, corresponding to the next-to-NNLO corrections to the inner solution and the NLO corrections to the outer solution, respectively.
From these expansions we obtain
\begin{align}
    g^2S[
    \mathcal{A},
    \mathcal{H};
    r_\mathrm{min},r_\mathrm{max}
    ]
    =
    \pi ^2 (\rho m_A)^4 \left(
    \frac{24}{m_A^4r_{\min }^4}
    -\frac{4}{m_A^2r_{\min }^2}
    -\frac{m_H^2}{m_A^2}\log \frac{r_{\max }}{r_{\min }}
    \right)
    \notag\\
    +f_\mathrm{in}(r_\mathrm{max})
    +f_\mathrm{in}^{(\mathrm{NNLO})}(r_\mathrm{max})
    +f_\mathrm{out}(r_\mathrm{min})
    +\order{\rho^6 m_{A,H}^6}\,,
    \label{eq: S overlapping in app}
\end{align}
valid for $\rho\ll r_\mathrm{min}<r_\mathrm{max}\ll m_{A,H}^{-1}$\,.
Here, $f_\mathrm{in}(r_\mathrm{max})$, 
$f_\mathrm{out}(r_\mathrm{min})$
and  $f_\mathrm{in}^{(\mathrm{NNLO})}(r_\mathrm{max})$ 
are the same functions as those appearing in Eqs.\,\eqref{eq: S inner in app}, \eqref{eq: S outer in app} and \eqref{eq: S inner NNLO in app},
respectively.

Finally, the action is obtained by adding the inner and outer contributions and subtracting the overlap contribution:
\begin{align}
    S
    =
    S[
    \mathcal{A}^{(\mathrm{in})}_{\le4}, 
    \mathcal{H}^{(\mathrm{in})}_{\le4};
    0,r_\mathrm{max}
    ]
    +
    S[
    \mathcal{A}^{(\mathrm{out})}_{\le2}, 
    \mathcal{H}^{(\mathrm{out})}_{\le2};
    r_\mathrm{min},\infty
    ]
    -
    S[
    \mathcal{A},
    \mathcal{H};
    r_\mathrm{min},r_\mathrm{max}
    ]+\order{\rho^6m_{A,H}^6}\,,
    \label{eq: action evaluation by subtracting overlapping}
\end{align}
which yields
\begin{align}
    g^2S&=8\pi^2+4\pi^2(\rho m_A)^2
    +(\rho m_A)^4\pi^2
    \qty[3\log{\frac{\rho m_A}{2}}+3\gamma_E-\frac{188333}{27720}]
    \notag\\&\quad
    -\rho^4m_A^2m_H^2\pi^2
    \qty[\log\frac{\rho m_H}{2}+\gamma_E+\frac{1}{2}]
    +\order{\rho^6 m_{A,H}^6}\,.
    \label{eq: action value from matching, 4th order in app}
\end{align}
As expected, the divergent terms in the limits $r_\mathrm{max}/\rho\to\infty$ and $m_{A,H}r_\mathrm{min}\to0$ cancel among the three contributions. 
Moreover, the finite cutoff-dependent pieces,  $f_\mathrm{in}(r_\mathrm{max})$, 
$f_\mathrm{out}(r_\mathrm{min})$
and  $f_\mathrm{in}^{(\mathrm{NNLO})}(r_\mathrm{max})$ appearing in Eqs.\,\eqref{eq: S inner in app}, \eqref{eq: S outer in app}, \eqref{eq: S inner NNLO in app}
and \eqref{eq: S overlapping in app}, also cancel automatically.

\section{Constrained Instanton and Minimization}
\label{app: remarks on constraint}
In Sec.\,\ref{sec: explicit construction in YM},
we numerically constructed the constrained instanton solution to Eqs.\,\eqref{eq: gauge field ansatz EOM} and \eqref{eq: scalar field ansatz EOM} by minimizing the modified action,
\begin{align}
    S_{\mathrm{YMH}+\sigma}
    =
    S + \frac{\sigma}{g^2}S_\mathrm{con}\,,
\end{align}
at fixed $\sigma$.
Note that $S_{\mathrm{YMH}+\sigma}$ differs from the constrained action $S_\mathrm{tot}$ merely by a constant shift $\sigma\lambda/g^2$.
Depending on the choice of $S_\mathrm{con}$,
however, a solution to Eqs.\,\eqref{eq: gauge field ansatz EOM} and \eqref{eq: scalar field ansatz EOM} can be a saddle point of $S_{\mathrm{YMH}+\sigma}$, rather than a minimum.
With this in mind, we choose $\mathcal{O}_\mathrm{con}$ such that the constrained instanton is obtained as a minimum of $S_{\mathrm{YMH}+\sigma}$.
In the following, we discuss the properties that $\mathcal{O}_\mathrm{con}$ should satisfy
so that the constrained instanton solutions constructed in Sec.\,\ref{sec: explicit construction in YM} are indeed obtainable by minimizing $S_{\mathrm{YMH}+\sigma}$.

Specifically, we assume that $\mathcal{O}_\mathrm{con}\ge0$ and that, on the constrained instanton profile
$(\bar{\mathcal{A}},\bar{\mathcal{H}})$, the corresponding functional behaves as
\begin{align}
    S_\mathrm{con}[\bar{\mathcal{A}},\bar{\mathcal{H}}]
    =C\rho^{-N}[1+\order{\rho^2v^2}]\,,
    \qquad
    N>0\,,
    \qquad
    C>0\,,
    \label{eq: typical behavior of constraint functional in app, minimization}
\end{align}
so that $S_\mathrm{con}$ decreases monotonically with the size $\rho$.
As an explicit example, for $\mathcal{O}_\mathrm{con}$ in Eq.\,\eqref{eq: Ocon = FFtilde} we find
\begin{align}
    S_\mathrm{con}[\bar{\mathcal{A}}]
    =
    S_\mathrm{con}[\mathcal{A}^{(\mathrm{in})}_0](1+\order{\rho^2m_A^2})
    =
    \frac{384\pi^2}{7}\rho^{-4}(1+\order{\rho^2m_A^2})\,,
    \label{eq: value of constraint functional, FFtilde}
\end{align}
which corresponds to the case $N=4$.

Under the above assumptions, the corresponding Lagrange multiplier 
$\bar\sigma^{(\lambda)}$ is positive.
Indeed,
\begin{align}
    \frac{\bar\sigma^{(\lambda)}}{g^2}
    =
    -\frac{\dd \bar{S}^{(\lambda)}}{\dd\lambda}
    =
    -\frac{\dd \bar{S}^{(\lambda)}}{\dd \rho}
    \qty(\frac{\dd S_\mathrm{con}}{\dd \rho})^{-1}>0\,.
\end{align}
Here, we used Eq.\,\eqref{eq: constraint value dependence of action}, together with $S_\mathrm{con}[\bar{\mathcal{A}}]=\lambda$ and $\bar{S}^{(\lambda)}-8\pi^2/g^2=4\pi^2\rho^2m_A^2/g^2+\order{\rho^4 m_{A,H}^4}$ in the last inequality.
Combined with $S_\mathrm{con}\ge0$,
the positivity of $\bar\sigma^{(\lambda)}$ ensures that $S_{\mathrm{YMH}+\sigma}$
evaluated at $\sigma=\bar\sigma^{(\lambda)}$
is bounded from below as $S_{\mathrm{YMH}+\sigma}\ge 8\pi^2/g^2$.

Note that $S_{\mathrm{YMH}+\sigma}$ evaluated at the constrained instanton constructed in Sec.\,\ref{sec: explicit construction in YM} satisfies
\begin{align}
    S_{\mathrm{YMH}+\sigma}[\mathcal{A},\mathcal{H}]
    =
    \frac{8\pi^2}{g^2}+\order{\rho^2 v^2}\,.
    \label{eq: Stot evaluated at the solution in app}
\end{align}
Here, we used Eq.\,\eqref{eq: action value from matching, 4th order} together with $\bar{\sigma}^{(\lambda)} S_\mathrm{con}=\order{\rho^2 v^2}$, which in turn follows from
Eq.\,\eqref{eq: typical behavior of constraint functional in app, minimization} and  $\bar{\sigma}^{(\lambda)}\times\rho^{-N}=\order{\rho^2 v^2}$ (see Eqs.\,\eqref{eq: Expansion of Lagrange multiplier}, \eqref{eq: sigma0} and \eqref{eq: sigma2 value for YMH}).
Since the minimum of $S_{\mathrm{YMH}+\sigma}$ does not exceed the value in Eq.\,\eqref{eq: Stot evaluated at the solution in app},
the latter provides the upper bound on the minimum of $S_{\mathrm{YMH}+\sigma}$.
This upper bound suggests
that the minimum configuration of $S_{\mathrm{YMH}+\sigma}$ can be obtained through a perturbative expansion in $\rho v\ll1$ around the instanton configuration in the unbroken phase.
Therefore, the minimum of $S_{\mathrm{YMH}+\sigma}$ is exactly the constrained instanton constructed in Sec.\,\ref{sec: explicit construction in YM}.

\bibliographystyle{apsrev4-1}
\bibliography{bibtex}

\begin{thebibliography}{33}%
\makeatletter
\providecommand \@ifxundefined [1]{%
 \@ifx{#1\undefined}
}%
\providecommand \@ifnum [1]{%
 \ifnum #1\expandafter \@firstoftwo
 \else \expandafter \@secondoftwo
 \fi
}%
\providecommand \@ifx [1]{%
 \ifx #1\expandafter \@firstoftwo
 \else \expandafter \@secondoftwo
 \fi
}%
\providecommand \natexlab [1]{#1}%
\providecommand \enquote  [1]{``#1''}%
\providecommand \bibnamefont  [1]{#1}%
\providecommand \bibfnamefont [1]{#1}%
\providecommand \citenamefont [1]{#1}%
\providecommand \href@noop [0]{\@secondoftwo}%
\providecommand \href [0]{\begingroup \@sanitize@url \@href}%
\providecommand \@href[1]{\@@startlink{#1}\@@href}%
\providecommand \@@href[1]{\endgroup#1\@@endlink}%
\providecommand \@sanitize@url [0]{\catcode `\\12\catcode `\$12\catcode `\&12\catcode `\#12\catcode `\^12\catcode `\_12\catcode `\%12\relax}%
\providecommand \@@startlink[1]{}%
\providecommand \@@endlink[0]{}%
\providecommand \url  [0]{\begingroup\@sanitize@url \@url }%
\providecommand \@url [1]{\endgroup\@href {#1}{\urlprefix }}%
\providecommand \urlprefix  [0]{URL }%
\providecommand \Eprint [0]{\href }%
\providecommand \doibase [0]{http://dx.doi.org/}%
\providecommand \selectlanguage [0]{\@gobble}%
\providecommand \bibinfo  [0]{\@secondoftwo}%
\providecommand \bibfield  [0]{\@secondoftwo}%
\providecommand \translation [1]{[#1]}%
\providecommand \BibitemOpen [0]{}%
\providecommand \bibitemStop [0]{}%
\providecommand \bibitemNoStop [0]{.\EOS\space}%
\providecommand \EOS [0]{\spacefactor3000\relax}%
\providecommand \BibitemShut  [1]{\csname bibitem#1\endcsname}%
\let\auto@bib@innerbib\@empty
%</preamble>
\bibitem [{\citenamefont {Belavin}\ \emph {et~al.}(1975)\citenamefont {Belavin}, \citenamefont {Polyakov}, \citenamefont {Schwartz},\ and\ \citenamefont {Tyupkin}}]{Belavin:1975fg}%
  \BibitemOpen
  \bibfield  {author} {\bibinfo {author} {\bibfnamefont {A.~A.}\ \bibnamefont {Belavin}}, \bibinfo {author} {\bibfnamefont {A.~M.}\ \bibnamefont {Polyakov}}, \bibinfo {author} {\bibfnamefont {A.~S.}\ \bibnamefont {Schwartz}}, \ and\ \bibinfo {author} {\bibfnamefont {Y.~S.}\ \bibnamefont {Tyupkin}},\ }\href {\doibase 10.1016/0370-2693(75)90163-X} {\bibfield  {journal} {\bibinfo  {journal} {Phys. Lett. B}\ }\textbf {\bibinfo {volume} {59}},\ \bibinfo {pages} {85} (\bibinfo {year} {1975})}\BibitemShut {NoStop}%
\bibitem [{\citenamefont {'t~Hooft}(1976{\natexlab{a}})}]{tHooft:1976snw}%
  \BibitemOpen
  \bibfield  {author} {\bibinfo {author} {\bibfnamefont {G.}~\bibnamefont {'t~Hooft}},\ }\href {\doibase 10.1103/PhysRevD.14.3432} {\bibfield  {journal} {\bibinfo  {journal} {Phys. Rev. D}\ }\textbf {\bibinfo {volume} {14}},\ \bibinfo {pages} {3432} (\bibinfo {year} {1976}{\natexlab{a}})},\ \bibinfo {note} {[Erratum: Phys.Rev.D 18, 2199 (1978)]}\BibitemShut {NoStop}%
\bibitem [{\citenamefont {'t~Hooft}(1976{\natexlab{b}})}]{tHooft:1976rip}%
  \BibitemOpen
  \bibfield  {author} {\bibinfo {author} {\bibfnamefont {G.}~\bibnamefont {'t~Hooft}},\ }\href {\doibase 10.1103/PhysRevLett.37.8} {\bibfield  {journal} {\bibinfo  {journal} {Phys. Rev. Lett.}\ }\textbf {\bibinfo {volume} {37}},\ \bibinfo {pages} {8} (\bibinfo {year} {1976}{\natexlab{b}})}\BibitemShut {NoStop}%
\bibitem [{\citenamefont {Callan}\ \emph {et~al.}(1976)\citenamefont {Callan}, \citenamefont {Dashen},\ and\ \citenamefont {Gross}}]{Callan:1976je}%
  \BibitemOpen
  \bibfield  {author} {\bibinfo {author} {\bibfnamefont {C.~G.}\ \bibnamefont {Callan}, \bibfnamefont {Jr.}}, \bibinfo {author} {\bibfnamefont {R.~F.}\ \bibnamefont {Dashen}}, \ and\ \bibinfo {author} {\bibfnamefont {D.~J.}\ \bibnamefont {Gross}},\ }\href {\doibase 10.1016/0370-2693(76)90277-X} {\bibfield  {journal} {\bibinfo  {journal} {Phys. Lett. B}\ }\textbf {\bibinfo {volume} {63}},\ \bibinfo {pages} {334} (\bibinfo {year} {1976})}\BibitemShut {NoStop}%
\bibitem [{\citenamefont {Callan}\ \emph {et~al.}(1978)\citenamefont {Callan}, \citenamefont {Dashen},\ and\ \citenamefont {Gross}}]{Callan:1977gz}%
  \BibitemOpen
  \bibfield  {author} {\bibinfo {author} {\bibfnamefont {C.~G.}\ \bibnamefont {Callan}, \bibfnamefont {Jr.}}, \bibinfo {author} {\bibfnamefont {R.~F.}\ \bibnamefont {Dashen}}, \ and\ \bibinfo {author} {\bibfnamefont {D.~J.}\ \bibnamefont {Gross}},\ }\href {\doibase 10.1103/PhysRevD.17.2717} {\bibfield  {journal} {\bibinfo  {journal} {Phys. Rev. D}\ }\textbf {\bibinfo {volume} {17}},\ \bibinfo {pages} {2717} (\bibinfo {year} {1978})}\BibitemShut {NoStop}%
\bibitem [{\citenamefont {Novikov}\ \emph {et~al.}(1983)\citenamefont {Novikov}, \citenamefont {Shifman}, \citenamefont {Vainshtein},\ and\ \citenamefont {Zakharov}}]{Novikov:1983ee}%
  \BibitemOpen
  \bibfield  {author} {\bibinfo {author} {\bibfnamefont {V.~A.}\ \bibnamefont {Novikov}}, \bibinfo {author} {\bibfnamefont {M.~A.}\ \bibnamefont {Shifman}}, \bibinfo {author} {\bibfnamefont {A.~I.}\ \bibnamefont {Vainshtein}}, \ and\ \bibinfo {author} {\bibfnamefont {V.~I.}\ \bibnamefont {Zakharov}},\ }\href {\doibase 10.1016/0550-3213(83)90340-1} {\bibfield  {journal} {\bibinfo  {journal} {Nucl. Phys. B}\ }\textbf {\bibinfo {volume} {229}},\ \bibinfo {pages} {407} (\bibinfo {year} {1983})}\BibitemShut {NoStop}%
\bibitem [{\citenamefont {Affleck}\ \emph {et~al.}(1985)\citenamefont {Affleck}, \citenamefont {Dine},\ and\ \citenamefont {Seiberg}}]{Affleck:1984xz}%
  \BibitemOpen
  \bibfield  {author} {\bibinfo {author} {\bibfnamefont {I.}~\bibnamefont {Affleck}}, \bibinfo {author} {\bibfnamefont {M.}~\bibnamefont {Dine}}, \ and\ \bibinfo {author} {\bibfnamefont {N.}~\bibnamefont {Seiberg}},\ }\href {\doibase 10.1016/0550-3213(85)90408-0} {\bibfield  {journal} {\bibinfo  {journal} {Nucl. Phys. B}\ }\textbf {\bibinfo {volume} {256}},\ \bibinfo {pages} {557} (\bibinfo {year} {1985})}\BibitemShut {NoStop}%
\bibitem [{\citenamefont {Novikov}\ \emph {et~al.}(1986)\citenamefont {Novikov}, \citenamefont {Shifman}, \citenamefont {Vainshtein},\ and\ \citenamefont {Zakharov}}]{Novikov:1985rd}%
  \BibitemOpen
  \bibfield  {author} {\bibinfo {author} {\bibfnamefont {V.~A.}\ \bibnamefont {Novikov}}, \bibinfo {author} {\bibfnamefont {M.~A.}\ \bibnamefont {Shifman}}, \bibinfo {author} {\bibfnamefont {A.~I.}\ \bibnamefont {Vainshtein}}, \ and\ \bibinfo {author} {\bibfnamefont {V.~I.}\ \bibnamefont {Zakharov}},\ }\href {\doibase 10.1016/0370-2693(86)90810-5} {\bibfield  {journal} {\bibinfo  {journal} {Phys. Lett. B}\ }\textbf {\bibinfo {volume} {166}},\ \bibinfo {pages} {329} (\bibinfo {year} {1986})}\BibitemShut {NoStop}%
\bibitem [{\citenamefont {Seiberg}(1994)}]{Seiberg:1994bz}%
  \BibitemOpen
  \bibfield  {author} {\bibinfo {author} {\bibfnamefont {N.}~\bibnamefont {Seiberg}},\ }\href {\doibase 10.1103/PhysRevD.49.6857} {\bibfield  {journal} {\bibinfo  {journal} {Phys. Rev. D}\ }\textbf {\bibinfo {volume} {49}},\ \bibinfo {pages} {6857} (\bibinfo {year} {1994})},\ \Eprint {http://arxiv.org/abs/hep-th/9402044} {arXiv:hep-th/9402044} \BibitemShut {NoStop}%
\bibitem [{\citenamefont {Seiberg}(1995)}]{Seiberg:1994pq}%
  \BibitemOpen
  \bibfield  {author} {\bibinfo {author} {\bibfnamefont {N.}~\bibnamefont {Seiberg}},\ }\href {\doibase 10.1016/0550-3213(94)00023-8} {\bibfield  {journal} {\bibinfo  {journal} {Nucl. Phys. B}\ }\textbf {\bibinfo {volume} {435}},\ \bibinfo {pages} {129} (\bibinfo {year} {1995})},\ \Eprint {http://arxiv.org/abs/hep-th/9411149} {arXiv:hep-th/9411149} \BibitemShut {NoStop}%
\bibitem [{\citenamefont {Affleck}(1981)}]{Affleck:1980mp}%
  \BibitemOpen
  \bibfield  {author} {\bibinfo {author} {\bibfnamefont {I.}~\bibnamefont {Affleck}},\ }\href {\doibase 10.1016/0550-3213(81)90307-2} {\bibfield  {journal} {\bibinfo  {journal} {Nucl. Phys. B}\ }\textbf {\bibinfo {volume} {191}},\ \bibinfo {pages} {429} (\bibinfo {year} {1981})}\BibitemShut {NoStop}%
\bibitem [{\citenamefont {Ringwald}(1990)}]{Ringwald:1989ee}%
  \BibitemOpen
  \bibfield  {author} {\bibinfo {author} {\bibfnamefont {A.}~\bibnamefont {Ringwald}},\ }\href {\doibase 10.1016/0550-3213(90)90300-3} {\bibfield  {journal} {\bibinfo  {journal} {Nucl. Phys. B}\ }\textbf {\bibinfo {volume} {330}},\ \bibinfo {pages} {1} (\bibinfo {year} {1990})}\BibitemShut {NoStop}%
\bibitem [{\citenamefont {Espinosa}(1990)}]{Espinosa:1989qn}%
  \BibitemOpen
  \bibfield  {author} {\bibinfo {author} {\bibfnamefont {O.}~\bibnamefont {Espinosa}},\ }\href {\doibase 10.1016/0550-3213(90)90473-Q} {\bibfield  {journal} {\bibinfo  {journal} {Nucl. Phys. B}\ }\textbf {\bibinfo {volume} {343}},\ \bibinfo {pages} {310} (\bibinfo {year} {1990})}\BibitemShut {NoStop}%
\bibitem [{\citenamefont {Arnold}\ and\ \citenamefont {Mattis}(1991)}]{Arnold:1990pe}%
  \BibitemOpen
  \bibfield  {author} {\bibinfo {author} {\bibfnamefont {P.~B.}\ \bibnamefont {Arnold}}\ and\ \bibinfo {author} {\bibfnamefont {M.~P.}\ \bibnamefont {Mattis}},\ }\href {\doibase 10.1103/PhysRevLett.66.13} {\bibfield  {journal} {\bibinfo  {journal} {Phys. Rev. Lett.}\ }\textbf {\bibinfo {volume} {66}},\ \bibinfo {pages} {13} (\bibinfo {year} {1991})}\BibitemShut {NoStop}%
\bibitem [{\citenamefont {Khoze}\ and\ \citenamefont {Ringwald}(1991)}]{Khoze:1990bm}%
  \BibitemOpen
  \bibfield  {author} {\bibinfo {author} {\bibfnamefont {V.~V.}\ \bibnamefont {Khoze}}\ and\ \bibinfo {author} {\bibfnamefont {A.}~\bibnamefont {Ringwald}},\ }\href {\doibase 10.1016/0550-3213(91)90118-H} {\bibfield  {journal} {\bibinfo  {journal} {Nucl. Phys. B}\ }\textbf {\bibinfo {volume} {355}},\ \bibinfo {pages} {351} (\bibinfo {year} {1991})}\BibitemShut {NoStop}%
\bibitem [{\citenamefont {Silvestrov}(1994)}]{Silvestrov:1992ct}%
  \BibitemOpen
  \bibfield  {author} {\bibinfo {author} {\bibfnamefont {P.~G.}\ \bibnamefont {Silvestrov}},\ }\href {\doibase 10.1016/0370-2693(94)00059-X} {\bibfield  {journal} {\bibinfo  {journal} {Phys. Lett. B}\ }\textbf {\bibinfo {volume} {323}},\ \bibinfo {pages} {25} (\bibinfo {year} {1994})},\ \Eprint {http://arxiv.org/abs/hep-ph/9212215} {arXiv:hep-ph/9212215} \BibitemShut {NoStop}%
\bibitem [{\citenamefont {Elliott}\ and\ \citenamefont {King}(1993)}]{Elliott:1992ut}%
  \BibitemOpen
  \bibfield  {author} {\bibinfo {author} {\bibfnamefont {T.}~\bibnamefont {Elliott}}\ and\ \bibinfo {author} {\bibfnamefont {S.~F.}\ \bibnamefont {King}},\ }\href {\doibase 10.1007/BF01553021} {\bibfield  {journal} {\bibinfo  {journal} {Z. Phys. C}\ }\textbf {\bibinfo {volume} {58}},\ \bibinfo {pages} {609} (\bibinfo {year} {1993})},\ \Eprint {http://arxiv.org/abs/hep-ph/9206202} {arXiv:hep-ph/9206202} \BibitemShut {NoStop}%
\bibitem [{\citenamefont {Redi}\ and\ \citenamefont {Sato}(2016)}]{Redi:2016esr}%
  \BibitemOpen
  \bibfield  {author} {\bibinfo {author} {\bibfnamefont {M.}~\bibnamefont {Redi}}\ and\ \bibinfo {author} {\bibfnamefont {R.}~\bibnamefont {Sato}},\ }\href {\doibase 10.1007/JHEP05(2016)104} {\bibfield  {journal} {\bibinfo  {journal} {JHEP}\ }\textbf {\bibinfo {volume} {05}},\ \bibinfo {pages} {104} (\bibinfo {year} {2016})},\ \Eprint {http://arxiv.org/abs/1602.05427} {arXiv:1602.05427 [hep-ph]} \BibitemShut {NoStop}%
\bibitem [{\citenamefont {Agrawal}\ and\ \citenamefont {Howe}(2018{\natexlab{a}})}]{Agrawal:2017ksf}%
  \BibitemOpen
  \bibfield  {author} {\bibinfo {author} {\bibfnamefont {P.}~\bibnamefont {Agrawal}}\ and\ \bibinfo {author} {\bibfnamefont {K.}~\bibnamefont {Howe}},\ }\href {\doibase 10.1007/JHEP12(2018)029} {\bibfield  {journal} {\bibinfo  {journal} {JHEP}\ }\textbf {\bibinfo {volume} {12}},\ \bibinfo {pages} {029} (\bibinfo {year} {2018}{\natexlab{a}})},\ \Eprint {http://arxiv.org/abs/1710.04213} {arXiv:1710.04213 [hep-ph]} \BibitemShut {NoStop}%
\bibitem [{\citenamefont {Agrawal}\ and\ \citenamefont {Howe}(2018{\natexlab{b}})}]{Agrawal:2017evu}%
  \BibitemOpen
  \bibfield  {author} {\bibinfo {author} {\bibfnamefont {P.}~\bibnamefont {Agrawal}}\ and\ \bibinfo {author} {\bibfnamefont {K.}~\bibnamefont {Howe}},\ }\href {\doibase 10.1007/JHEP12(2018)035} {\bibfield  {journal} {\bibinfo  {journal} {JHEP}\ }\textbf {\bibinfo {volume} {12}},\ \bibinfo {pages} {035} (\bibinfo {year} {2018}{\natexlab{b}})},\ \Eprint {http://arxiv.org/abs/1712.05803} {arXiv:1712.05803 [hep-ph]} \BibitemShut {NoStop}%
\bibitem [{\citenamefont {Gaillard}\ \emph {et~al.}(2018)\citenamefont {Gaillard}, \citenamefont {Gavela}, \citenamefont {Houtz}, \citenamefont {Quilez},\ and\ \citenamefont {Del~Rey}}]{Gaillard:2018xgk}%
  \BibitemOpen
  \bibfield  {author} {\bibinfo {author} {\bibfnamefont {M.~K.}\ \bibnamefont {Gaillard}}, \bibinfo {author} {\bibfnamefont {M.~B.}\ \bibnamefont {Gavela}}, \bibinfo {author} {\bibfnamefont {R.}~\bibnamefont {Houtz}}, \bibinfo {author} {\bibfnamefont {P.}~\bibnamefont {Quilez}}, \ and\ \bibinfo {author} {\bibfnamefont {R.}~\bibnamefont {Del~Rey}},\ }\href {\doibase 10.1140/epjc/s10052-018-6396-6} {\bibfield  {journal} {\bibinfo  {journal} {Eur. Phys. J. C}\ }\textbf {\bibinfo {volume} {78}},\ \bibinfo {pages} {972} (\bibinfo {year} {2018})},\ \Eprint {http://arxiv.org/abs/1805.06465} {arXiv:1805.06465 [hep-ph]} \BibitemShut {NoStop}%
\bibitem [{\citenamefont {Cs{\'a}ki}\ \emph {et~al.}(2020)\citenamefont {Cs{\'a}ki}, \citenamefont {Ruhdorfer},\ and\ \citenamefont {Shirman}}]{Csaki:2019vte}%
  \BibitemOpen
  \bibfield  {author} {\bibinfo {author} {\bibfnamefont {C.}~\bibnamefont {Cs{\'a}ki}}, \bibinfo {author} {\bibfnamefont {M.}~\bibnamefont {Ruhdorfer}}, \ and\ \bibinfo {author} {\bibfnamefont {Y.}~\bibnamefont {Shirman}},\ }\href {\doibase 10.1007/JHEP04(2020)031} {\bibfield  {journal} {\bibinfo  {journal} {JHEP}\ }\textbf {\bibinfo {volume} {04}},\ \bibinfo {pages} {031} (\bibinfo {year} {2020})},\ \Eprint {http://arxiv.org/abs/1912.02197} {arXiv:1912.02197 [hep-ph]} \BibitemShut {NoStop}%
\bibitem [{\citenamefont {Aoki}\ \emph {et~al.}(2024)\citenamefont {Aoki}, \citenamefont {Ibe}, \citenamefont {Shirai},\ and\ \citenamefont {Watanabe}}]{Aoki:2024usv}%
  \BibitemOpen
  \bibfield  {author} {\bibinfo {author} {\bibfnamefont {T.}~\bibnamefont {Aoki}}, \bibinfo {author} {\bibfnamefont {M.}~\bibnamefont {Ibe}}, \bibinfo {author} {\bibfnamefont {S.}~\bibnamefont {Shirai}}, \ and\ \bibinfo {author} {\bibfnamefont {K.}~\bibnamefont {Watanabe}},\ }\href {\doibase 10.1007/JHEP07(2024)269} {\bibfield  {journal} {\bibinfo  {journal} {JHEP}\ }\textbf {\bibinfo {volume} {07}},\ \bibinfo {pages} {269} (\bibinfo {year} {2024})},\ \Eprint {http://arxiv.org/abs/2404.19342} {arXiv:2404.19342 [hep-ph]} \BibitemShut {NoStop}%
\bibitem [{\citenamefont {Frishman}\ and\ \citenamefont {Yankielowicz}(1979)}]{Frishman:1978xs}%
  \BibitemOpen
  \bibfield  {author} {\bibinfo {author} {\bibfnamefont {Y.}~\bibnamefont {Frishman}}\ and\ \bibinfo {author} {\bibfnamefont {S.}~\bibnamefont {Yankielowicz}},\ }\href {\doibase 10.1103/PhysRevD.19.540} {\bibfield  {journal} {\bibinfo  {journal} {Phys. Rev. D}\ }\textbf {\bibinfo {volume} {19}},\ \bibinfo {pages} {540} (\bibinfo {year} {1979})}\BibitemShut {NoStop}%
\bibitem [{\citenamefont {Klinkhamer}(1992)}]{Klinkhamer:1991pq}%
  \BibitemOpen
  \bibfield  {author} {\bibinfo {author} {\bibfnamefont {F.~R.}\ \bibnamefont {Klinkhamer}},\ }\href {\doibase 10.1016/0550-3213(92)90125-U} {\bibfield  {journal} {\bibinfo  {journal} {Nucl. Phys. B}\ }\textbf {\bibinfo {volume} {376}},\ \bibinfo {pages} {255} (\bibinfo {year} {1992})}\BibitemShut {NoStop}%
\bibitem [{\citenamefont {Klinkhamer}(1993)}]{Klinkhamer:1993kn}%
  \BibitemOpen
  \bibfield  {author} {\bibinfo {author} {\bibfnamefont {F.~R.}\ \bibnamefont {Klinkhamer}},\ }\href {\doibase 10.1016/0550-3213(93)90275-T} {\bibfield  {journal} {\bibinfo  {journal} {Nucl. Phys. B}\ }\textbf {\bibinfo {volume} {407}},\ \bibinfo {pages} {88} (\bibinfo {year} {1993})},\ \Eprint {http://arxiv.org/abs/hep-ph/9306208} {arXiv:hep-ph/9306208} \BibitemShut {NoStop}%
\bibitem [{\citenamefont {Aoyama}\ \emph {et~al.}(1996{\natexlab{a}})\citenamefont {Aoyama}, \citenamefont {Harano}, \citenamefont {Sato},\ and\ \citenamefont {Wada}}]{Aoyama:1995zt}%
  \BibitemOpen
  \bibfield  {author} {\bibinfo {author} {\bibfnamefont {H.}~\bibnamefont {Aoyama}}, \bibinfo {author} {\bibfnamefont {T.}~\bibnamefont {Harano}}, \bibinfo {author} {\bibfnamefont {M.}~\bibnamefont {Sato}}, \ and\ \bibinfo {author} {\bibfnamefont {S.}~\bibnamefont {Wada}},\ }\href {\doibase 10.1142/S0217732396000072} {\bibfield  {journal} {\bibinfo  {journal} {Mod. Phys. Lett. A}\ }\textbf {\bibinfo {volume} {11}},\ \bibinfo {pages} {43} (\bibinfo {year} {1996}{\natexlab{a}})},\ \Eprint {http://arxiv.org/abs/hep-th/9507111} {arXiv:hep-th/9507111} \BibitemShut {NoStop}%
\bibitem [{\citenamefont {Aoyama}\ \emph {et~al.}(1996{\natexlab{b}})\citenamefont {Aoyama}, \citenamefont {Harano}, \citenamefont {Sato},\ and\ \citenamefont {Wada}}]{Aoyama:1995ca}%
  \BibitemOpen
  \bibfield  {author} {\bibinfo {author} {\bibfnamefont {H.}~\bibnamefont {Aoyama}}, \bibinfo {author} {\bibfnamefont {T.}~\bibnamefont {Harano}}, \bibinfo {author} {\bibfnamefont {M.}~\bibnamefont {Sato}}, \ and\ \bibinfo {author} {\bibfnamefont {S.}~\bibnamefont {Wada}},\ }\href {\doibase 10.1016/0550-3213(96)00066-1} {\bibfield  {journal} {\bibinfo  {journal} {Nucl. Phys. B}\ }\textbf {\bibinfo {volume} {466}},\ \bibinfo {pages} {127} (\bibinfo {year} {1996}{\natexlab{b}})},\ \Eprint {http://arxiv.org/abs/hep-th/9512064} {arXiv:hep-th/9512064} \BibitemShut {NoStop}%
\bibitem [{\citenamefont {Nielsen}\ and\ \citenamefont {Nielsen}(2000)}]{Nielsen:1999vq}%
  \BibitemOpen
  \bibfield  {author} {\bibinfo {author} {\bibfnamefont {M.}~\bibnamefont {Nielsen}}\ and\ \bibinfo {author} {\bibfnamefont {N.~K.}\ \bibnamefont {Nielsen}},\ }\href {\doibase 10.1103/PhysRevD.61.105020} {\bibfield  {journal} {\bibinfo  {journal} {Phys. Rev. D}\ }\textbf {\bibinfo {volume} {61}},\ \bibinfo {pages} {105020} (\bibinfo {year} {2000})},\ \Eprint {http://arxiv.org/abs/hep-th/9912006} {arXiv:hep-th/9912006} \BibitemShut {NoStop}%
\bibitem [{\citenamefont {Elder}\ \emph {et~al.}(2025)\citenamefont {Elder}, \citenamefont {Gawrych},\ and\ \citenamefont {Rajantie}}]{Elder:2025kya}%
  \BibitemOpen
  \bibfield  {author} {\bibinfo {author} {\bibfnamefont {B.}~\bibnamefont {Elder}}, \bibinfo {author} {\bibfnamefont {K.}~\bibnamefont {Gawrych}}, \ and\ \bibinfo {author} {\bibfnamefont {A.}~\bibnamefont {Rajantie}},\ }\href@noop {} {\  (\bibinfo {year} {2025})},\ \Eprint {http://arxiv.org/abs/2510.21922} {arXiv:2510.21922 [hep-th]} \BibitemShut {NoStop}%
\bibitem [{\citenamefont {Vandoren}\ and\ \citenamefont {van Nieuwenhuizen}(2008)}]{Vandoren:2008xg}%
  \BibitemOpen
  \bibfield  {author} {\bibinfo {author} {\bibfnamefont {S.}~\bibnamefont {Vandoren}}\ and\ \bibinfo {author} {\bibfnamefont {P.}~\bibnamefont {van Nieuwenhuizen}},\ }\href@noop {} {\  (\bibinfo {year} {2008})},\ \Eprint {http://arxiv.org/abs/0802.1862} {arXiv:0802.1862 [hep-th]} \BibitemShut {NoStop}%
\bibitem [{\citenamefont {Shifman}(2012)}]{Shifman:2012zz}%
  \BibitemOpen
  \bibfield  {author} {\bibinfo {author} {\bibfnamefont {M.}~\bibnamefont {Shifman}},\ }\href {\doibase 10.1017/9781108885911} {\emph {\bibinfo {title} {{Advanced topics in quantum field theory.}: {A lecture course}}}}\ (\bibinfo  {publisher} {Cambridge Univ. Press},\ \bibinfo {address} {Cambridge, UK},\ \bibinfo {year} {2012})\BibitemShut {NoStop}%
\bibitem [{\citenamefont {Gervais}\ \emph {et~al.}(1978)\citenamefont {Gervais}, \citenamefont {Neveu},\ and\ \citenamefont {Virasoro}}]{Gervais:1977me}%
  \BibitemOpen
  \bibfield  {author} {\bibinfo {author} {\bibfnamefont {J.-L.}\ \bibnamefont {Gervais}}, \bibinfo {author} {\bibfnamefont {A.}~\bibnamefont {Neveu}}, \ and\ \bibinfo {author} {\bibfnamefont {M.~A.}\ \bibnamefont {Virasoro}},\ }\href {\doibase 10.1016/0550-3213(78)90156-6} {\bibfield  {journal} {\bibinfo  {journal} {Nucl. Phys. B}\ }\textbf {\bibinfo {volume} {138}},\ \bibinfo {pages} {45} (\bibinfo {year} {1978})}\BibitemShut {NoStop}%
\end{thebibliography}%

\end{document}